\documentclass[twocolumn]{aastex631}
\usepackage[varg]{txfonts}
\usepackage{graphicx}
\usepackage{rotating}
\usepackage[printonlyused,withpage]{acronym}

\usepackage{ulem}
\usepackage{amsmath}
\usepackage{amssymb}
\usepackage{graphicx}
\usepackage{inputenc}
\usepackage{longtable}
\usepackage{float}
\usepackage{hyperref}
\usepackage{xcolor}
\accepted{28 July 2026 -- The Astrophysical Journal}
\shorttitle{The GPSP Atlas}
\shortauthors{E.O.~Angüner}
\begin{document}

\title{The GPSP Atlas: High-resolution Galactic Photon Survival Probability Atlas for Very-High- and Ultra-High-Energy Gamma-Ray Astronomy}
\author[0000-0002-4712-4292]{Ekrem Oğuzhan Angüner}
\affiliation{TÜBİTAK Research Institute for Fundamental Sciences, 41470 Gebze, Kocaeli, Türkiye}
\email{oguzhan.anguner@tubitak.gov.tr \\ oguzhananguner@gmail.com}
%% Mark off the abstract in the ``abstract'' environment. 
\begin{abstract}
The emergence of Galactic PeV astronomy makes photon--photon ($\gamma\gamma$) absorption an increasingly important component of gamma-ray data analysis. In this work, the Galactic Photon Survival Probability (GPSP) atlas is presented as the first Galactic framework describing gamma-ray survival probabilities over the full parameter space $P(l,b,d,E_\gamma)$ between 1 TeV and 10 PeV. The atlas is constructed using two independent GALPROP-based interstellar radiation field (ISRF) models, allowing direct assessment of ISRF-related systematic uncertainties, and is released in a FITS format compatible with standard analysis frameworks. A novel gamma-ray survival phase space representation is introduced, extending 1D survival profiles into a 2D energy--distance framework that visualizes Galactic transparency and gamma-ray horizons. Using the full atlas, Galactic transparency and Galactic plane survival maps are constructed. These reveal that Galactic opacity is highly structured, with enhanced attenuation along spiral-arm tangent directions in the $\sim$100--300 TeV regime, while attenuation becomes nearly isotropic and cosmic microwave background (CMB) dominated near PeV energies. The framework naturally accommodates attenuation calculations for extended and diffuse sources. The atlas is further extended to subluminal Lorentz invariance violation (LIV) scenarios through the GPSP-LIV atlas, incorporating LIV-modified pair production kinematics and cross sections. Sources emitting in the 2--3 PeV regime beyond gamma-ray horizons are identified as promising targets for probing LIV effects. Finally, practical applications are presented for the Galactic Center PeVatron and the Cygnus X-3 regions. The GPSP and GPSP-LIV atlases establish a publicly available foundation for studies of Galactic PeVatrons, diffuse gamma-ray emission, and fundamental physics searches in the emerging era of Galactic PeV astronomy.
\end{abstract}


\section{Introduction} \label{sec:intro}
Very-high-energy (VHE; 0.1~TeV$<E<$100~TeV) gamma-ray astronomy has rapidly developed since the first detection of TeV emission from the Crab Nebula \citep{crab_first_TeV}, evolving from individual source discoveries into population studies of Galactic particle accelerators.~Currently, there are 287 TeV sources\footnote{This number is taken from the TeVCat catalog on 25.05.2026. Please see \url{https://www.tevcat.org/}},~most of which are located along the Galactic plane, detected largely through systematic survey programs and pointed observations performed by current-generation Imaging Atmospheric Cherenkov Telescopes (IACTs), including H.E.S.S.~\citep{hess_Crab_paper,hess_gps}, MAGIC~\citep{magic_crab}, and VERITAS~\citep{veritas_main_paper}. Together, these instruments have revealed a wide range of Galactic gamma-ray sources, such as pulsars and pulsar wind nebulae (PWN), supernova remnants (SNRs), gamma-ray binaries, microquasars, star-forming regions (SFRs), and globular clusters. 

Despite their excellent angular and energy resolution, IACTs are limited at the highest energies, particularly above $\sim$50~TeV. Their effective collection areas, together with limited duty cycles restricted to clear and moonless nights, do not provide sufficient photon statistics for the steeply falling gamma-ray fluxes, particularly in the sub-PeV regime. Exploration of the ultra-high-energy (UHE;~$E>100$~TeV) domain therefore requires extensive air-shower arrays with effective areas exceeding 1~$\mathrm{km}^{2}$.~Such capabilities are provided by wide-field observatories based on water Cherenkov detectors (WCDs) and scintillator arrays, which operate with high duty cycles and large instantaneous sky coverage, at the expense of reduced angular resolution compared to IACTs.~Early progress toward the sub-PeV regime was achieved by the Tibet AS$\gamma$ experiment~\citep{tibet_as_subPeV}, followed by the HAWC Galactic plane survey extending sensitivity above 56~TeV~\citep{hawc_56_TeV}. A major breakthrough was later achieved by the LHAASO observatory~\citep{lhaaso_main_crab}, which reported 12 sources emitting beyond 100~TeV, including the first detection of a 1.4~PeV photon from the Cygnus Cocoon region~\citep{lhaaso_UHE_paper}. The subsequent release of the first LHAASO source catalog~\citep{lhaaso_1st_catalog}, containing 43 sources detected above 100~TeV, now allows population studies of extreme Galactic accelerators.

Interpretation of VHE and UHE gamma-ray observations is fundamentally limited by the opacity of the Galaxy itself. During propagation through the interstellar medium (ISM), gamma rays undergo attenuation via electron--positron pair production ($\gamma\gamma \rightarrow e^{+}e^{-}$) through interactions with background photon fields~\citep{Zhang2006, Moskalenko_2006, vernetto, Zhang_2026}, a process also known as '$\gamma\gamma$ absorption'. The interstellar radiation field (ISRF), consisting of stellar and infrared (IR) dust emission, introduces a spatially dependent absorption that can significantly suppress observed spectra between $\sim$30 and 200~TeV. Early calculations by \citet{Zhang2006} demonstrated that attenuation toward the Galactic Center may become observable above $\sim$20~TeV and can reach the $\sim$10$\%$ level around 50~TeV. These absorption effects produce the characteristic ``infrared shoulder'' feature in gamma-ray survival probability profiles (see also, e.g., Fig.~12 in \citealt{vernetto}). At higher energies, the cosmic microwave background (CMB) becomes the dominant source of attenuation for sub-PeV and PeV photons. Together, these radiation fields form a ``Galactic fog'' that masks the intrinsic spectral properties of Galactic accelerators. Similar absorption processes can in principle take place within source environments, for instance, in compact binary systems such as Cygnus~X-3, where dense ultraviolet (UV) and X-ray radiation fields can strongly attenuate gamma rays produced in the inner regions~\citep{cygnus_x3_lhaaso}.

Correcting observed spectra for $\gamma\gamma$ absorption is particularly important for identifying Galactic PeVatrons, namely sources capable of accelerating particles to PeV energies and beyond (see \citealt{deOnaWilhelmi2024,pev_annual_rev,ozi_review} for reviews). The spectral signatures of these extreme accelerators appear predominantly in the UHE regime \citep{Celli_2020, ozi_GCR_mnras}. However, both leptonic and hadronic mechanisms can produce UHE gamma rays. In the leptonic scenario, relativistic electrons produce gamma rays through inverse Compton (IC) scattering of ambient photon fields, while in the hadronic scenario gamma rays originate from proton--proton ($pp$) or proton--photon ($p\gamma$) photohadronic interactions followed by subsequent neutral pion decay \citep{kelner_2006,kafexhiu2014}. Consequently, the detection of photons above 100~TeV alone does not necessarily imply the presence of a PeVatron~\citep{cta_pevatrons}. Reconstruction of the intrinsic gamma-ray spectrum, corrected for propagation effects, is therefore required to constrain the underlying parent particle population and determine whether acceleration beyond PeV energies is achieved. However, distinguishing between hadronic and leptonic scenarios additionally requires multiwavelength and multimessenger observations, including X-ray synchrotron emission, dense gas tracers for detecting molecular clouds, and simultaneous neutrino detections.

Beyond individual accelerators, Galactic gamma-ray absorption is also relevant for studies of the Galactic diffuse emission (GDE)~\citep{lipari_2018}. The GDE originates primarily from interactions between Galactic cosmic rays (CRs) and interstellar gas, IC scattering of CR electrons on the ISRF, and unresolved populations of gamma-ray sources~\citep{Vecchiotti_2025, Zhang_2023, He_2024, Yan_2023}. Recently, the LHAASO Collaboration reported measurements of the GDE between 10~TeV and 1~PeV from two different regions of the Galactic plane~\citep{lhaaso_GDE}. Since the mean free path of UHE photons becomes comparable to Galactic scales, $\gamma\gamma$ absorption can significantly modify the observed diffuse spectrum. Accurate modeling of Galactic opacity is therefore essential for reconstructing the intrinsic diffuse emission and comparing it with Galactic CR propagation models and local CR measurements.

The propagation of UHE gamma rays also provides a unique laboratory for testing fundamental physics beyond the Standard Model~\citep{lhaaso_LIV_2022, lhaaso_LIV_2024}. Lorentz invariance violation (LIV), predicted in several quantum gravity scenarios, introduces modifications to the photon dispersion relation at energies approaching the Planck scale~\citep{LIV_1989, LIV_2001, test_of_QG_1998, qg_mms_review}. Depending on the sign of the modification, LIV scenarios are generally classified as superluminal or subluminal\footnote{For the superluminal case, the photon phase velocity exceeds the standard relativistic limit, while for the subluminal case, it is reduced.}. In the subluminal case, the kinematic threshold for pair production is shifted, modifying the transparency of the Galaxy to UHE gamma rays~\citep{carmona_2024, Abdalla_2019, Li_2023}. Since Galactic attenuation is dominated by the $\gamma\gamma \rightarrow e^{+}e^{-}$ process, subluminal LIV effects can emerge as measurable deviations from the expected absorption features. Incorporating modified pair production kinematics into a Galactic photon survival probability (GPSP) framework is therefore important for distinguishing purely astrophysical signatures from possible new physics effects.

In order to reconstruct the intrinsic spectra of Galactic UHE gamma-ray sources, it is necessary to determine the survival probability $P(E,l,b,d)$, which depends on the photon energy $E$ and the source position in Galactic coordinates $(l,b,d)$, assuming an observer located at the solar position.~These corrections are becoming increasingly important with the growing number of VHE and UHE detections by wide-field observatories such as LHAASO and HAWC, and will be essential for future measurements with next-generation facilities including the Cherenkov Telescope Array (CTA;~\citealt{science_with_CTA, cta_gps}) and the Southern Wide-field Gamma-ray Observatory (SWGO;~\citealt{swgo_science, swgo_pevatrons}). Although individual absorption corrections have been applied in specific source studies (see, e.g., \citealt{cygnus_x3_lhaaso, Zhang_2026}), a unified, high-resolution GPSP framework is required to standardize the analysis of the rapidly growing UHE source population.

In this work, publicly available Galactic radiation field models implemented in the GALPROP framework~\citep{Porter_2022}, namely the R12~\citep{Robitaille_2012} and F98~\citep{Freudenreich_1998} ISRF models, are used to construct a high-resolution GPSP atlas. The atlas is designed as a ready-to-use, community-oriented data product for VHE and UHE gamma-ray astronomy, providing precomputed 4D survival probability lookup tables in FITS format as functions of photon energy and source position in Galactic coordinates. The GPSP atlas covers the Galactic plane within $|b|\le5^\circ$, distances up to 20~kpc, and photon energies between 1~TeV and 10~PeV. In addition, a LIV-modified version of the atlas (GPSP-LIV) is constructed to investigate the impact of subluminal LIV scenarios on Galactic gamma-ray propagation.

The paper is organized as follows.~Section~\ref{gama_ray_opacity} describes the pair production formalism, including the discussion of LIV effects, together with the ISRF models. Section~\ref{gpsp_atlas_construction} presents the construction and structure of the GPSP atlas data product. The resulting Galactic transparency maps, survival probability phase spaces, and gamma-ray horizons are discussed in Section~\ref{gpsp_atlas_results}. Section~\ref{gpsp_liv_section} introduces the GPSP-LIV framework and presents the corresponding results for subluminal LIV scenarios. Finally, Section~\ref{gpsp_application} demonstrates practical applications of both GPSP and GPSP-LIV atlases to the Galactic Center PeVatron and Cygnus~X-3 regions.

\section{Physics of Gamma-Ray Opacity}
\label{gama_ray_opacity}

The key process responsible for the absorption of VHE and UHE gamma rays during their propagation through the ISM is photon--photon pair production ($\gamma\gamma \rightarrow e^+e^-$), in which gamma rays interact with low-energy diffuse background photons from the Galactic ISRFs and the CMB. The attenuation is conveniently described through the photon survival probability concept, $P_{\rm surv}$, which represents the probability that a gamma ray emitted at a given location reaches the observer without undergoing pair production interactions. For a gamma ray of energy $E_{\gamma}$ propagating over a distance $L$, the survival probability is given by
\begin{equation}
\label{eq1}
P_{\text{surv}}(E_{\gamma}, L) = \exp\left[-\tau(E_{\gamma}, L)\right],
\end{equation}
where $\tau(E_{\gamma}, L)$ is the dimensionless optical depth accumulated along the propagation path. This framework allows the transparency of the Galaxy to be directly mapped as a function of energy and position, making the survival probability the fundamental quantity for constructing all-sky transparency maps. 

The optical depth provides a macroscopic measure of the medium's opacity\footnote{$\tau \ll 1$ corresponds to a transparent medium, while $\tau \gg 1$ indicates strong attenuation of gamma rays.} to gamma rays, and depends strongly on the density and spatial distribution of target photons, the interaction geometry, and the pair production cross section. Assuming a locally isotropic radiation field, the optical depth for a gamma ray of energy $E_\gamma$ propagating over a distance $L$ can be written as
\begin{equation}
\label{opt_depth}
\tau(E_\gamma, L) = \int_0^{L} d\ell \int_{\epsilon_{\rm thr}}^{\infty} d\epsilon \int_{-1}^{1} d\mu \frac{1}{2}(1-\mu) n(\epsilon,\mathbf{r}) \sigma_{\gamma\gamma}(E_\gamma,\epsilon,\mu),
\end{equation}
where $n(\epsilon,\mathbf{r})$ is the differential number density of target photons, $\epsilon$ is the target photon energy, $\sigma_{\gamma\gamma}(E_\gamma,\epsilon,\mu)$ is the pair production cross section, and $\mu=\cos\theta$, with $\theta$ representing the collision angle between the gamma ray and the target photon. The factor $(1-\mu)$ accounts for the collision geometry between gamma-rays and target photons, while the factor $1/2$ arises from averaging over an isotropic distribution of target photon directions. Throughout this work, the ISRF is represented by the angle-integrated spectral photon density provided by GALPROP, implicitly adopting the locally isotropic approximation.

\begin{figure*}[ht!]
\centering
\includegraphics[width=18.0cm]{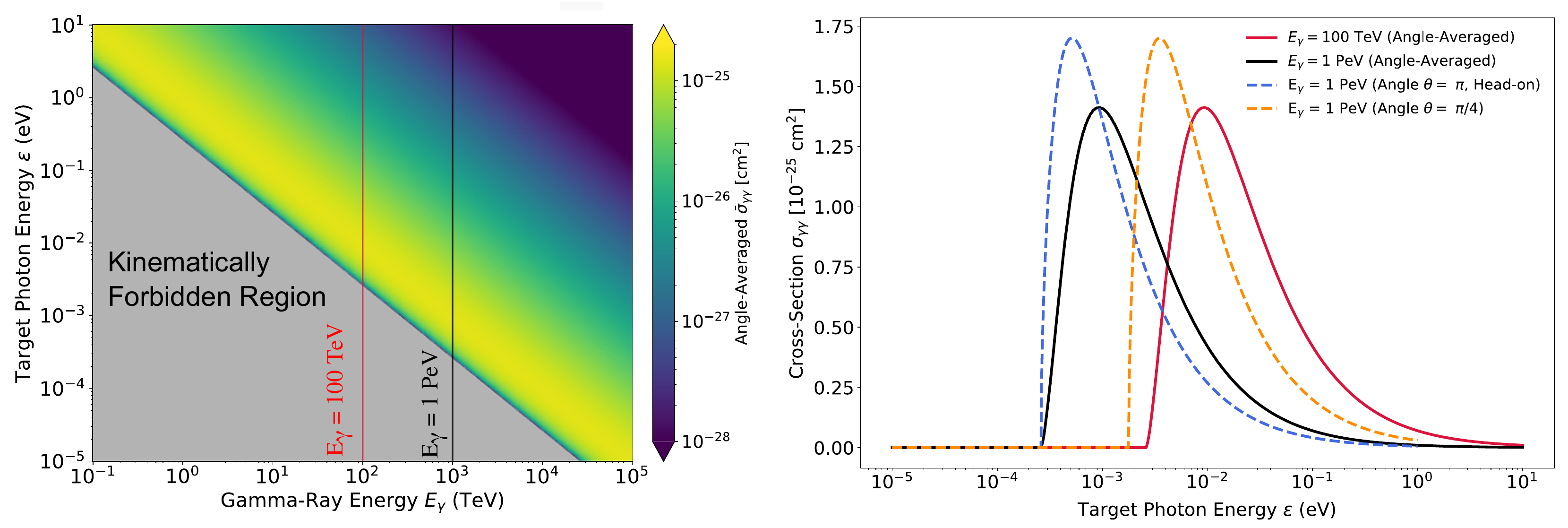}
\caption{Breit--Wheeler cross section for $\gamma\gamma$ interactions. \textbf{Left:} 2D distribution of the angle-averaged cross section ($\sigma_{\gamma\gamma}$) as a function of gamma-ray energy $E_\gamma$ and target photon energy $\epsilon$. The gray region indicates the kinematically forbidden region below the pair production threshold ($s<1$). Vertical markers indicate the selected energy slices at $E_\gamma = 100$~TeV (solid red) and $1$~PeV (solid black). \textbf{Right:} Cross section profiles corresponding to these slices. The red and black solid lines show the isotropic-averaged cross sections for 100~TeV and 1~PeV gamma rays, respectively. For comparison, the blue and orange dashed lines represent the cross section for a 1~PeV gamma ray at fixed collision angles of $\theta=\pi$ (head-on) and $\theta=\pi/4$, respectively.}
\label{bw_cross_sections}
\end{figure*}

Although the Galactic ISRF is intrinsically anisotropic, its angular structure introduces only a secondary correction to the large-scale opacity calculations.~As discussed by \citet{vernetto}, deviations from the isotropic approximation typically modify the interaction rate at the level of $\sim$10$\%$--20$\%$, with the largest effects occurring in the innermost Galactic regions where strong radiation gradients are present. Outside these regions, the impact of anisotropy is significantly reduced, and the isotropic approximation provides a reliable description of the target photon field.~Furthermore, the importance of ISRF anisotropy is energy dependent. In the $\sim$30--200~TeV range, where interactions with IR and optical photons dominate, angular variations in the ISRF can introduce moderate systematic effects. However, at higher energies approaching the PeV level, the dominant target field becomes the CMB, which is intrinsically isotropic. Overall, the isotropic approximation is well justified for the purpose of constructing large-scale GPSP maps.

It is important to note that gamma-ray propagation is assumed to follow straight-line trajectories, and secondary electromagnetic cascade effects are neglected in this work. Consequently, the survival probabilities given in Eq.~\ref{eq1} describe only the attenuation of the primary gamma-ray component through pair production interactions. In principle, $\gamma\gamma$ interactions produce electron--positron pairs that subsequently generate secondary IC emission. This secondary emission may partially reinject photons into the observable spectrum and modify the resulting spectral shape \citep{Aharonian_1994,DiMarco2025}. For lines of sight with modest optical depth ($\tau_\gamma \lesssim 1$), the effect is expected to be negligible. However, for large optical depths ($\tau_\gamma \gg 1$) particularly in the inner Galaxy and at energies approaching the PeV regime, electromagnetic cascades may become increasingly important. The resulting secondary $e^\pm$ pairs are expected to experience deflections in the Galactic magnetic field, potentially redistributing part of the reprocessed emission over extended angular scales rather than along the original line of sight \citep{Elyiv2009}. A quantitative assessment of these effects requires a self-consistent treatment of electromagnetic cascades and is beyond the scope of this work.

The calculation of the optical depth in Eq.~\ref{opt_depth} requires a detailed description of both the target photon fields and the pair production interaction cross section. In the following subsections, the Breit--Wheeler interaction formalism, LIV-induced modifications to the cross section, and ISRF models used in the GPSP atlas are described in detail.

\subsection{Standard Breit-Wheeler Cross Section}
\label{br_cs}

Photon--photon pair production is a purely quantum electrodynamic (QED) process in which two photons interact to produce an electron--positron pair once the center-of-mass energy exceeds the kinematic threshold. The interaction process was first derived by Breit and Wheeler in 1934 \citep{breit_wheeler_CS}, and provides the fundamental mechanism governing the attenuation of VHE and UHE gamma rays in diffuse Galactic radiation fields.

This microscopic interaction between the photons is described by the Breit--Wheeler cross section for pair production and depends on the dimensionless invariant
\begin{equation}
s = \frac{E_{\gamma}\,\epsilon}{2 m_e^2 c^4}(1-\cos\theta),
\end{equation}
where $E_{\gamma}$ is the gamma-ray energy, $\epsilon$ is the target photon energy, and $\theta$ is the collision angle between the photons. It is important to note that pair production is kinematically allowed only for $s \geq 1$. Above this threshold, the Breit--Wheeler cross section is given by
\begin{equation}
\sigma_{\gamma\gamma}(\beta) = \frac{1}{2} \pi r_e^2 (1-\beta^2) 
\left[ (3-\beta^4)\ln\left(\frac{1+\beta}{1-\beta}\right) - 2\beta(2-\beta^2) \right],
\end{equation}
where $r_e$ is the classical electron radius and
\begin{equation}
\beta = \sqrt{1-\frac{1}{s}}
\end{equation}
is the velocity of the produced leptons in the center-of-mass frame.

The main characteristics of the Breit--Wheeler cross section are illustrated in Fig.~\ref{bw_cross_sections}. The left panel shows the 2D distribution of the angle-averaged cross section over the gamma-ray energy range from 0.1~TeV to 100~PeV, and target photon energies between $10^{-5}$~eV and 10~eV, assuming isotropic interactions. The sharp diagonal threshold separating the forbidden and allowed regions directly visualizes the pair production threshold condition. Above threshold, the cross section rises rapidly, reaches a maximum near $s\sim2$, and decreases approximately as $\ln(s)/s$ at higher energies. The right panel presents cross section profiles for gamma-ray energies of 100~TeV and 1~PeV. These profiles show that pair production is most efficient for target photon energies close to the threshold condition, leading to a strong coupling between the gamma-ray energy and the characteristic energy of the target photon field. In particular, inverse scaling is evident, and interactions of 1~PeV gamma rays peak within the CMB energy range ($\sim10^{-3}$~eV), whereas 100~TeV gamma rays interact most efficiently with IR photons.

A key feature of the Breit--Wheeler cross section is that its maximum value is determined only by the invariant $s$. For fixed collision angles, the Breit--Wheeler cross section reaches a characteristic peak value of $\sim$1.7$\times10^{-25}~\mathrm{cm}^{2}$ once the invariant $s$ approaches its optimal value, independent of the specific combination of $E_\gamma$, $\epsilon$, and $\theta$. In contrast, the isotropic-averaged cross section exhibits a lower maximum value of $\sim$1.45$\times10^{-25}~\mathrm{cm}^{2}$. This reduction arises from the angle averaging over different interaction geometries, which kinematically broadens the interaction profile across a wider range of target photon energies. Indeed, this isotropic peak value remains approximately constant across the VHE and UHE regimes, ensuring a consistent maximum interaction strength over the full energy range considered.

\subsection{LIV-induced Modifications to the Pair production Cross Section}
\label{liv_cross_section}

Possible deviations from Lorentz invariance (LI) provide a potential source of modifications to gamma-ray propagation at extreme energies. Several quantum gravity scenarios suggest that the standard relativistic dispersion relation is modified at energies approaching the Planck scale, leading to observable effects in the propagation and interaction of UHE gamma rays.

In the subluminal, quadratic ($n=2$) case considered in this study, the modified photon dispersion relation can be written as
\begin{equation}
E^2 - p^2 \simeq -\frac{E^4}{\Lambda^2},
\end{equation}
where $\Lambda$ denotes the effective LIV energy scale. The focus on the quadratic  subluminal scenario is motivated by both observational constraints and its relevance for UHE gamma-ray propagation. In contrast to the subluminal case, superluminal LIV scenarios lead to qualitatively different effects, mostly the decay of high-energy photons into electron--positron pairs in vacuum, leading to sharp spectral cutoffs observable at the highest energies~\citep{superluminal_LIV_2016,superluminal_LIV_2017}. The observations of PeV photons by the LHAASO experiment have placed strong limits on such scenarios. In particular, linear ($n=1$) corrections are constrained to energy scales $E_{\rm LIV}^{(1)} \gtrsim 10^{5}\,E_{\rm Pl}$ ($\sim 10^{24}$~GeV), effectively pushing any observable effects far beyond the Planck scale. Quadratic superluminal corrections are also constrained, but much less strongly, with limits at the level of $E_{\rm LIV}^{(2)} \gtrsim 10^{-3}\,E_{\rm Pl}$ ($\sim10^{16}$~GeV) \citep{lhaaso_LIV_2022}. As a result, for studies of gamma-ray propagation and absorption, the quadratic subluminal scenario remains the most promising regime in which observable deviations from standard QED predictions may still arise. 

\begin{figure*}
\centering
\includegraphics[width=18.0cm]{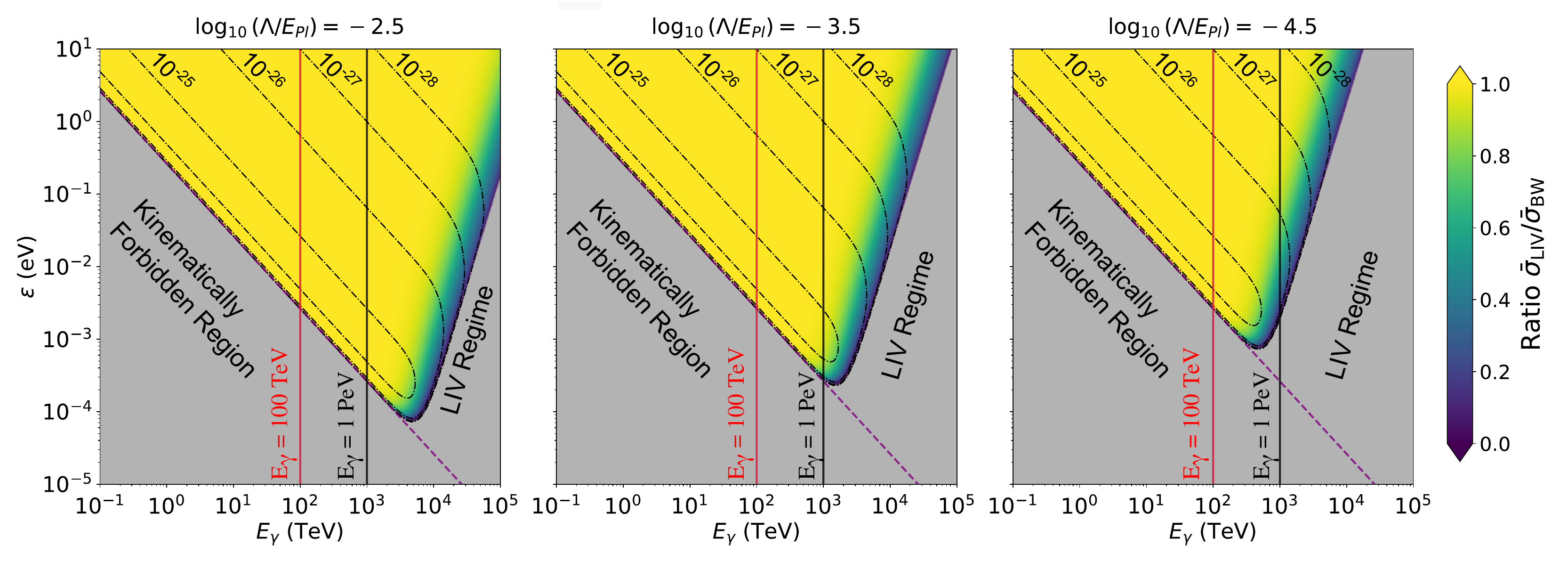}
\caption{Ratio maps of the LIV-modified cross section relative to the standard Breit--Wheeler cross section ($\bar{\sigma}_{\rm LIV}/\bar{\sigma}_{\rm BW}$) for different LIV strengths of $\log_{10}(\Lambda/E_{\rm Pl})=-2.5$ (left), $-3.5$ (center), and $-4.5$ (right). The horizontal and vertical axes show the gamma-ray energy and target photon energy, respectively. Gray areas correspond to kinematically forbidden regions, where the pair production is suppressed. Black dashed contours indicate absolute LIV cross section values, while the dashed diagonal purple lines mark the standard LI pair production threshold. The red and black solid lines indicate reference gamma-ray energies of 100~TeV and 1~PeV, respectively. Increasing LIV strength progressively suppresses the pair production cross section and expands the kinematically forbidden regions marked as ``LIV regime'' in the plots.}
\label{liv_2d_cross_sections}
\end{figure*}

Within standard QED, the Breit--Wheeler interaction is fully determined by the invariant $s$. However, in the presence of LIV effects, the interaction additionally depends on a new parameter. Following the explicit calculation derived in~\citep{carmona_2024}, denoted therein as $\mathcal{F}_{LIV}^{(expl)}$, the LIV effects enter the cross section through a dimensionless parameter $\bar{\mu}$, which quantifies the modification of the interaction dynamics
\begin{equation}
\bar{\mu} = \frac{E_{\gamma}^4}{4 m_e^2 c^4 \Lambda^2},
\end{equation}
where $E_{\gamma}$ is the gamma-ray energy. In this framework, the standard dimensionless invariant $s$ is replaced by a modified dynamical variable $\bar{\tau}$, defined as
\begin{equation}
\bar{\tau} = s - \bar{\mu}.
\end{equation}
The kinematic threshold condition for pair production is consequently shifted to $\bar{\tau} \geq 1$, while the velocity of the produced leptons in the center-of-mass frame is now given by
\begin{equation}
\beta = \sqrt{1 - \frac{1}{\bar{\tau}}}.
\end{equation}
Above this modified threshold ($\bar{\tau} > 1$), the explicit cross section $\sigma_{\rm LIV}$ is governed by the function $\mathcal{F}_{LIV}^{(expl)}(\bar{\tau}, \bar{\mu})$ (see Eq.~22 in~\cite{carmona_2024}), which accounts for both the threshold shift and the dynamical suppression of the peak intensity, and is given by
\begin{equation}
\sigma_{\rm LIV}(s, E_{\gamma}) = \frac{\pi \alpha^2 \hbar^2}{2 m_e^2 c^2 s} \left[ \mathcal{A}_1 \ln\left(\frac{1+\beta}{1-\beta}\right) - \mathcal{A}_2 \beta \right],
\label{eq_liv_cs_carmona}
\end{equation}
where $\alpha$ is the fine-structure constant. The dimensionless coefficients $\mathcal{A}_1$ and $\mathcal{A}_2$ are expressed as
\begin{equation}
\mathcal{A}_1 = 2 + \frac{2\bar{\tau}(1-2\bar{\mu}) - (1-\bar{\mu})}{(\bar{\tau}+\bar{\mu})^2},
\end{equation}
\begin{equation}
\mathcal{A}_2 = 2 + \frac{2\bar{\tau}(1-4\bar{\mu})}{(\bar{\tau}+\bar{\mu})^2}.
\end{equation}

\begin{figure}
\centering
\includegraphics[width=8.5cm]{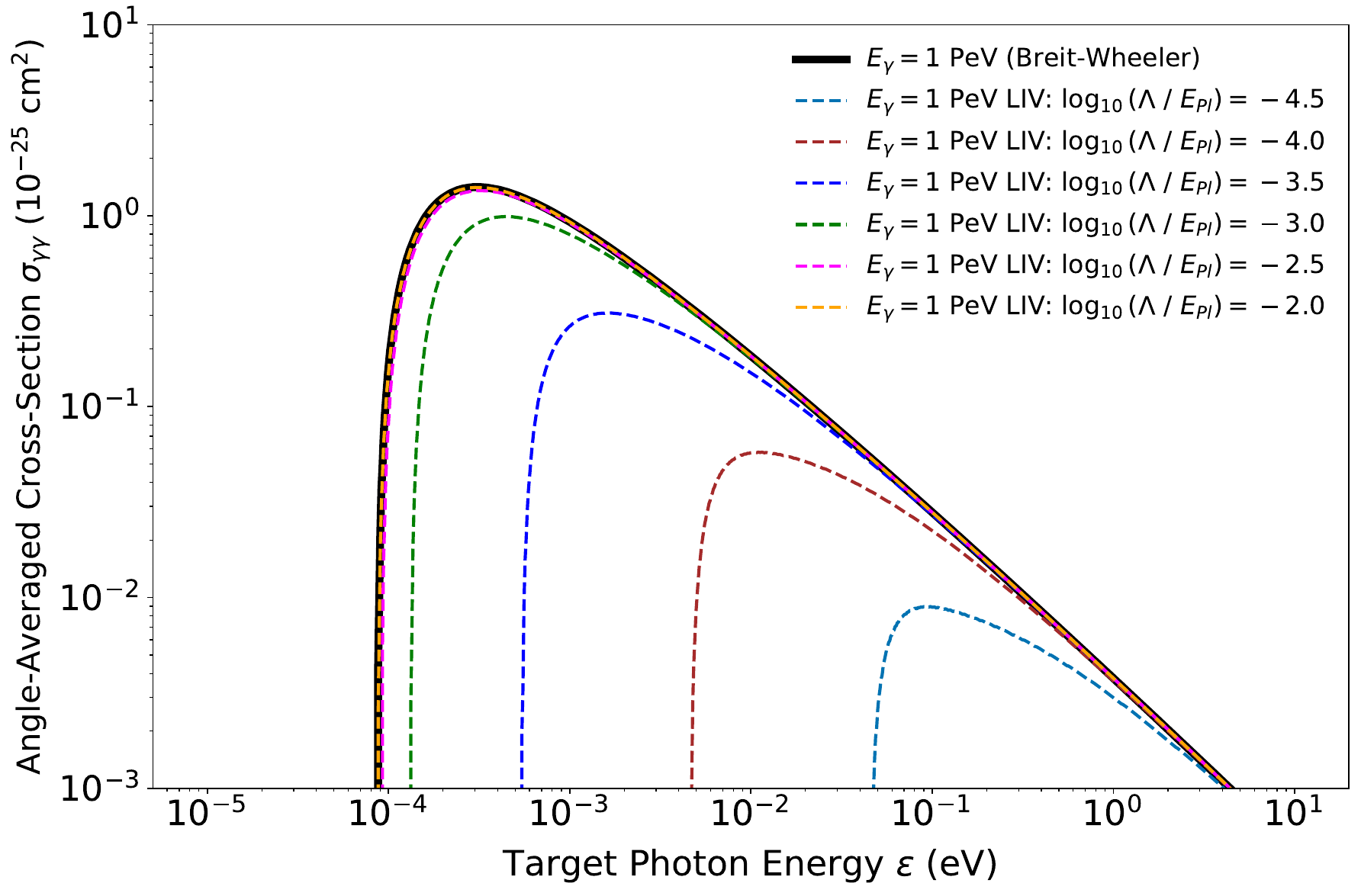}
\caption{Angle-averaged pair production cross section profiles as a function of target photon energy $\epsilon$ at a fixed gamma-ray energy of $E_\gamma=1$~PeV. The solid black line shows the standard Breit--Wheeler cross section, while the dashed colored curves represent LIV-modified cross sections for LIV strengths $\log_{10}(\Lambda/E_{\rm Pl})=-4.5$ to $-2.0$ in steps of 0.5.}
\label{liv_profiles}
\end{figure}

As discussed in detail in \citet{carmona_2024}, this explicit formulation introduces two main physical modifications to the interaction physics.~First, as $\Lambda$ becomes smaller (i.e.,the LIV effect becomes stronger), the value of $\bar{\mu}$ increases rapidly with $E_{\gamma}$, causing a significant shift of the $\gamma\gamma$ interaction threshold toward higher target photon energies.~This means that the threshold condition $\bar{\tau}\geq$~1 implies that interactions that would be kinematically allowed in the Breit--Wheeler case ($s\geq$~1) can now become forbidden, therefore effectively opening a new transparency window for UHE gamma rays. This behavior is clearly visualized in the ratio maps of the LIV-modified cross section to the standard Breit-Wheeler cross section ($\bar{\sigma}_{\mathrm{LIV}} / \bar{\sigma}_{\mathrm{BW}}$) shown in Fig.~\ref{liv_2d_cross_sections} for a given set of $\log_{10}(\Lambda/E_{Pl})$ values of -2.5 (left panel), -3.5 (middle panel), and -4.5 (right panel).~As it can be seen from the plots, as the LIV effect becomes stronger, a new kinematically forbidden LIV regime opens up for PeV gamma rays. Second, the interaction is no longer governed by the characteristic peak value determined by the invariant $s$. Even above threshold, the peak value and shape of the cross section become energy dependent through the new parameter $\bar{\mu}$. As shown in Fig.~\ref{liv_profiles}, the peak cross section at $E_{\gamma}=1$~PeV can be reduced by tens of percent depending on the LIV strength. 

In principle, both effects act in the same direction, reducing the overall opacity and increasing the transparency of the Galaxy to UHE gamma rays.~Particularly, the LIV effects for $\log_{10}(\Lambda/E_{Pl})$~$<$~-2.5 cases may produce observable signatures that can be detected with current LHAASO and future SWGO experiments, especially at PeV energies. 

\subsection{The Interstellar Radiation Fields}
\label{isrf_sect}

\begin{figure*}[ht!]
\centering
\includegraphics[width=18.0cm]{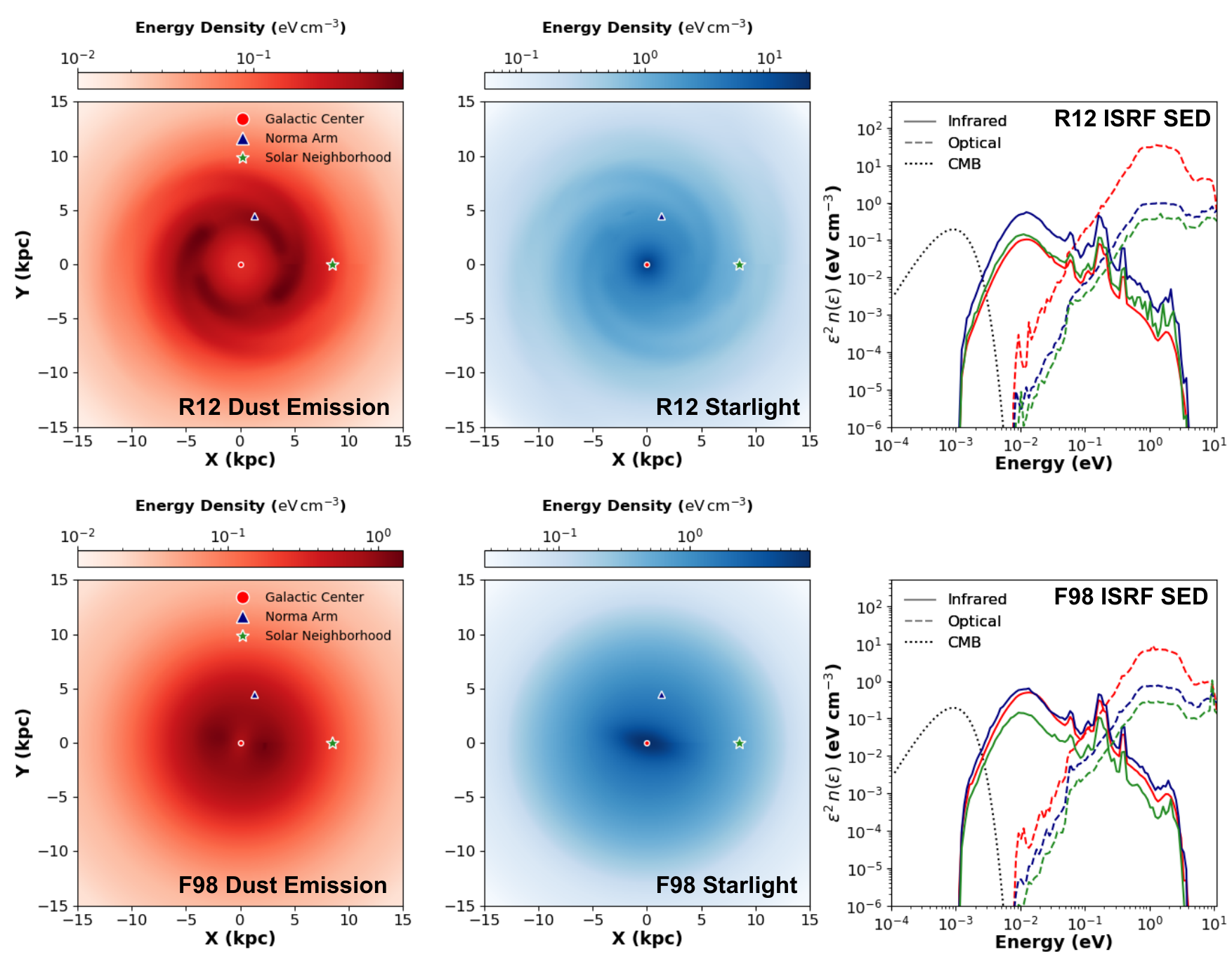}
\caption{Comparison of the R12 (top row) and F98 (bottom row) ISRF models. The left and center panels show the spatial distributions of the IR dust-emission and optical starlight energy densities, respectively, in Galactocentric coordinates ($X,Y$). Colored markers indicate the locations used for spectral energy distribution (SED) extraction: the Galactic Center (red circle), an example position in the Norma Arm (blue triangle), and the solar neighborhood (green star). The right panels show the corresponding SEDs, expressed as $\epsilon^{2}n(\epsilon)$ ($\mathrm{eV\,cm^{-3}}$) as a function of target photon energy $\epsilon$, including the CMB component. Solid, dashed, and dotted lines represent the IR, optical, and CMB components, respectively, with colors corresponding to the locations indicated in the spatial maps.}
\label{isrfmaps}
\end{figure*}

The optical depth calculation in Eq.~\ref{opt_depth} requires a detailed description of the low-energy target photon fields within the Milky Way. These radiation fields, denoted as $n(\epsilon,\mathbf{r})$, are primarily composed of the CMB, IR emission from interstellar dust, and stellar radiation (starlight) spanning optical and UV wavelengths. The extragalactic background light (EBL) is relevant only for extragalactic propagation \citep{Franceschini2008, Dominguez2011}, contributing negligibly to gamma-ray attenuation on Galactic scales, and is therefore excluded in this work.

The CMB is the dominant target photon field at the highest energies. With a blackbody temperature of $T_{\rm CMB}=2.7255$~K \citep{Fixsen2009}, it has an average photon energy of $\langle\epsilon\rangle\simeq6.3\times10^{-4}$~eV and a number density of $\sim$410~$\mathrm{cm}^{-3}$. Due to its high photon density and isotropic nature, the CMB dominates the attenuation of gamma rays above $\sim$300~TeV and reaches maximum absorption efficiency at $\sim$2--3~PeV \citep{Gould1967}.

At lower energies, gamma-ray propagation is primarily governed by interactions with the Galactic ISRF. The IR component originates from thermal dust emission seen in the $\lambda$$\sim$10--500~$\mu{\rm m}$ range, corresponding to dust temperatures of $\sim$15--30~${\rm K}$ \citep{Draine2003}. This component provides the dominant target field for $\gamma$ rays in the $\sim$10--300~TeV range, with the strongest absorption typically occurring around $E_{\gamma}$$\sim$100--200~TeV depending on the local target photon spectrum \citep{vernetto}. The optical and UV components arise from stellar populations distributed throughout the Galactic disk and bulge, with temperatures ranging from $3000$ to $7500$~K. Although its number density is lower than that of the CMB and the IR component, it remains the main source of absorption for gamma rays in the lower TeV range.

There is currently no unique description of the Galactic ISRF structure, and different models adopt distinct assumptions regarding the spatial distribution of stars and dust. In particular, the IR component can differ significantly depending on the treatment of the inner Galaxy.~The analytical model of \citet{vernetto} assumes a smooth exponentially decreasing dust density that peaks toward the GC, resulting in a monotonic increase of the IR photon density toward $R=0$ (e.g.~see Fig.~11 of their paper).~In contrast, the 3D ISRF models, implemented within the publicly available\footnote{The full R12 and F98 ISRF model data are provided on \url{https://galprop.stanford.edu/download.php}.} GALPROP framework \citep{Porter_2017, Porter_2022}, include more sophisticated geometries constrained by observational survey data such as \textit{IRAS} and \textit{COBE/DIRBE}.

In this work, two independent GALPROP ISRF models are adopted: the R12 model \citep{Robitaille_2012} and the F98 model \citep{Freudenreich_1998}. The R12 model includes pronounced spiral-arm structures together with an inner dust ``hole'' introduced to reproduce observations of the central Galaxy. As a result, the IR energy density near the GC becomes comparable to that of the solar neighborhood (e.g.~see Fig.~6 left of \cite{Porter_2017}), while enhanced IR densities appear along spiral-arm regions such as the Norma Arm. In contrast, the F98 model predicts a smoother and more centrally concentrated dust distribution, including a nonaxisymmetric Galactic bar structure. Consequently, the IR energy density in the F98 model remains high both in the GC region and along the spiral arms (e.g.~see Fig.~6 right of \cite{Porter_2017}). These differences become particularly important above $\sim$100~TeV, where the opacity is dominated by the IR photon density.

Similarly, the spatial distribution of starlight differs between models, affecting $\gamma$-ray absorption particularly at lower TeV energies.~In \citet{vernetto}, the optical and UV components are treated as subdominant compared to the IR field, and their contribution to the total opacity is considered negligible for $E_{\gamma} \gtrsim 30$~TeV. In contrast, the R12 and F98 models include detailed stellar distributions. The R12 model assumes a continuous stellar disk and a high-luminosity bulge, leading to an enhanced optical energy density in the inner Galaxy, while the F98 model employs a holed stellar disk and a nonaxisymmetric bar structure, accounting for localized density variations. Although these differences have a limited impact at $\sim$100~TeV, they may become important when modeling $\gamma$-ray absorption in the lower VHE regime below several tens of TeV.

The structural differences between the R12 and F98 models are illustrated in Fig.~\ref{isrfmaps}.~For the IR component (left panels), the R12 model exhibits clear spiral-arm features together with a suppressed central region due to the imposed dust hole. Consequently, the IR energy density at the GC is comparable to that of the solar neighborhood, while higher values are observed in the Norma Arm region. In contrast, the F98 model shows a smoother and more centrally concentrated IR distribution with less pronounced spiral-arm structure. For the optical component (middle panels), both models display increasing energy densities toward the inner Galaxy, although the morphologies differ substantially. The F98 model exhibits a horizontally elongated bulge associated with the Galactic bar, whereas the R12 model shows clearer spiral-arm signatures in the stellar component. The right panels present the corresponding SEDs extracted at the GC, the Norma Arm, and the solar neighborhood, demonstrating how the local target photon spectrum varies spatially between models. The CMB contribution remains identical in all cases due to its isotropic and homogeneous nature.

It is important to note that the optical depth $\tau_{\gamma\gamma}$ is evaluated along the line of sight. Consequently, axisymmetric models assuming cylindrical symmetry, such as those of \citet{vernetto} and \citet{Popescu_2017}, cannot fully capture the impact of nonaxisymmetric structures including spiral arms and the Galactic bar, which host significant concentrations of stars and dust. The absence of these structures may therefore introduce systematic deviations in the predicted attenuation, particularly for sources located along spiral-arm tangents or in the inner Galaxy.

For the construction of the high-resolution GPSP atlas, the use of the R12 and F98 models is well motivated by their detailed 3D descriptions of the Galactic stellar and dust distributions, including spiral-arm structure, a central bar, and spatially varying dust components. Both models are implemented within the GALPROP framework and are publicly available, providing a transparent and reproducible basis for modeling Galactic $\gamma$-ray attenuation. Although they reproduce the same main IR and optical components of the Galactic radiation field relevant for Galactic $\gamma\gamma$ attenuation, they differ in their treatment of the underlying stellar and dust distributions. Therefore, constructing separate GPSP atlases based on both models allows the impact of the adopted ISRF model on the resulting survival probabilities to be directly evaluated and provides an estimate of the associated model dependence.

\section{Construction of the GPSP Atlas}
\label{gpsp_atlas_construction}

The formalism introduced in the previous section provides a continuous description of the gamma-ray survival probability as a function of energy and spatial coordinates. However, practical applications to observational data require a computationally efficient numerical representation. For this purpose, a high-resolution GPSP atlas is constructed as a 4D lookup table, allowing direct evaluation of attenuation effects for arbitrary source configurations across the Galactic plane. This section describes the construction of the atlas, the adopted grid configuration, and the computational framework used in its production.

\begin{figure}
\centering
\includegraphics[width=8.5cm]{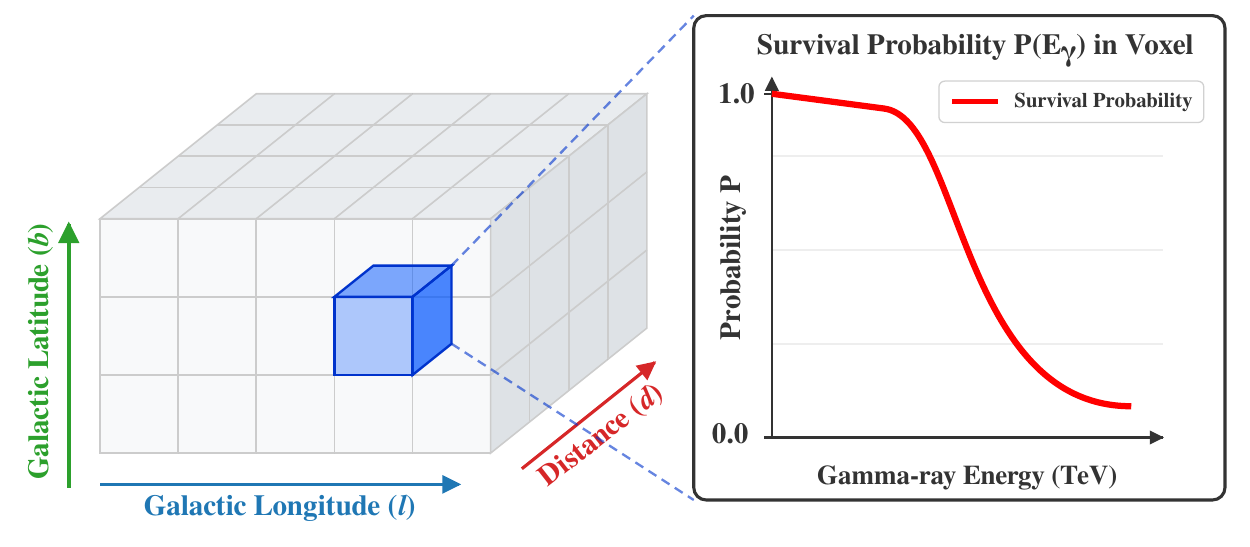}
\caption{Schematic representation of the 4D GPSP atlas structure. The left panel illustrates the 3D Galactic grid defined by longitude ($l$), latitude ($b$), and distance ($d$). Each spatial voxel contains an energy-dependent survival probability profile, $P(E_{\gamma})$, spanning the range from 1~TeV to 10~PeV, as shown schematically in the right panel.}
\label{atlas_structure}
\end{figure}

The GPSP atlas is defined on a 4D grid in Galactic coordinates and energy $(l,b,d,E)$. The spatial component $(l,b,d)$ contains  $\sim$7.27 $\times$10$^{7}$ voxels representing Galactic longitude, latitude, and source distance from the Sun. The grid configuration is designed to match the requirements of current and next-generation VHE and UHE gamma-ray observations:
\begin{itemize}
    \item \textbf{Galactic Longitude ($l$):} Covers the full Galactic circle from $0.0^{\circ}$ to $359.9^{\circ}$ with a step size of $0.1^{\circ}$ (3600 bins).
    \item \textbf{Galactic Latitude ($b$):} Focuses on the Galactic plane from $-5.0^{\circ}$ to $+5.0^{\circ}$ with a step size of $0.1^{\circ}$ (101 bins).
    \item \textbf{Distance ($d$):} Extends from $0.1$~kpc to $20.0$~kpc with a linear step of $0.1$~kpc (200 bins).
    \item \textbf{Energy ($E$):} Covers the range from $1$~TeV to $10$~PeV, sampled with 20 logarithmically equal bins per decade (81 energy bins).
\end{itemize}
By evaluating 81 energy bins for each voxel, the resulting 4D atlas contains $\sim$5.89$\times$10$^{9}$ survival probability data entries. The selected spatial range configuration covers the Galactic plane, where most known VHE and UHE gamma-ray sources are located, while the adopted angular resolution is sufficient to resolve large-scale ISRF variations, particularly in the inner Galaxy and along spiral-arm structures, and matches the angular resolution of current and future VHE/UHE gamma-ray observatories.~A schematic representation of the atlas structure is shown in Fig.~\ref{atlas_structure}.

The energy range spans both the ISRF- and CMB-dominated absorption regimes and adequately resolves the transition between them. The adopted logarithmic binning is also sufficient to sample the rapid energy dependence of the Breit--Wheeler pair production cross section near threshold. In addition to the primary gamma-ray energy grid, the numerical evaluation of the optical depth uses an internal target photon energy grid consisting of 500 logarithmically spaced bins between $10^{-6}$~eV and $100$~eV. 

The distance axis extends from 0.1 to 20.0~kpc using a uniform spacing of 0.1~kpc (200 bins). The choice of a maximum distance of 20~kpc was motivated by the spatial distribution of currently known Galactic VHE and UHE $\gamma$-ray sources. Existing source catalogs indicate that the most distant Galactic objects with distance estimates are located within $\sim$15~kpc of the Sun\footnote{For example, HESS J1640$-$465 ($13.0\pm1.4$~kpc), HESS J1641$-$463 ($\sim$11--13~kpc), and W49B/HESS J1911+090 ($11.1\pm0.8$~kpc). Distance estimates compiled in the TeVCat catalog (\url{https://www.tevcat.org/}) indicate that no currently known Galactic VHE/UHE source lies beyond $\sim$15~kpc from the Sun.}. Therefore, a maximum distance of 20~kpc provides sufficient coverage for essentially all currently known Galactic VHE and UHE $\gamma$-ray sources while maintaining a computationally manageable atlas size. Although regions on the far side of the Galactic disk beyond 20~kpc from the Sun are not included in the current atlas, no currently known Galactic VHE/UHE source is located at such distances. The adopted distance resolution of 0.1~kpc is comparable to or smaller than the typical distance uncertainties of Galactic sources and is sufficient to capture spatial ISRF variations. Beyond the current atlas boundary, the contribution of the spatially varying Galactic ISRF is expected to decrease, particularly outside the main stellar and dust disk of the Milky Way, while attenuation becomes increasingly dominated by the isotropic CMB. Extension of the GPSP atlas to larger distances, together with a dedicated investigation of such regimes, will be considered in future releases of the framework.

The optical depth defined in Eq.~\ref{opt_depth} is evaluated numerically along each line of sight connecting the observer, assumed to be located at the solar position, to a given source location. For each $(l,b,d)$ grid point, the propagation path is discretized using steps of $\Delta \ell = 0.1$~kpc.~At each step, the local target photon density $n(\epsilon,\mathbf{r})$ is obtained from the GALPROP ISRF models (R12 and F98; see Sect.~\ref{isrf_sect}) using trilinear interpolation in cylindrical Galactic coordinates. The ISRF spectral photon densities are interpolated in logarithmic space to maintain numerical stability across the wide dynamic range in photon energy.

The total target photon field is constructed by combining the ISRF with the CMB, modeled as a blackbody spectrum with temperature $T = 2.73$~K, as discussed in Sect.~\ref{isrf_sect}. The integration over target photon energy $\epsilon$ is performed on a logarithmic grid using a precomputed interaction kernel that incorporates both the Breit--Wheeler cross section and the corresponding integration weights. The angular dependence is treated through explicit numerical integration over $\mu = \cos\theta$ using a fixed grid of 50 points, providing an accurate evaluation of the angle-averaged cross section under the isotropic approximation described in Sect.~\ref{br_cs}. The Breit--Wheeler cross section is evaluated with the explicit kinematic threshold condition $s > 1$, excluding subthreshold contributions.

To improve computational efficiency, the angle-integrated interaction kernel\footnote{The precomputed Breit--Wheeler interaction kernel corresponds to the 2D distribution shown in the left panel of Fig.~\ref{bw_cross_sections}.} is precomputed as a function of gamma-ray and target photon energies, reducing the dimensionality of the numerical integration during the line of sight calculation. The total optical depth is accumulated over all propagation steps, and the survival probability is computed using Eq.~\ref{eq1}. The resulting atlas is applicable to both point-like and extended gamma-ray sources. For extended emission, the survival probability can be integrated over the region of interest (ROI) weighted by the intrinsic source morphology.

Because of the large size of the ISRF datasets ($\sim$17~GB per model), the computation is parallelized over Galactic longitude. Each longitude slice is processed independently using a dedicated {\tt C++} pipeline, allowing efficient scaling across multiple computing cores. The final atlas is stored in FITS (Flexible Image Transport System) format for compatibility with standard astrophysical analysis tools. The uncompressed atlas size is approximately $\sim$23.5~GB per ISRF model and can be reduced to $\sim$11.2~GB using compressed binary storage (\texttt{.fits.gz}) without loss of numerical precision. The FITS structure consists of the following:

\begin{itemize}
    \item \textbf{Primary HDU:} Four-dimensional data cube $(l,b,d,E)$ containing photon survival probabilities, stored in 32-bit floating-point format.
    \item \textbf{BinTableHDU Extension:} Energy binning table associated with the energy axis.
    \item \textbf{Header Metadata:} Coordinate definitions, physical units (kpc, TeV), and standard FITS keywords.
\end{itemize}
This structure allows straightforward integration of the GPSP atlas into existing analysis frameworks such as \texttt{Gammapy}~\citep{gammapy}, providing a consistent high-resolution framework for modeling gamma-ray attenuation across the Galactic plane, suitable for current and next-generation VHE and UHE gamma-ray observations.

\subsection{Systematic Uncertainties of the GPSP Atlas}

The main systematic uncertainties of the GPSP atlas arise from the choice of ISRF model, the isotropic approximation of the radiation field, and the adopted grid resolution. Among these, the uncertainty associated with the adopted ISRF model can be directly estimated by comparing the survival probabilities obtained using the GPSP R12 and GPSP F98 ISRF atlases. The following analysis provides a quantitative characterization of these model-dependent uncertainties as a function of Galactic direction, energy, and source distance. To provide a practical estimate of the model-dependent uncertainties associated with the GPSP atlas, the systematic differences between the R12 and F98 models were quantified using the fractional deviation
\begin{equation}
\delta_{P} = \frac{P_{\rm R12}-P_{\rm F98}}
{\frac{1}{2}(P_{\rm R12}+P_{\rm F98})},
\label{frac_diff}
\end{equation}
where $P_{\rm R12}$ and $P_{\rm F98}$ are the survival probabilities obtained using the respective ISRF models. Table~\ref{systematics_table} summarizes the median and the central 68$\%$ intervals of $|\delta_{P}|$ for four representative Galactic regions and three different energy intervals. As expected, the largest deviations are observed toward the inner Galaxy, where the median model-dependent uncertainty reaches $\sim$9$\%$ in the 50--200 TeV energy interval and increases to $\sim$13$\%$ when only the central Galactic plane latitude range of $|b|<1^\circ$ is considered. In contrast, uncertainties toward the Scutum--Crux and Norma spiral-arm tangent directions remain at the level of $\sim$3$\%$--4$\%$, while the Galactic anticenter (AC) shows deviations below 1$\%$ across all investigated energy ranges. A clear decrease in the model-dependent uncertainty is also observed with increasing energy. This behavior is consistent with the transition from attenuation dominated by the Galactic IR field, where the R12 and F98 models differ most strongly, to regimes where the interaction is increasingly dominated by the CMB component.

\begin{table}[h!]
\footnotesize
\caption{Model-dependent systematic uncertainties of the GPSP atlas estimated from the fractional difference between the R12 and F98 ISRF models using Eq.~\ref{frac_diff}. For each Galactic region and energy interval, the reported values correspond to the median and the central 68$\%$ intervals (16th and 84th percentile) of the absolute fractional deviation $|\delta_{P}|$ distribution extracted from the atlas. Values shown in bold are obtained for the full latitude range of $|b|<$5$^{\circ}$, while the values given in parentheses correspond to the central Galactic plane region of $|b|<$1$^{\circ}$. For entries with uncertainties below the displayed numerical precision, only the median value is reported.}
\centering
\renewcommand{\arraystretch}{1.2}
\begin{tabular}{cccc}
\hline\hline
Galactic Region & 50--200 TeV & 300--600 TeV & 1--3 PeV \\
\hline
Inner Galaxy & \textbf{9.3$^{+3.9}_{-2.6}$} (13.3$^{+2.8}_{-3.2}$) & \textbf{7.3$^{+2.6}_{-1.8}$} (10.0$^{+1.9}_{-1.8}$) & \textbf{3.0$^{+1.3}_{-0.9}$} (4.1$^{+1.4}_{-1.0}$) \\
$l$=[-20$^{\circ}$,20$^{\circ}$] & & &\\
\hline
Scutum--Crux & \textbf{3.3$^{+1.8}_{-1.1}$} (3.7$^{+2.8}_{-1.3}$) & \textbf{3.0$^{+1.4}_{-0.7}$} (3.4$^{+2.2}_{-0.6}$) & \textbf{1.3$^{+0.6}_{-0.4}$} (1.6$^{+0.8}_{-0.5}$)  \\
$l$=[20$^{\circ}$,40$^{\circ}$] & & &\\
\hline
Norma        & \textbf{3.5$^{+2.0}_{-1.2}$} (4.1$^{+3.1}_{-1.4}$) & \textbf{3.2$^{+1.5}_{-0.7}$} (3.7$^{+2.5}_{-0.7}$) & \textbf{1.4$^{+0.7}_{-0.4}$} (1.7$^{+0.9}_{-0.5}$)  \\
$l$=[320$^{\circ}$,340$^{\circ}$] & & &\\
\hline
Anti Center  & \textbf{0.4$\pm$0.1} (0.4$\pm$0.1) & \textbf{0.3} (0.3) & \textbf{0.1} (0.1)\\
$l$=[160$^{\circ}$,200$^{\circ}$] & & &\\
\hline
\end{tabular}
\label{systematics_table}
\end{table}

\begin{figure}
\centering
\includegraphics[width=8.5cm]{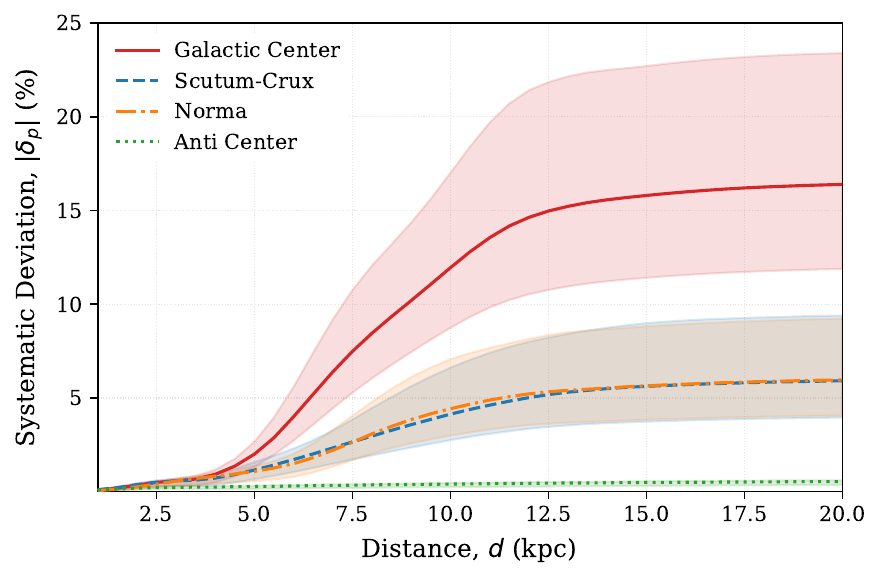}
\caption{Distance dependence of the model-dependent systematic uncertainties of the GPSP atlas in the 50--200 TeV energy interval. Different lines show the median absolute fractional deviation $|\delta_{P}|$ between the R12 and F98 ISRF models for the Galactic Center (red solid), Scutum--Crux (blue dashed), Norma (orange dot--dashed line), and Galactic anticenter (green dotted) regions, respectively. Shaded bands indicate the corresponding 68$\%$ intervals.}
\label{systematics_distance}
\end{figure}
The dependence of the ISRF model-based uncertainty on source distance is illustrated in Fig.~\ref{systematics_distance} for the 50--200 TeV energy range. For all investigated directions, the deviation between the R12 and F98 models increases monotonically with distance, reflecting the cumulative nature of the optical depth along the line of sight. The strongest effect is observed toward the inner Galaxy, where the median deviation increases from below 1$\%$ for nearby sources (d~$\lesssim$~4~kpc) to $\sim$15$\%$--16$\%$ for distances beyond $\sim$13 kpc. The uncertainty then exhibits a gradual saturation, suggesting that a large fraction of the model-dependent differences accumulate within the high-density ISRF environment of the Galactic disk and inner Galaxy. In comparison, the Scutum--Crux and Norma directions asymptotically approach uncertainties of $\sim$6$\%$, while the AC direction remains below $\sim$0.6$\%$ throughout the entire distance range. These results indicate that the dominant source of model-dependent uncertainty is confined to distant sources observed through the inner Galactic regions, where differences in the underlying ISRF distributions have the largest impact on the predicted attenuation. The values provided here may therefore be adopted as representative model-dependent uncertainty estimates when applying the GPSP atlas to Galactic gamma-ray sources located in the corresponding sky regions and energy ranges.\\
It is important to note that the systematic uncertainty estimates presented in Table~\ref{systematics_table} and Fig.~\ref{systematics_distance} quantify only the model-dependent differences between the R12 and F98 ISRF models. The values provided here should therefore be regarded as lower-limit estimates of the total systematic uncertainty. Additional systematic uncertainty is expected from the isotropic approximation adopted in this work. Previous studies suggest that anisotropy effects may modify the attenuation at the level of $\sim$10$\%$--20$\%$, depending on energy and line of sight direction \citep{Moskalenko_2006, porter_2018_pev}. A dedicated quantification of this contribution would require fully anisotropic radiative transfer calculations and is beyond the scope of the current atlas release.

\section{The GPSP Atlas Results and Maps}
\label{gpsp_atlas_results}

\begin{figure*}[ht!]
\centering
\includegraphics[width=18.0cm]{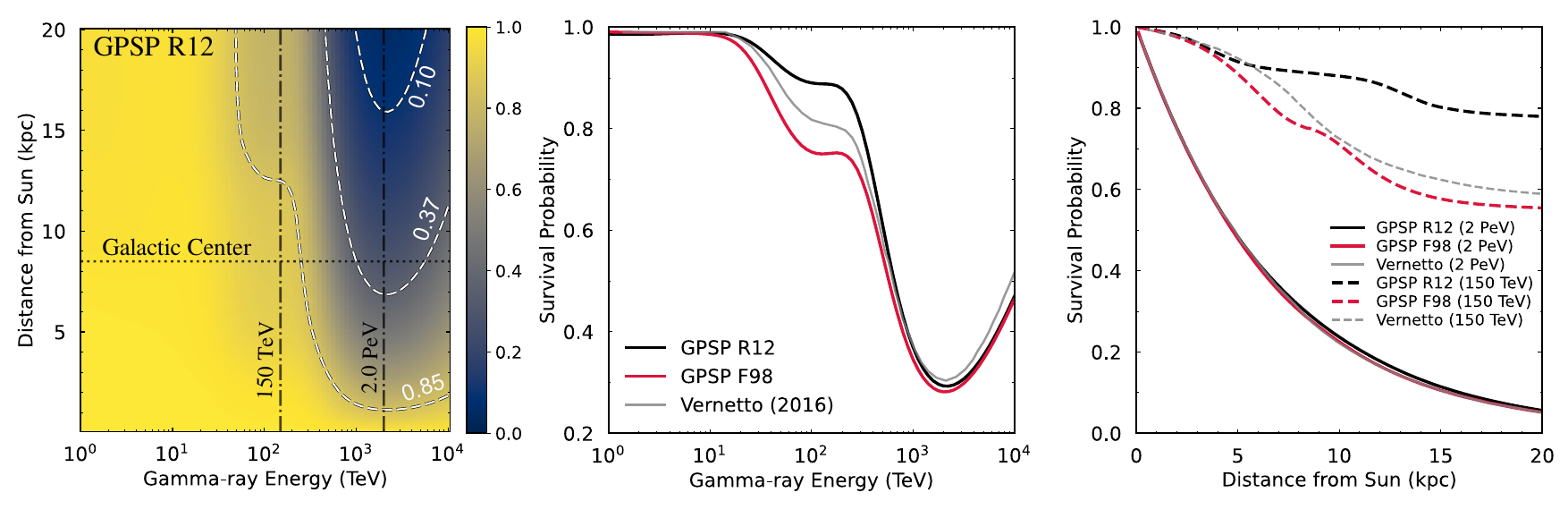}
\caption{Two-dimensional visualization of the survival phase space toward the GC ($l = 0^{\circ},\, b = 0^{\circ}$) direction. \textbf{Left:} Survival probability distribution as a function of gamma-ray energy and distance from the Sun, derived from the GPSP R12 atlas. White dashed contours indicate survival probability levels of 0.10, 0.37 and 0.85. The horizontal dotted and vertical dash--dotted black lines mark reference values corresponding to the GC distance ($d = 8.5$~kpc), and $E_{\gamma} = 150$~TeV and 2~PeV energies, respectively. \textbf{Middle:} Survival probability as a function of energy for gamma rays originating from the GC at 8.5~kpc. The GPSP R12 (solid black) GPSP F98 (solid red) results are compared with the results of \citet{vernetto} (solid gray). \textbf{Right:} Survival probability as a function of distance along the line of sight toward the GC for $E_{\gamma} = 150$~TeV (dashed lines) and $E_{\gamma} = 2$~PeV (solid lines), shown up to 20~kpc, comparing the GPSP R12 (black), GPSP F98 (red) and \citet{vernetto} (gray) profiles.}
\label{survival_phase_space}
\end{figure*}

This section presents the main results and survival probability maps derived from the GPSP atlas\footnote{The GPSP atlas files are published together with accompanying Python helper modules, example Jupyter notebooks, and a dedicated online documentation website. All results and maps presented in this paper can be reproduced using the provided ready-to-use analysis functions. The documentation includes installation instructions, API reference pages, worked examples, and tutorials covering common analysis tasks such as loading the atlas FITS files, extracting survival-probability profiles, applying attenuation corrections, and performing extended-source analyses. Further details are provided in Sect.~\ref{data_section}.}, focusing on the visualization and interpretation of gamma-ray attenuation across the Galactic plane. Owing to its 4D structure, the atlas allows a unified exploration of the survival probability as a function of gamma-ray energy, source distance, and position in the Galaxy. First, a detailed case study toward the GC is presented, introducing a novel 2D energy--distance representation of the survival probability phase space. This is followed by global bird's-eye view maps of the Galaxy at representative energies, highlighting the evolution of the gamma-ray horizon. Finally, Galactic plane survival probability maps are discussed to characterize localized attenuation features, in particular associated with Galactic spiral-arm tangents.~Unless otherwise stated explicitly, the main results presented throughout this paper are derived using the GPSP atlas constructed with the R12 ISRF model (hereafter GPSP~R12). Results obtained using the F98 ISRF model-based atlas (hereafter GPSP~F98) are provided for cross-checks and comparative assessment of model-dependent systematic uncertainties.

\subsection{The Gamma-Ray Survival Phase Space}

Gamma-ray attenuation is traditionally characterized using 1D survival probability profiles, evaluated either as a function of energy for a fixed distance or as a function of distance at a fixed energy. While such representations are useful for individual sources with well-constrained distances, they do not fully capture the complex interplay between spatial and spectral dependence of the attenuation. Exploiting the 4D structure of the GPSP atlas, a more general representation can be introduced in the form of "gamma-ray survival phase space", as demonstrated in the left panel of Fig.~\ref{survival_phase_space}. This 2D representation visualizes the survival probability distribution in the $E_{\gamma}$--distance plane for a given Galactic ($l$, $b$) direction.

One important advantage of this representation is its applicability to sources with uncertain or poorly constrained distances, which is a common challenge in Galactic astronomy. Instead of evaluating multiple discrete survival profiles for different assumed distances, the survival phase-space map directly visualizes the entire attenuation regime and the transition into optically thick regions. In this framework, gamma-ray horizons naturally emerge as survival probability contours, with the canonical horizon defined by $P_{\rm surv}$=$e^{-1}$$\approx$0.37 \citep{Dominguez_2013, Arsioli_2025}, corresponding to an optical depth of $\tau=1$.

The GPSP atlas also allows a direct assessment of model-dependent uncertainties through comparison between the R12 and F98 ISRF models. As shown in the middle panel of Fig.~\ref{survival_phase_space} for the GC region, differences between the models (GPSP R12, GPSP F98, and \citet{vernetto}) become increasingly visible above several tens of TeV and reach their maximum in the $\sim$100--200~TeV range, where attenuation is dominated by the IR component of different ISRF models. Consequently, the atlas can provide a direct way to assess whether observed spectral softening or cut-off features are robust against ISRF systematics or instead fall within a regime dominated by propagation uncertainties. This distinction is particularly important when separating intrinsic acceleration limits from absorption-induced spectral features.

The right panel of Fig.~\ref{survival_phase_space} validates the spatial resolution of the GPSP atlas through survival probability profiles toward the GC at two representative energies. At 2~PeV, all models exhibit very similar attenuation behavior, as expected when the optical depth is dominated by interactions with the CMB. In this regime, attenuation depends primarily on propagation distance and is largely insensitive to the detailed structure of the Galactic ISRF. In contrast, at 150~TeV, the interaction threshold shifts toward the IR component of the ISRF and clear differences emerge between the models. The bumpy structures visible in the GPSP R12 and GPSP F98 atlas profiles arise from localized variations in photon density resolved by the high spatial resolution of the atlas as the propagation path intersects the major Galactic features discussed in Sect.~\ref{isrf_sect}. 

\subsection{Galactic Transparency Maps and Gamma-Ray Survival Horizons: A Top-down Perspective}

\begin{figure*}[ht!]
\centering
\includegraphics[width=18.0cm]{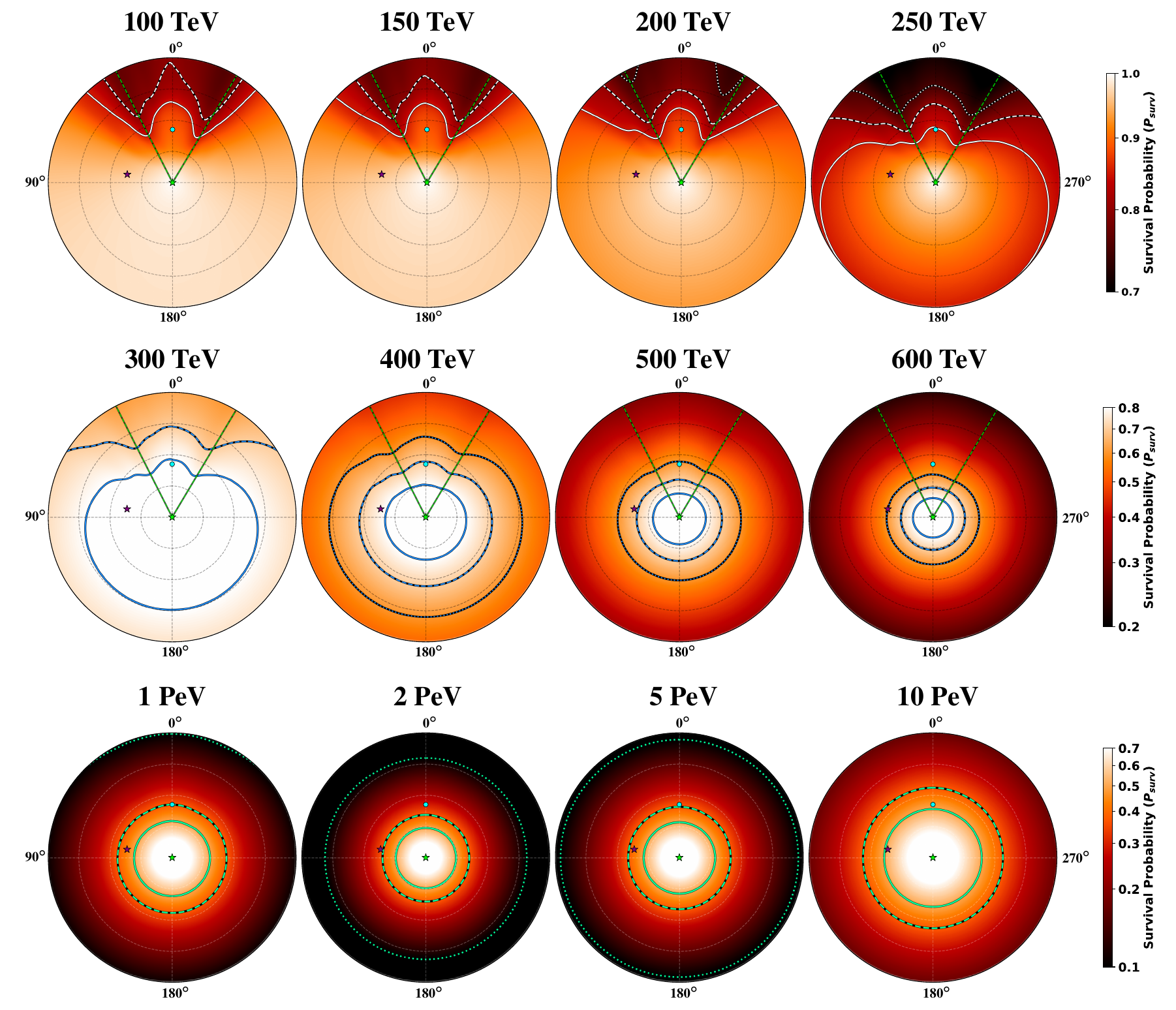}
\caption{Bird's-eye view maps of Galactic gamma-ray survival probability ($P_{\mathrm{Surv}}$) at twelve representative energies. In all panels, markers denote the Sun (lime star), Galactic Center (cyan circle), and Cygnus X-3 (purple star). Tangential directions for the Norma ($l=329^{\circ}$) and Scutum--Crux ($l=30^{\circ}$) arms are indicated by dashed green lines. \textbf{Top Row:} Survival probability maps for 100, 150, 200, and 250~TeV. Solid, dashed, and dotted white lines represent the 0.85, 0.80, and 0.75 contours, respectively. \textbf{Middle Row:} Maps for 300, 400, 500, and 600~TeV. Solid, dashed, and dotted blue lines indicate the 0.80, 0.70, and 0.60 contours, respectively. \textbf{Bottom Row:} Maps for 1, 2, 5, and 10~PeV. Solid, dashed, and dotted lime lines mark the 0.50, 0.37, and 0.10 contours, respectively. Color scales are adjusted independently per row to optimize visual contrast across varying opacity regimes.}
\label{birdviewmaps}
\end{figure*}

The 4D structure of the GPSP atlas allows the reconstruction of bird’s-eye view maps of the Milky Way at fixed gamma-ray energies, providing a global picture of Galactic transparency, as shown in Fig.~\ref{birdviewmaps}. In particular, $P=0.37$ survival probability contours provide crucial information on the gamma-ray horizon\footnote{It is important to note that, due to the exponential nature of photon attenuation, the gamma-ray horizon is not a sharp physical boundary but is conventionally defined as a contour marking the gradual transition into the opaque regime.} and its evolution with energy.

The top-row maps of Fig.~\ref{birdviewmaps} show that the survival probability contours are strongly shaped by the structured IR component of the Galactic ISRF at energies between 100 and 250~TeV. In this regime, the contours ($P_{\rm surv}$=0.75, 0.80, 0.85) exhibit clear anisotropies rather than smooth radial behavior. The most prominent features are the reduced transparency regions toward the Galactic tangent directions associated with the Norma ($l=329^{\circ}$) and Scutum--Crux ($l=30^{\circ}$) arms~\citep{Vallee_2014,Hou_2014}, indicated by the dashed green lines. Along these directions, the contours contract significantly toward the Sun's position.

These structures originate from both the enhanced IR photon densities within spiral-arm and molecular-ring environments and the longer effective path lengths through these regions. Consequently, attenuation accumulates more efficiently along the tangent directions than toward the GC itself. Although the central Galactic regions also contain intense radiation fields, the integrated propagation length through the densest IR structures is shorter than along the tangential-arm lines of sight. In contrast, the Galactic AC ($l = 180^{\circ}$) remains the most transparent direction. In this outward direction, the line of sight rapidly exits the dense radiation environment of the inner Galaxy, allowing a deeper gamma-ray reach than for trajectories toward the inner Galaxy.

A gradual transition in the absorption regime becomes visible in the middle-row maps. Around 400--500~TeV, attenuation increasingly becomes dominated by interactions with the CMB, and the imprint of the structured Galactic ISRF begins to fade. This transition is reflected in the morphology of the survival contours ($P_{\rm surv}$=0.60, 0.70, 0.80), which gradually evolve toward smoother and more circular shapes.

At PeV energies (bottom-row maps), the contours ($P_{\rm surv}$=0.10, 0.37, 0.50) become nearly circular, reflecting the dominance of CMB-induced attenuation. The gamma-ray horizon distance defined by the $P_{\rm surv}=0.37$ contour (dashed lime line) reaches a minimum at $\sim$2--3~PeV, corresponding to the peak efficiency of pair production on the CMB. At even higher energies (5--10~PeV), the horizon expands again as the number density of CMB photons satisfying the interaction threshold decreases. Although an additional diffuse radio background could contribute to the opacity at these extreme energies \citep{Protheroe_1996}, it is not included in the current GPSP atlas.

\begin{figure}
\centering
\includegraphics[width=8.5cm]{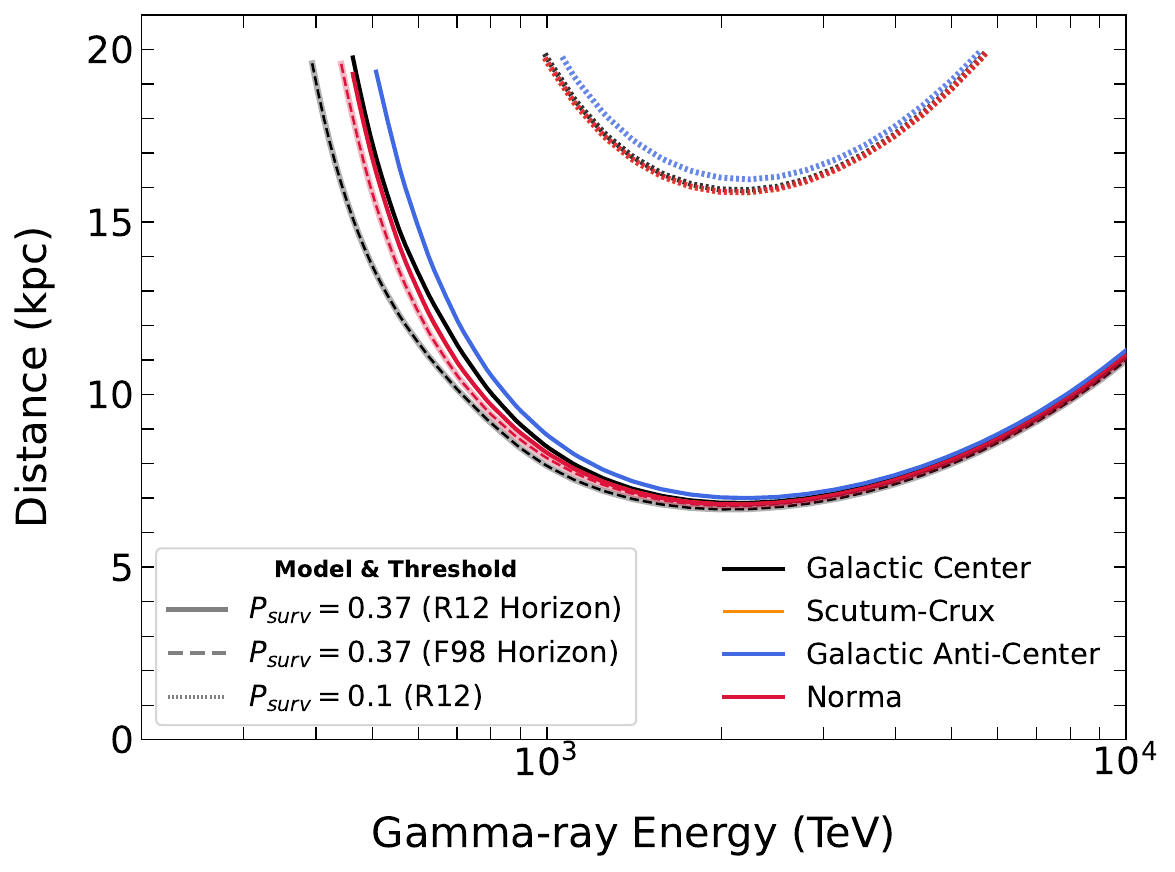}
\caption{Gamma-ray horizon distances as a function of energy for four representative Galactic directions: GC ($l$=$0^{\circ}$, $b$=$0^{\circ}$; black), Scutum--Crux tangent ($l$=$30^{\circ}$, $b$=$0^{\circ}$; red), Norma ($l$=$329^{\circ}$, $b$=$0^{\circ}$; orange), and AC ($l$=$180^{\circ}$, $b=$0$^{\circ}$; blue). Horizons are defined by survival probability thresholds $P_{\mathrm{surv}}$=0.37 (solid) and $0.10$ (dotted) from the GPSP R12 atlas. Dashed lines indicate the $P_{\mathrm{surv}}$=0.37 horizon for the GC and Norma directions from the GPSP F98 atlas for comparison.}
\label{horizon}
\end{figure}

Complementary to the bird's-eye view maps, Fig.~\ref{horizon} presents the corresponding 1D gamma-ray horizons extracted along representative Galactic directions.~The canonical $P_{\rm surv}$=0.37 horizon shows a nonmonotonic energy dependence in all directions, shrinking toward a minimum distance of $\sim$7~kpc at 2--3~PeV before expanding again at higher energies. At lower energies, the horizons retain the strong anisotropy visible in the bird’s-eye maps, with reduced reach along the tangential-arm directions and larger horizon distances toward the AC. Above several hundred TeV, the horizons from different directions converge as attenuation becomes increasingly dominated by the isotropic CMB component.

The deeper transmission horizon defined by $P_{\rm surv}$=0.1 (corresponding to $\tau$$\approx$2.3) reaches a minimum distance of $\sim$16~kpc.~Even at this extreme limit, the Galaxy does not become perfectly isotropic and small directional differences remain visible. In particular, the AC direction consistently exhibits a slightly larger horizon distance than the GC and tangential-arm directions.~This behavior originates from small residual differences in the accumulated optical depth. For example, at $\sim$2~PeV the survival probability at 20~kpc differs at the level of $P_{\rm surv}\sim$0.095 for the AC direction compared to $\sim$0.088 toward the GC, corresponding to a difference in optical depth of $\Delta\tau\sim$0.08. Although this difference is small, it translates into a horizon shift of $\sim$0.7--0.8~kpc because $\tau$ increases very slowly with distance in the outer Galactic regions where the ISRF density is low. Consequently, even small residual opacity differences can produce noticeable shifts in the horizon distance. The effect is therefore amplified for the $P_{\rm surv}=0.1$ contour, which probes larger optical depths of $\tau\approx2.3$, and is therefore more sensitive to accumulated residual ISRF contributions than the canonical $\tau=1$ horizon. The comparison with the alternative GPSP F98 atlas horizon further shows that although moderate differences remain at sub-PeV energies, both atlas horizons converge toward the same CMB-dominated behavior at the highest energies.

\subsection{Galactic Plane Survival Maps}

Another key advantage of the GPSP atlas is its ability to provide a spatially resolved view of gamma-ray attenuation across the entire Galactic plane for any given energy and distance slices. Unlike traditional 1D representations, the atlas captures both large-scale transparency gradients and localized structures associated with the underlying ISRF distribution. This is particularly important for interpreting extended gamma-ray sources, as well as GDE.

\begin{figure*}[ht!]
\centering
\includegraphics[width=18.0cm]{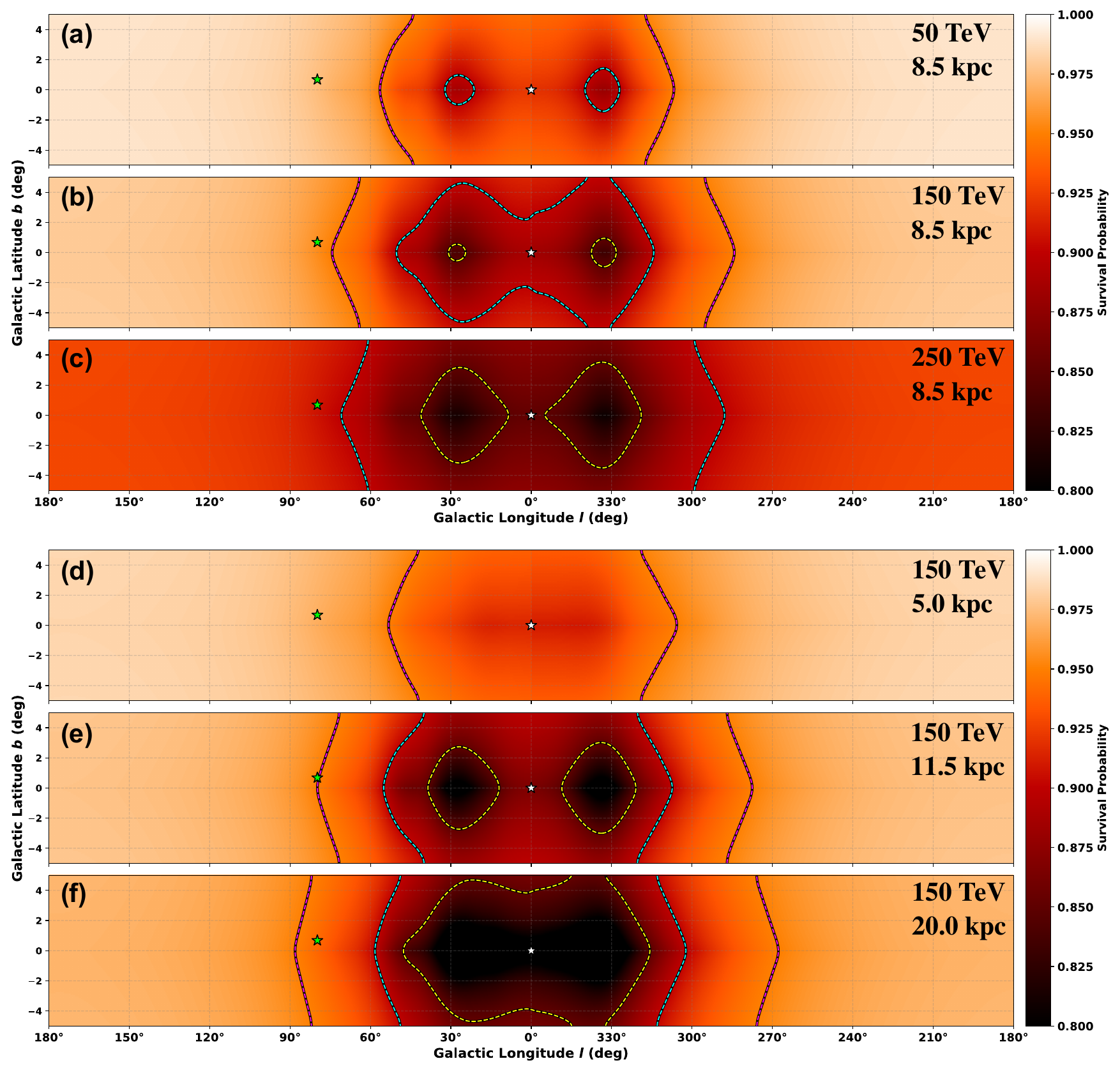}
\caption{The Galactic Plane survival probability maps reconstructed from the GPSP atlas. In all panels, the white and green stars indicate the positions of the GC and the Cygnus X-3 region, respectively. The dashed magenta, cyan, and yellow lines represent the 0.95, 0.90, and 0.85 survival probability contours. The top row (a--c) presents maps at a fixed source distance of $d = 8.5$~kpc for gamma-ray energies of 50, 150, and 250~TeV, respectively. The bottom row (d--f) presents maps at a fixed energy of $E = 150$~TeV for source distances of 5.0, 11.5, and 20.0~kpc.}
\label{gp_maps}
\end{figure*}

The Galactic plane survival probability maps reconstructed from the GPSP atlas are shown in Fig.~\ref{gp_maps}. The top panels (a--c) present fixed-distance slices at $d = 8.5$~kpc for gamma-ray energies of 50, 150, and 250~TeV, respectively, while the bottom panels (d--f) show fixed-energy slices at $E = 150$~TeV for source distances of 5, 11.5, and 20~kpc. Several important morphological features emerge from these maps. In particular, the attenuation is evidently nonuniform along the Galactic plane, with localized regions of reduced survival probability emerging near the tangential directions of the spiral arms. These hole-like structures correspond to the line of sight projection of the spiral-arm tangents visible in the bird's-eye view maps (see the top row of Fig.~\ref{birdviewmaps}). They arise where the line of sight intersects regions of enhanced IR photon density. Consequently, these localized minima trace the geometry of the high-density ISRF regions, effectively projecting the 3D structure of the Galactic disk onto the 2D Galactic plane skymap.

At a fixed distance (a--c), the survival probability decreases systematically with increasing gamma-ray energy as pair production interactions become more efficient. At 50~TeV, attenuation remains relatively mild. However, increasingly pronounced structures emerge along the Galactic plane at higher energies. A similar trend is observed in the bottom panels (d--f), where increasing source distance leads to a cumulative growth of the optical depth and a global reduction of the survival probability. At the same time, the contrast between different Galactic longitudes becomes more prominent because longer propagation paths increase sensitivity to structured ISRF components. In particular, the imprint of the tangential spiral-arm regions becomes progressively stronger at larger distances.

\begin{figure}
\centering
\includegraphics[width=8.5cm]{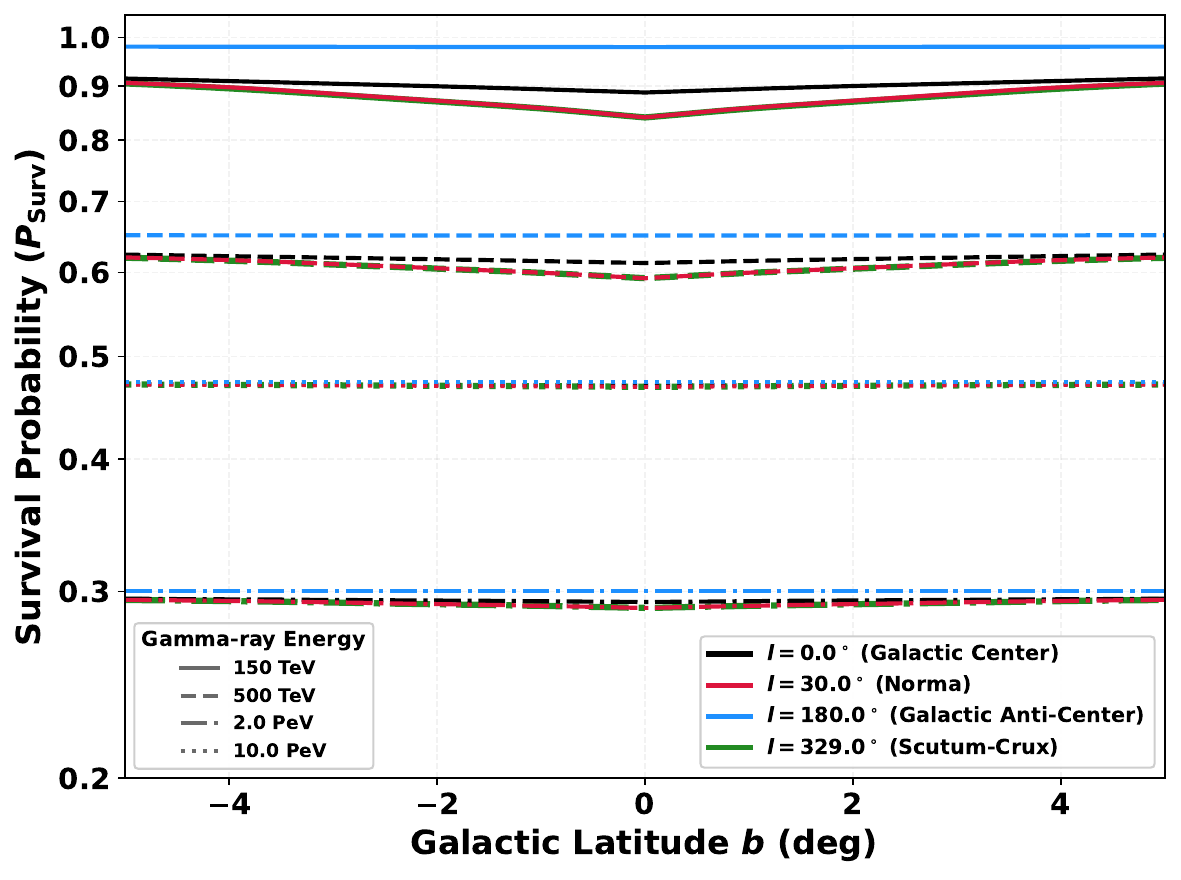}
\caption{Survival probability as a function of Galactic latitude ($b$) for representative longitudes and gamma-ray energies. Profiles are shown for four Galactic directions: $l = 0.0^{\circ}$ (GC, black), $l = 30.0^{\circ}$ (Norma, red), $l = 180.0^{\circ}$ (AC, blue), and $l = 329.0^{\circ}$ (Scutum--Crux, green). Four energy levels are represented by different line styles: 150~TeV (solid), 500~TeV (dashed), 2.0~PeV (dash--dotted), and 10.0~PeV (dotted).}
\label{lat_profile}
\end{figure}

These structural differences are further quantified in the latitude profiles of the survival probability shown in Fig.~\ref{lat_profile} for representative Galactic longitudes and energies. At 150 and 500~TeV, the GC and tangential-arm directions show clear U-shaped absorption profiles centered around $b$=0$^{\circ}$. This reflects the concentration of the IR radiation field near the Galactic plane and toward the inner Galaxy. Directions intersecting dense ISRF regions therefore show stronger attenuation and steeper latitude gradients. In contrast, the AC direction remains comparatively flat because of the lower photon densities in the outer Galaxy. As the energy increases into the PeV regime, these directional differences progressively diminish and the latitude profiles converge toward a nearly uniform distribution. This marks the transition to a regime dominated by interactions with the CMB, where the influence of localized Galactic ISRF structures disappears. Consequently, the attenuation pattern evolves toward a smoother and more isotropic behavior.

It should be noted that the current GPSP atlas is defined within the latitude range $-$5$^\circ$$\le$b$\le$+5$^\circ$. A small number of Galactic VHE sources, such as the Crab Nebula ($l$=184.56$^\circ$, $b$=$-$5.79$^\circ$) \citep{hess_Crab_paper}, therefore lie slightly outside the formal GPSP atlas boundary. However, the latitude profiles presented in Fig.~\ref{lat_profile} show that the survival probability becomes nearly constant at $|b|\gtrsim$~4$^\circ$, indicating that the difference between the atlas boundary value and the exact source position is negligible for practical applications. Therefore, for sources located outside the tabulated latitude range, the survival probability can be approximated using the corresponding boundary value at $b=\pm5^\circ$ for the same Galactic longitude and distance.

Overall, results discussed in this section demonstrate that the GPSP atlas provides a powerful framework for interpreting gamma-ray attenuation in a fully spatial context, revealing both the imprint of Galactic structure at intermediate energies and the emergence of isotropic behavior at the highest energies.

\section{Propagation of UHE Gamma-Rays under Subluminal LIV Scenarios}
\label{gpsp_liv_section}

The GPSP atlas presented in the previous sections was constructed under the standard LI framework, where pair production between VHE/UHE gamma rays and the ISRF follows the standard Breit--Wheeler formalism described in Sect.~\ref{br_cs}. However, as discussed in Sect.~\ref{liv_cross_section}, subluminal LIV scenarios can significantly modify the pair production process at the highest energies by suppressing the interaction cross section and shifting the effective kinematic threshold.~Such effects can substantially alter the transparency of the Galaxy to UHE gamma rays, particularly in the PeV regime where standard LI attenuation is expected to be highest.

Motivated by this, a LIV-extended version of the GPSP atlas (GPSP-LIV) was constructed by incorporating the modified LIV pair production cross section directly into the line of sight propagation calculations. This framework allows a systematic investigation of how different LIV scales modify the Galactic gamma-ray horizon, reshape the survival probability phase space, and alter the expected visibility of Galactic UHE sources.

\subsection{LIV-modified Galactic Photon Survival Probability Atlas} 

The GPSP-LIV atlas was generated using the same Galactic geometry, ISRF models, and propagation framework adopted for the standard GPSP atlas, ensuring direct comparability between the LI and LIV scenarios. The atlas construction follows the same line of sight integration procedure described previously (see Sect.~\ref{gpsp_atlas_construction}), with the only modification being the replacement of the standard pair production cross section by the subluminal LIV formalism introduced in Sect.~\ref{liv_cross_section}.

The calculations were performed using both the R12 and F98 ISRF models, combined with the CMB component. For each line of sight, the radiation field energy density was interpolated in cylindrical Galactic coordinates over the $(R,\phi,z)$ ISRF grid using trilinear interpolation. The gamma-ray optical depth was then accumulated along the propagation path according to
\begin{equation}
\tau(E_\gamma,d) = \int_0^d \int n(\epsilon,\mathbf{r})\, \bar{\sigma}_{\rm LIV} ,d\epsilon\,dl,
\end{equation}
where $n(\epsilon,\mathbf{r})$ is the total target photon density including both the Galactic ISRF and the CMB, and $\bar{\sigma}_{\rm LIV}$ represents the angular-averaged LIV-modified pair production cross section given in Eq.~\ref{eq_liv_cs_carmona}.

For computational efficiency, the interaction kernel\footnote{The precomputed interaction kernel for the LIV framework corresponds to a 2D phase space structurally similar to that shown in Fig.~\ref{liv_2d_cross_sections}, calculated directly for $\bar{\sigma}_{\rm LIV}$ rather than the ratio ($\bar{\sigma}_{\rm LIV}/\bar{\sigma}_{\rm BW}$), across the full spectrum of $\log_{10}(\lambda/E_{\rm Pl})$ grid steps implemented in the GPSP-LIV atlas.} was generated separately for each LIV scale parameter. The angular integration over the collision angle $\theta$ was performed numerically using a discretized angular grid, while the target photon integration was evaluated over a logarithmically spaced energy grid extending from $10^{-6}$~eV to $10^{2}$~eV. The ISRF spectra were mapped onto this common integration grid using logarithmic interpolation in photon energy.

The GPSP-LIV atlas was constructed over a 5D parameter space consisting of Galactic longitude $l$, Galactic latitude $b$, source distance $d$, gamma-ray energy $E_\gamma$, and the LIV scale parameter $\log_{10}(\lambda/E_{\rm Pl})$. The adopted grid configuration is summarized below:
\begin{itemize}
\item Galactic longitude: $0^\circ \leq l < 360^\circ$ with $0.5^\circ$ resolution ($720$ bins),
\item Galactic latitude: $-3^\circ \leq b \leq 3^\circ$ with $1^\circ$ resolution\\ ($7$ bins),
\item Source distance: $0.1 \leq d \leq 20$~kpc with $0.1$~kpc resolution ($200$ bins),
\item Gamma-ray energy: $10^{14} \leq E_\gamma \leq 10^{16}$~eV using $41$ logarithmically spaced bins,
\item LIV parameter: $-4.5 \leq \log_{10}(\lambda/E_{\rm Pl}) \leq -2.0$ with steps of $0.05$ ($51$ bins).
\end{itemize}
For each grid point, the cumulative optical depth was converted into the survival probability using Eq.~\ref{eq1}. Compared to the original GPSP atlas, the angular binning was reduced in order to keep the full 5D GPSP-LIV atlas computationally manageable after introducing the additional LIV dimension. In particular, the Galactic longitude resolution was reduced from $\Delta l = 0.1^\circ$ to $\Delta l = 0.5^\circ$, which remains comparable to the angular resolution currently achievable by wide-field UHE gamma-ray observatories. Similarly, the latitude range was reduced from $|b| \leq 5^\circ$ to $|b| \leq 3^\circ$ along with $\Delta b = 1.0^\circ$. As shown previously in the latitude profiles of Fig.~\ref{lat_profile}, the latitude dependence of the attenuation becomes weaker in sub-PeV and PeV regimes due to the dominant contribution of the CMB, while the remaining ISRF-induced variations remain concentrated near the Galactic plane. Therefore, the adopted latitude range is sufficient to capture the physically relevant attenuation structure in the LIV regime. In contrast, the distance axis was kept identical to the original GPSP atlas, covering distances from $0.1$ to $20$~kpc with $0.1$~kpc resolution. Preserving the original distance resolution remains important because the accumulated optical depth strongly depends on the propagation distance in both the LI and LIV scenarios.

The selected gamma-ray energy range between 100~TeV and 10~PeV fully covers the domain where LIV effects become physically relevant.~As shown in Fig.~\ref{liv_2d_cross_sections}, even for the most extreme LIV scenario considered ($\log_{10}(\lambda/E_{\rm Pl})=-4.5$), significant deviations from the standard pair production cross section emerge only above several hundred TeV. The lower energy threshold of $100$~TeV therefore provides a smooth transition between the standard attenuation regime and the LIV-dominated regime while retaining the ISRF-induced absorption features at lower energies.

The explored LIV parameter interval, -4.5$\leq\log_{10}(\lambda/E_{\rm Pl})\leq$-2.0, was chosen to cover the transition from strong LIV scenarios to the quasi-standard LI regime. The adopted parameter resolution of $\Delta\log_{10}(\lambda/E_{\rm Pl})$ = 0.05 corresponds to a multiplicative factor of $10^{0.05}$$\approx$1.12 in the effective LIV scale, providing sufficiently fine sampling of the gradual LIV-induced evolution of the pair production opacity and Galactic transparency.

The resulting atlas provides the Galactic gamma-ray survival probability as a function of propagation direction, distance, energy, and LIV scale. In total, the GPSP-LIV atlas contains more than $2.1\times10^{9}$ survival probability values stored in compressed FITS format. The final atlas size corresponds to approximately $8.4$~GB in uncompressed form and $\sim5.5$~GB in compressed \texttt{fits.gz} format, providing an efficient lookup-table-based analysis of Galactic UHE gamma-ray propagation under modified LIV scenarios. The GPSP-LIV atlas enables a direct exploration of how the Galactic gamma-ray horizon evolves as a function of the LIV scale, allowing the transition between standard attenuation and increasingly transparent propagation regimes to be studied within a unified framework.

\subsection{LIV-corrected Gamma-Ray Survival Phase Spaces and Gamma-Ray Horizons}
\label{liv_horizons}

\begin{figure*}[ht!]
\centering
\includegraphics[width=18.0cm]{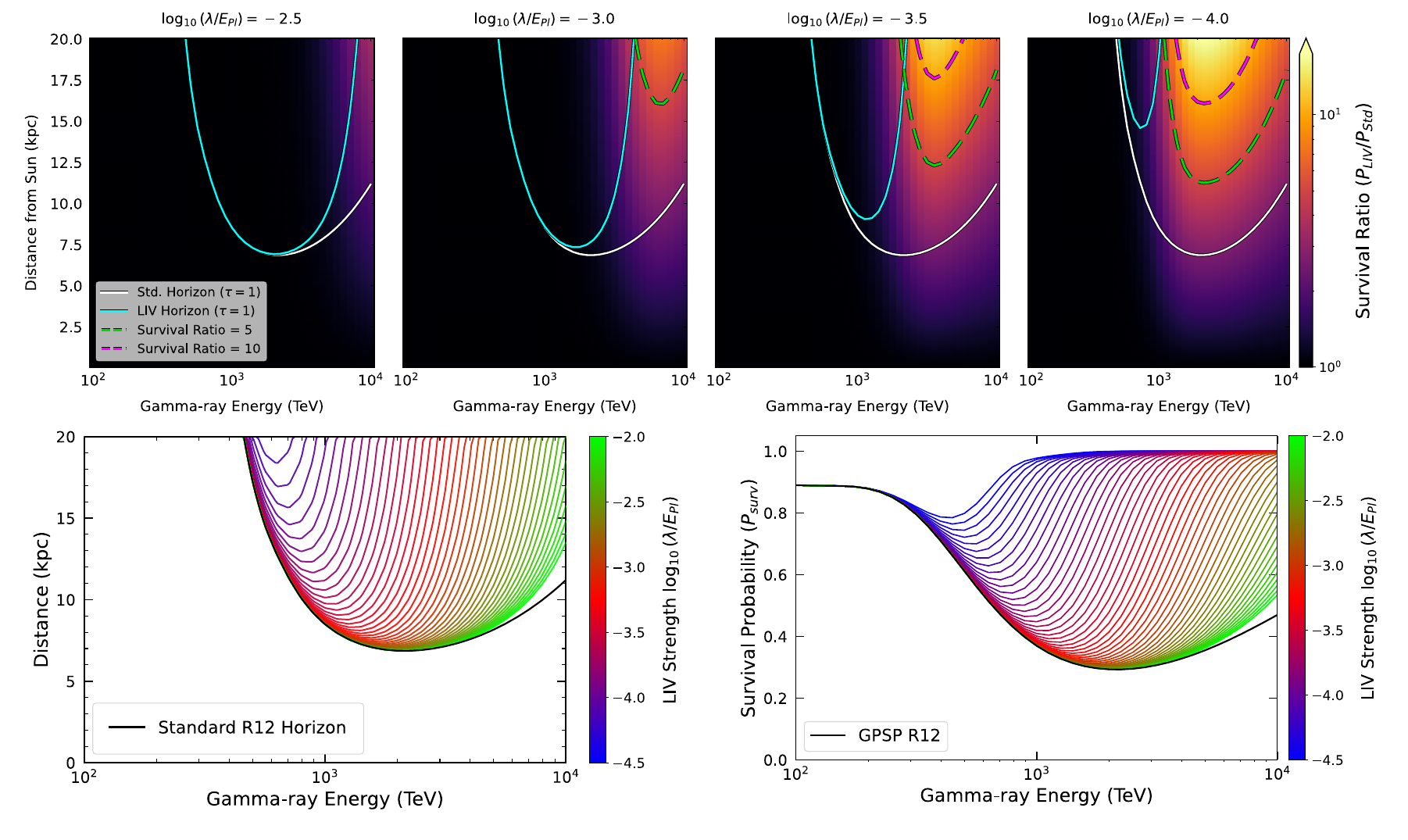}
\caption{Impact of subluminal LIV on the Galactic propagation of UHE gamma rays. \textbf{Top panels:} Survival probability ratio maps ($P_{\rm LIV}/P_{\rm Std}$) as a function of gamma-ray energy and source distance toward the GC direction for $\log_{10}(\lambda/E_{\rm Pl})=-2.5$, $-3.0$, $-3.5$, and $-4.0$. White and cyan curves denote the $\tau=1$ gamma-ray horizons for the standard and LIV scenarios, respectively. Dashed contours indicate survival ratio levels of 5 and 10. \textbf{Bottom left:} Gamma-ray horizon distance as a function of energy for the full LIV parameter range explored in the GPSP-LIV atlas ($\log_{10}(\lambda/E_{\rm Pl})=-4.5$ to $-2.0$). The standard GPSP R12 prediction is shown by the black curve. \textbf{Bottom right:} Survival probability profiles for a source at the GC distance of $d=8.5$~kpc for the same LIV parameter range, compared with the standard GPSP R12 atlas (black curve). Increasing LIV strength progressively enhances Galactic transparency and extends the gamma-ray horizon at PeV energies.}
\label{liv_phase_space}
\end{figure*}

Figure~\ref{liv_phase_space} illustrates the impact of subluminal LIV effects on the propagation of UHE gamma rays within the Galaxy using the GPSP-LIV atlas framework.~The figure summarizes how the modification of the pair production cross section alters both the survival probability phase space and the corresponding Galactic gamma-ray horizon. The upper panels show the ratio between the LIV-modified and standard survival probabilities, $(P_{\rm LIV}/P_{\rm Std}$), for four representative LIV strengths toward the GC direction. Several important features are immediately visible. First, the deviation from standard propagation is strongly energy dependent. Below several hundred TeV, the survival ratio remains close to unity even for strong LIV scenarios, indicating that the propagation remains indistinguishable from the standard LI case. This behavior reflects the LIV-modified pair production cross section discussed in Sect.~\ref{liv_cross_section}, where significant suppression of the interaction rate emerges only once the LIV correction term becomes comparable to the standard kinematic threshold scale.

As the energy approaches the PeV regime, the attenuation gradually weakens and the Galaxy becomes increasingly transparent to UHE gamma rays. This transition appears as a rapid increase of the survival-ratio values at high energies and large propagation distances. The enhancement becomes stronger for strong LIV scenarios, corresponding to larger deviations from LI. In the most extreme scenario shown (top-right panel), the LIV-modified survival probability exceeds the standard expectation by more than an order of magnitude over substantial regions of the $(E_\gamma,d)$ phase space.

The cyan $\tau=1$ contours in the upper panels demonstrate that the Galactic gamma-ray horizon shifts systematically toward larger distances under LIV scenarios. This behavior directly reflects the reduction of the pair production opacity. While the standard horizon (white contours) remains tightly restricted in the PeV regime and only begins to reopen above 2–3 PeV, the LIV-modified horizons (cyan contours) expand much earlier, particularly for strong LIV scenarios. The displacement becomes especially visible above $\sim$1~PeV, where the suppression of the pair production cross section substantially increases the effective attenuation length.

The lower-left panel of Fig.~\ref{liv_phase_space} presents the evolution of the gamma-ray horizon for the full range of LIV scales included in the GPSP-LIV atlas. A smooth transition from the standard propagation regime to increasingly transparent scenarios is observed as $\log_{10}(\lambda/E_{\rm Pl})$ decreases. Under standard LI propagation, the gamma-ray horizon decreases rapidly with energy due to the increasing efficiency of pair production on the Galactic radiation fields. In contrast, the LIV-modified horizons flatten progressively at high energies because the optical depth grows more slowly once the interaction cross section is suppressed, allowing gamma rays to propagate over significantly larger Galactic distances before reaching $\tau=1$.

The lower-right panel shows the corresponding survival probability profiles for a source located at the GC distance of 8.5~kpc. The standard survival probability exhibits the expected strong attenuation above several hundred TeV. In comparison, the LIV-modified profiles develop progressively earlier attenuation turnovers as the LIV strength increases. For sufficiently strong LIV scenarios, the survival probability remains close to unity even in the multi-PeV regime, indicating that pair production becomes strongly suppressed over Galactic propagation distances.

An additional feature visible in both lower panels is the smooth ordering of the curves with respect to the LIV scale parameter. Since the GPSP-LIV atlas samples the LIV parameter space with 0.05 step between $\log_{10}(\lambda/E_{\rm Pl})=-4.5$ and $-2.0$, the propagation behavior can be traced clearly between the standard LI and LIV-dominated regimes without requiring independent recalculations for each individual scenario.

These results indicate that observable LIV signatures are expected primarily in the PeV regime, where the standard Galactic opacity reaches its maximum. In particular, Galactic sources located beyond the standard gamma-ray horizon and showing significant emission around 2--3~PeV represent promising targets for testing LIV scenarios. Since this energy range corresponds to the strongest attenuation expected under standard propagation, any systematic excess of observable PeV emission from distant Galactic sources could provide evidence for reduced pair production opacity. The effect would be especially intriguing for sources located well beyond the traditional gamma-ray horizon, where standard models predict substantial attenuation of the emitted gamma-ray flux.

Nevertheless, the detection of multi-PeV emission from a distant source should not be regarded as direct evidence for LIV. Even under standard propagation, sufficiently luminous accelerators may remain detectable despite strong attenuation. Consequently, robust LIV constraints are unlikely to emerge from individual sources alone, particularly when their intrinsic luminosities, distances, or underlying gamma-ray emission mechanisms are uncertain. More compelling tests are expected from population studies of Galactic PeV emitters with well-characterized source properties and independently constrained emission models. In such cases, the observed gamma-ray fluxes can be compared with the expected attenuated spectra predicted under standard propagation, allowing statistically meaningful constraints to be placed on LIV-induced transparency effects.

In this context, the GPSP-LIV atlas provides a practical framework for current and future wide-field UHE gamma-ray observatories targeting Galactic PeVatrons and other extreme particle accelerators. By allowing direct comparisons between standard and LIV-modified propagation, the atlas offers a systematic approach for investigating potential deviations from standard gamma-ray absorption and for placing indirect constraints on LIV scenarios.

\section{Observable Spectral Signatures of $\gamma\gamma$ Attenuation in Galactic Sources}
\label{gpsp_application}

The attenuation of gamma rays introduces energy-dependent spectral modifications that become increasingly important toward PeV energies. While the overall effect remains modest for nearby Galactic sources, the attenuation can become important for distant sources, particularly those located toward the inner Galaxy or tangential directions. Consequently, the observed spectra of Galactic sources, and in particular Galactic PeVatron candidates, may differ significantly from their intrinsic emission spectra in the UHE regime.

The GPSP atlas developed in this work provides a practical framework for quantitatively evaluating these effects for arbitrary source configurations. In contrast to traditional line of sight calculations, the atlas allows attenuation estimates not only for point-like sources but also for spatially extended emission regions where the survival probability may vary significantly across the source extent. This capability is particularly relevant for the GC environment, giant molecular cloud complexes, and large SFRs, where UHE gamma-ray emission is generally extended.

In the following subsections, the GPSP atlas is applied to representative Galactic sources. These examples are used to investigate how Galactic $\gamma\gamma$ attenuation modifies observable gamma-ray spectra, influences the interpretation of potential Galactic PeVatrons, and can be incorporated into studies of LIV using the GPSP-LIV atlas framework.

\subsection{GPSP Atlas Application to the Galactic Center PeVatron Region}
\label{gc_region_example}

The GC region represents one of the most intriguing Galactic PeVatron candidates currently known. Using deep observations with H.E.S.S., \citet{hess_gc_pevatron} reported diffuse gamma-ray emission from an annular region surrounding the GC, defined by inner and outer radii of $0.15^\circ$ and $0.45^\circ$, respectively, while excluding known gamma-ray sources. The measured diffuse gamma-ray spectrum was found to be compatible with a power-law (PL) distribution of the form 
\begin{equation}
\frac{dN}{dE}=N_0\left(\frac{E}{E_0}
\right)^{-\Gamma},
\end{equation}
with spectral index $\Gamma = 2.32 \pm 0.05_{\rm stat} \pm 0.11_{\rm sys}$, extending up to several tens of TeV without evidence for a significant spectral cutoff. Under the assumption of a hadronic origin, the corresponding 95$\%$ confidence level lower limit on the parent proton cutoff energy was estimated to be $\sim$0.4~PeV, suggesting the presence of CRs accelerated close to PeV energies. The inferred radial distribution of emission further suggested the existence of a quasi-continuous accelerator operating near or at the GC, making this region one of the strongest Galactic PeVatron candidates. However, due to the lack of observations above $\sim$100~TeV, the PeVatron nature of the source cannot yet be conclusively established.

\begin{figure*}[ht!]
\centering
\includegraphics[width=18.0cm]{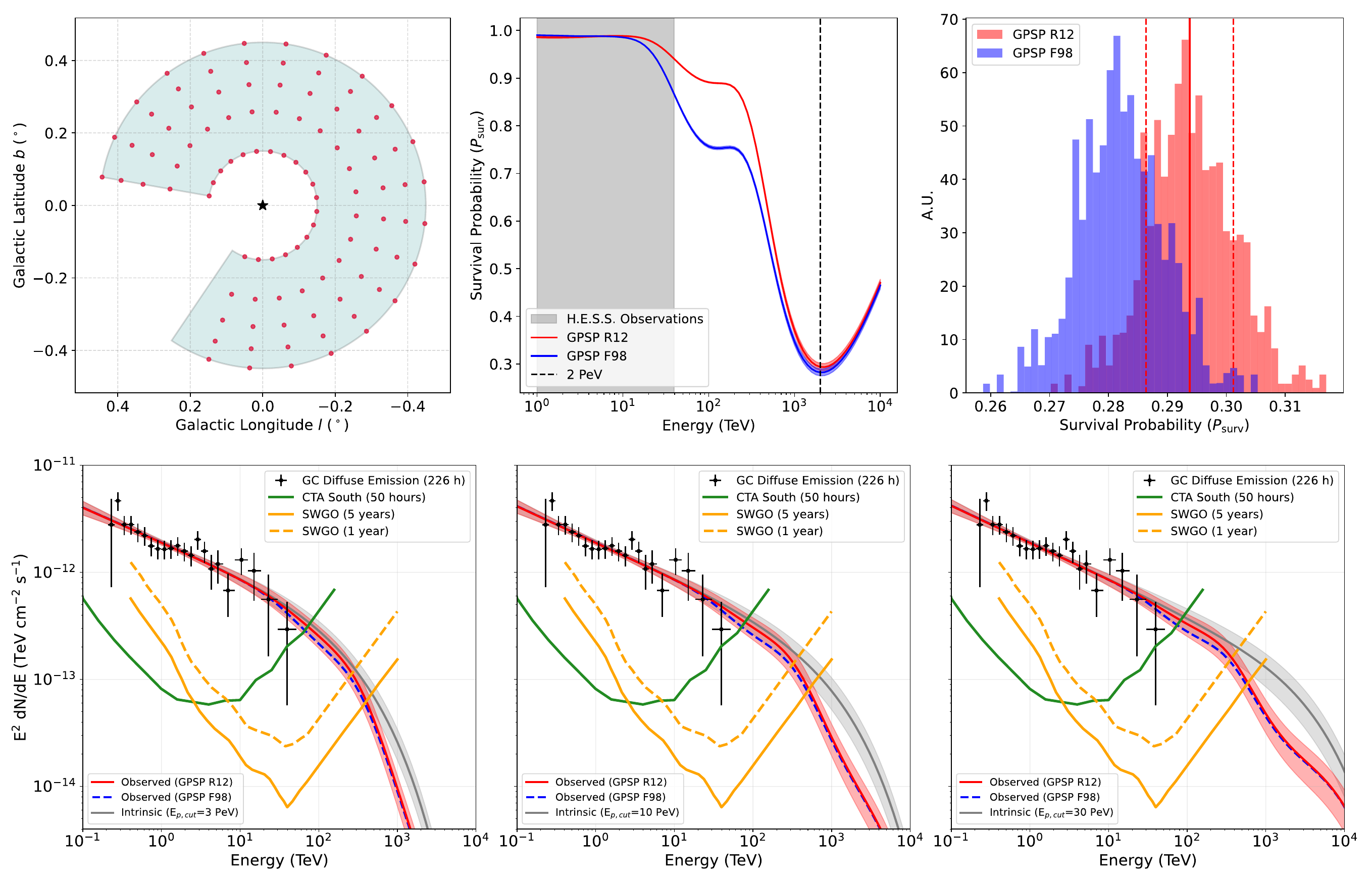}
\caption{\textbf{Top row:} Diagnostics of the survival probability for the GC diffuse-emission region. \textbf{Top left:} Spatial sampling geometry for the H.E.S.S. ROI (0.15$^\circ$$<$r$<$0.45$^\circ$, blue shaded region). Red markers indicate the 100 sampling locations used for the region-averaged calculations, while the black star marks Sgr A*. \textbf{Top middle:} Region-averaged survival probability profiles extracted from the GPSP R12 (red) and F98 (blue) atlases. Shaded bands show the spread of attenuation realizations resulting from spatial and distance variations within the ROI. The gray region indicates the H.E.S.S. observation range ($\lesssim40$ TeV), and the vertical dashed line marks 2~PeV. \textbf{Top right:} Distributions of sampled survival probabilities at 2~PeV for the R12 and F98 models. Solid lines indicate the weighted median values, while dashed lines denote the central 68$\%$ intervals. \textbf{Bottom row:} Comparison of intrinsic and absorbed gamma-ray spectra for proton cutoff energies of 3~PeV (left), 10~PeV (center), and 30~PeV (right). Black points show the H.E.S.S. diffuse-emission flux data. Sensitivity curves for CTA South (50 hr, prod5 v0.1) \citep{ctao_sens}, SWGO (1 yr), and SWGO (5 yr) \citep{swgo_science} are shown by the green solid, orange dashed, and orange solid lines, respectively. The intrinsic spectra are shown in gray, while attenuation-corrected spectra derived using the GPSP R12 and GPSP F98 atlases are shown by the red solid and blue dashed lines, respectively. The GPSP R12 predictions include the corresponding $1\sigma$ uncertainty bands.}
\label{gc_pev_diag}
\end{figure*}

The extended morphology of the GC diffuse emission region provides a useful test case for the GPSP atlas framework.~Unlike point-like sources, the attenuation profile across an extended emission region is spatially nonuniform due to variations in both the ISRF density and the line of sight geometry. Consequently, the observable attenuation signature depends not only on gamma-ray energy and source distance, but also on the spatial distribution of the emission itself. A realistic attenuation treatment therefore requires averaging the survival probability over the full emitting region rather than relying on a single line of sight approximation.

In this work, the GC diffuse emission region reported by H.E.S.S. is represented by spatial sampling points distributed uniformly across the ROI. For each sampled position, the corresponding survival probability profile is extracted from the GPSP atlas through multidimensional interpolation within the underlying $(l,b,d,E)$ parameter space. The resulting survival probability realizations are then combined using weighted averaging to construct an effective region-averaged attenuation profile. Distance uncertainties are incorporated through Monte Carlo (MC) sampling around the nominal GC distance of $8.5 \pm 0.2$~kpc, assuming a Gaussian distribution. For each realization, the sampled distance is propagated independently through the interpolation framework, naturally accounting for the nonlinear dependence of $\gamma\gamma$ attenuation on both source distance and ISRF structure. The top-left panel of Fig.~\ref{gc_pev_diag} illustrates the adopted spatial sampling geometry, assuming uniform spatial weighting and 100 sampling locations. The number of sampling locations was chosen to exceed the number of independent atlas resolution elements contained within the extraction region, using the approximate criterion
\begin{equation}
N_{\rm points} \gtrsim \frac{A}{\Delta l \times \Delta b}
\label{n_points_eq}
\end{equation}
where A is the angular area of the region and $\Delta l$~=~$\Delta b$ = 0.1$^{\circ}$ correspond to the GPSP atlas angular resolution. For the Galactic Center PeVatron region considered here, the extraction region contains $\sim$50 independent resolution elements, while $N_{\rm points}$=100 sampling points were adopted in the averaging procedure. A dedicated convergence test was performed by repeating the profile extraction procedure using sampling numbers between $N_{\rm points}$=10 and $N_{\rm points}$=10$^{4}$. The resulting region-averaged survival probability profiles were found to be numerically stable, with the adopted value of $N_{\rm points}$=100 yielding median fractional deviations below 10$^{-4}$ over the full investigated energy range relative to a highly sampled reference solution obtained with $N_{\rm points}$=10$^{4}$. However, for significantly larger source regions, such as Galactic diffuse-emission studies, the number of sampling locations should be increased to ensure sufficient coverage of the underlying atlas resolution elements according to Eq.~\ref{n_points_eq}. Uniform spatial weighting was adopted because no prior information on the intrinsic gamma-ray surface brightness distribution within the extraction region was assumed. Although uniform spatial weighting was adopted in this analysis, the GPSP framework can readily accommodate alternative morphology-dependent weighting schemes\footnote{For example, centrally concentrated emission profiles may be approximated using Gaussian weighting functions with widths corresponding to the measured source extension.}. The top-middle panel shows the corresponding region-averaged survival probability profiles extracted from the GPSP R12 and GPSP F98 atlases. The uncertainty bands shown in the figure do not represent observational uncertainties; instead, they reflect the spread of attenuation realizations originating from the combined effects of the source extent and distance uncertainty across the ROI. The top-right panel presents the distribution of sampled survival probabilities at 2~PeV for both ISRF models.

To investigate the observable impact of Galactic $\gamma\gamma$ absorption, the measured H.E.S.S. diffuse gamma-ray spectrum is interpreted within a hadronic scenario in which gamma rays are produced through $pp$ interactions. Since Galactic attenuation remains relatively weak over the H.E.S.S. energy range\footnote{The attenuation is below $\sim$5$\%$ between 10 and 40~TeV for the R12 model, which is well below the statistical and systematic uncertainties of the H.E.S.S. flux measurements; see the top-middle panel of Fig.~\ref{gc_pev_diag}}, the observed spectrum provides a reasonable approximation of the intrinsic source spectrum for the purpose of reconstructing the parent proton distribution. The intrinsic proton spectrum is modeled using an exponential cutoff power-law (ECPL) distribution
\begin{equation}
\frac{dN_p}{dE_p}=N_0\left(\frac{E_p}{E_0}
\right)^{-\alpha_p}\exp\left(-\frac{E_p}{E_{\mathrm{p, cut}}}
\right),
\end{equation}
where the normalization $N_0$ and proton spectral index $\alpha_p$ are treated as free parameters, while the proton cutoff energy $E_{\mathrm{p, cut}}$ is fixed. Three representative cutoff energies of 3, 10, and 30~PeV are considered in order to explore observable signatures associated with different maximum acceleration energies relevant to Galactic PeVatron scenarios and the origin of CRs around the knee region. For each scenario, the proton model is fitted to the H.E.S.S. flux data points using the \texttt{NaimaSpectralModel} implementation within \texttt{Gammapy}, and the corresponding intrinsic gamma-ray spectrum is calculated.

The GPSP atlas attenuation correction is then applied to the intrinsic gamma-ray spectra using the region-averaged survival probability profile derived for the GC diffuse emission region. This procedure allows the prediction of the observable UHE gamma-ray spectra expected after propagation, making comparison with the projected sensitivities of CTA South and SWGO possible in order to investigate the detectability of attenuation-induced spectral suppression features at UHE energies. 

The bottom row of Fig.~\ref{gc_pev_diag} illustrates the impact of Galactic $\gamma\gamma$ attenuation on the expected observable gamma-ray spectra for proton cutoff energies of $E_{\rm p,cut}=3$~PeV (left panel), $10$~PeV (middle panel), and $30$~PeV (right panel). For all three scenarios, the nominal 50~h CTA South sensitivity appears insufficient to significantly probe the predicted UHE emission, suggesting that deeper observations of $\mathcal{O(\text{100})}$ hr would be required to achieve robust detections above $\sim$100~TeV in this region. Although the R12 and F98 ISRF models produce noticeable differences in the survival probability profiles around $\sim$100--200~TeV, the corresponding differences in the attenuated gamma-ray spectra remain comparatively small. In this energy range the attenuation is still moderate, with survival probabilities of approximately 0.9 and 0.75 for the R12 and F98 models, respectively. Consequently, the resulting spectral differences correspond primarily to flux variations at the $\sim$15$\%$--20$\%$ level, making them challenging to distinguish observationally given the expected statistical and systematic uncertainties of current and near-future instruments.

The projected SWGO sensitivities will likely provide improved access to the UHE regime. Even under the conservative 3~PeV cutoff scenario, the attenuated spectra remain detectable up to several hundred TeV within 1 yr of observations, while longer observations may extend the detectable energy range further depending on the intrinsic proton cutoff energy.

One of the most intriguing features emerging from these spectra is the increasing suppression above several hundred TeV caused by strong $\gamma\gamma$ attenuation in the CMB-dominated regime. In this energy range, attenuation introduces a propagation-induced steepening of the observable spectra, producing apparent "cutoff-like" features even when the intrinsic source spectra extend to significantly higher energies. As a consequence, the observable spectra corresponding to different intrinsic proton cutoff energies become gradually more similar at the highest energies. The UHE emission therefore becomes increasingly propagation dominated rather than source dominated, illustrating how Galactic $\gamma\gamma$ attenuation can partially mask the intrinsic maximum acceleration energy of Galactic PeVatrons.

The spectra further indicate that the direct detection of PeV photons from the GC diffuse emission region remains challenging within hadronic $pp$ scenarios, even for proton cutoff energies as high as 30~PeV. Under the assumptions adopted here, the detection of photons at or above PeV energies would likely require very long ($\gg$~5 yr) SWGO observation times. Consequently, a future detection of gamma rays extending deeply into the PeV regime from the GC region may point either to exceptionally efficient particle acceleration or to the presence of additional gamma-ray production channels beyond standard $pp$ interactions. One possibility involves photohadronic $p\gamma$ interactions of ultrarelativistic protons with the intense UV radiation fields surrounding the GC environment, similar to interpretations proposed for Cygnus X-3. Such processes may enhance the production of observable PeV gamma rays despite the strong Galactic attenuation effects~\citep{Kachelriess2025,Zdziarski2026}.

\subsection{Correction of the Cygnus X-3 Spectrum for $\gamma\gamma$ Absorption Effects}
\label{cygnus_std_corr}

\begin{figure*}[ht!]
\centering
\includegraphics[width=18.0cm]{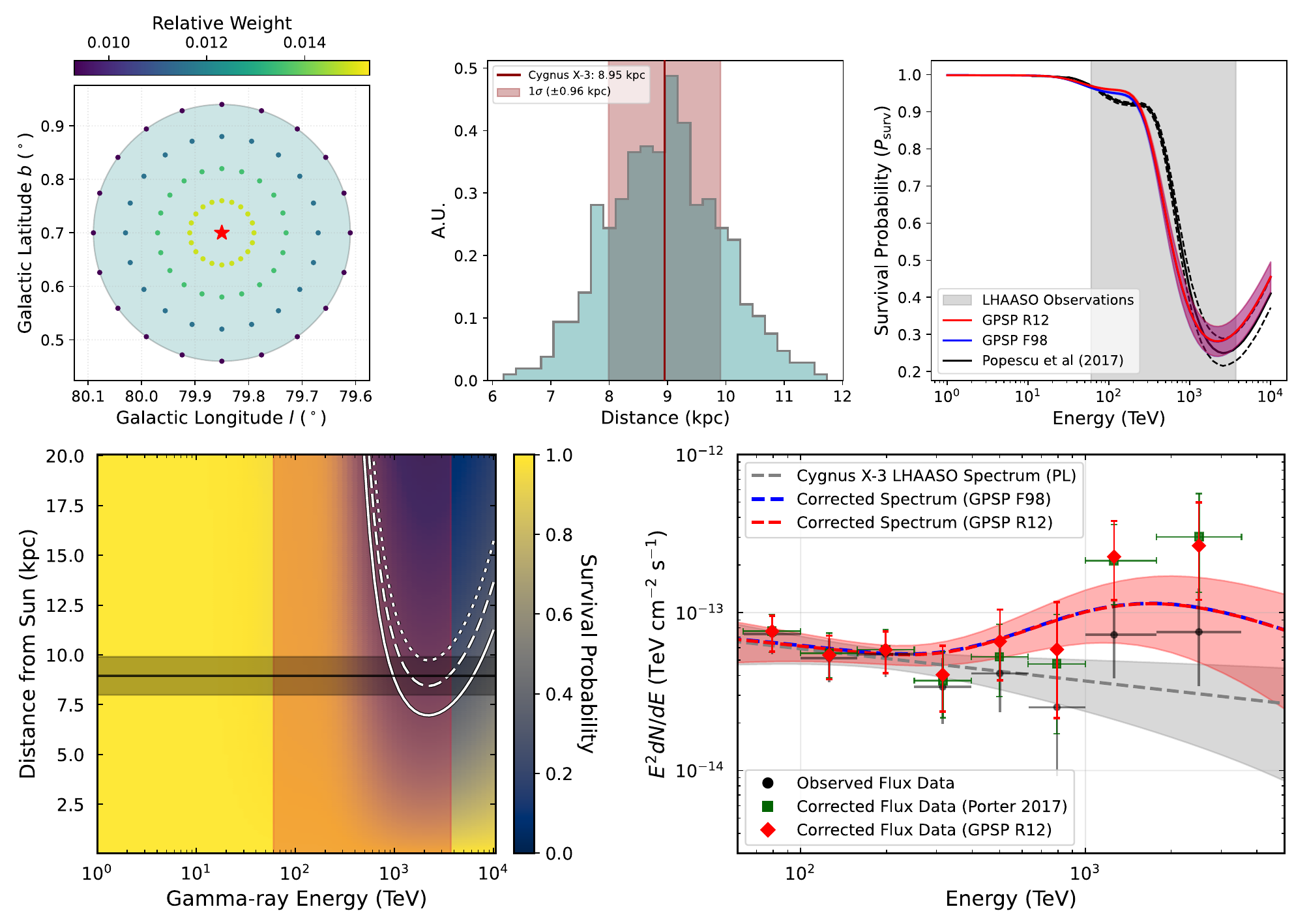}
\caption{Multipanel diagnostics of the Galactic $\gamma\gamma$ attenuation analysis for Cygnus X-3. \textbf{Top left:} Spatial sampling distribution of 100 positions in Galactic coordinates centered on Cygnus X-3 (red star). Colors indicate the Gaussian spatial weights assigned to the sampling locations. \textbf{Top middle:} MC distance distribution adopted for Cygnus X-3. The solid dark-red line marks the nominal distance of 8.95~kpc, while the shaded region denotes the corresponding $1\sigma$ uncertainty of $\pm0.96$~kpc. \textbf{Top right:} Survival probability profiles derived from the GPSP R12 (red) and F98 (blue) atlases, with shaded bands indicating the corresponding uncertainties. The solid black line shows the attenuation profile adopted by \citet{cygnus_x3_lhaaso} based on the P17 ISRF model, while the dashed black lines correspond to source distances of 7.99 and 9.91~kpc (Fig.~M5 of \citealt{cygnus_x3_lhaaso}). The gray shaded region indicates the LHAASO energy range. \textbf{Bottom left:} Survival probability phase space toward the Cygnus X-3 direction. White solid, dashed, and dotted curves indicate survival probability contours of 0.37, 0.30, and 0.25, respectively. The horizontal black line and shaded band mark the nominal source distance and its uncertainty, while the vertical red band indicates the LHAASO energy range. \textbf{Bottom right:} Comparison of the observed and attenuation-corrected SEDs. Dashed lines show the best-fit observed spectrum (gray) and the spectra corrected using the GPSP R12 (red) and F98 (blue) atlases. The red shaded band indicates the combined uncertainty of the GPSP R12-corrected spectrum. Flux points correspond to the measured LHAASO data (black circles), the P17-corrected spectrum (green squares), and the GPSP R12-corrected spectrum (red diamonds).}
\label{cygnus_x3_diag}
\end{figure*}

In the previous subsection, the GPSP atlas was used to estimate the observable UHE spectrum of the GC diffuse emission after propagation, assuming that the intrinsic source spectrum is known. In this subsection, the reverse procedure is demonstrated by reconstructing the intrinsic spectrum of the Cygnus X-3 region from the observed UHE gamma-ray data through correction for Galactic $\gamma\gamma$ absorption.

Cygnus X-3 is one of the most extensively studied high-mass X-ray binaries in the Galaxy \citep{Giacconi_1967}, showing complex multiwavelength variability \citep{Becklin_1973,Molnar_1984}. The system consists of a compact object orbiting a Wolf--Rayet (WR) companion star embedded within a dense stellar wind environment \citep{Kerkwijk_1992}, producing strong nonthermal activity across the spectrum \citep{Tavani_2009}. The discovery of relativistic radio jets established Cygnus X-3 as a microquasar \citep{Mioduszewski_2001,Marti_2001,MJ_2004}.

Recently, the LHAASO Collaboration reported variable gamma-ray emission from the direction of Cygnus X-3 extending up to $\sim$4~PeV with a significance of $\sim$10$\sigma$, providing strong evidence for particle acceleration to up to at least tens of PeV \citep{cygnus_x3_lhaaso}. The observed emission shows month-scale variability together with indications of orbital modulation, suggesting that the UHE emission region is located within, or in close proximity to, the binary system itself. The measured spectrum was described by a PL model, where the reference energy was fixed at $E_0=50$~TeV. The reported best-fit parameters were $N_0=(2.6\pm0.6)\times10^{-17}$~TeV$^{-1}$~cm$^{-2}$~s$^{-1}$ and $\Gamma=2.18\pm0.14$. After correcting for attenuation effects, the intrinsic SED derived by the LHAASO Collaboration shows a clear hardening toward PeV energies. Such behavior is difficult to reproduce within standard hadronic $pp$ interactions alone and has been interpreted as evidence for additional photohadronic $p\gamma$ processes taking place within the intense radiation environment of the system. In particular, interactions between accelerated protons and UV photons from the WR companion star have been proposed as a natural explanation for the spectral hardening at PeV energies, while interactions involving X-ray photons from the accretion flow, jet environment, or stellar wind may contribute to the lower-energy component below $\sim$1~PeV \citep{cygnus_x3_lhaaso}.

Given its Galactic location ($l$=79.85$^\circ$, $b$=0.70$^\circ$), estimated distance of $\sim$8.95$\pm$0.96~kpc, and gamma-ray spectrum extending beyond PeV energies, Cygnus X-3 provides an excellent target for investigating Galactic $\gamma\gamma$ attenuation effects outside the GC direction. Unlike the inner Galaxy, where most ISRF models predict strong attenuation, the Cygnus X-3 line of sight passes through an intermediate-opacity Galactic environment, making it particularly suitable for comparing attenuation predictions derived from different ISRF models.

The top-left panel of Fig.~\ref{cygnus_x3_diag} shows the spatial sampling distribution used for the Cygnus X-3 region. Although the source appears point-like in the LHAASO observations, the corresponding emission region remains spatially extended\footnote{The LHAASO point spread function at 0.1~PeV is approximately $0.24^\circ$; therefore, the corresponding region remains spatially extended with respect to the $0.1^\circ$ angular resolution of the GPSP atlas.} for the GPSP atlas. Consequently, 100 sampling locations were generated according to a Gaussian spatial weighting profile with $\sigma=0.24^\circ$, corresponding to the LHAASO point-spread function at 0.1~PeV as shown in the top-left panel of Fig.~\ref{cygnus_x3_diag}. In addition, the source distance uncertainty of $\pm0.96$~kpc was taken into account through MC sampling of 500 distance realizations (see top-middle panel of Fig.~\ref{cygnus_x3_diag}). The combined spatial and distance sampling procedure therefore yields $100\times500=50,000$ survival probability profiles that are subsequently extracted from the atlas and used to estimate uncertainties on the reconstructed attenuation profile.

The resulting survival probability profiles are shown in the top-right panel of Fig.~\ref{cygnus_x3_diag}, together with the attenuation profile adopted by \citet{cygnus_x3_lhaaso} based on the ISRF model of \citet{Popescu_2017} (hereafter P17). For completeness, the corresponding 2D survival probability phase space along the Cygnus X-3 line of sight is presented in the bottom-left panel. Overall, all models predict qualitatively similar behavior, with the survival probability remaining close to unity below several hundred TeV before gradually decreasing toward a minimum in the PeV regime. Nevertheless, several systematic differences are visible.

Indeed, the most noticeable difference is the IR shoulder feature. In the P17 model, this feature is broader, extends to energies of $\sim$300~TeV, and produces stronger attenuation than in the GPSP-based profiles. In contrast, the R12- and F98-based attenuation profiles exhibit a smoother and less pronounced feature extending only to $\sim$200~TeV. As a consequence, the attenuation minimum occurs near $\sim$3~PeV in the P17 model, whereas it appears closer to $\sim$2.2~PeV in the R12 and F98 models. These differences likely originate from the distinct ISRF descriptions used by the underlying models. The P17 framework is based on an axisymmetric radiative transfer treatment of the Galaxy, whereas the R12 and F98 models incorporate a three-dimensional ISRF structure, as discussed in Sect.~\ref{isrf_sect}. Along the Cygnus X-3 line of sight, which passes close to the Galactic plane without directly intersecting the most intense IR environments, the attenuation profile becomes particularly sensitive to the detailed spatial distribution of IR photons. In this regime, localized variations in the 3D ISRF structure can modify and reduce the line-of-sight-integrated photon density relative to smoother axisymmetric models. Since pair production interactions with IR photons dominate the attenuation process around $\sim$100--$200$~TeV, moderate differences in the effective IR photon density can influence both the depth and width of the attenuation feature, as well as the energy at which the survival probability reaches its minimum. Although the differences between the models remain modest and well within plausible systematic uncertainties associated with Galactic ISRF modeling, the comparison illustrates that UHE attenuation predictions outside the GC direction are not entirely model independent.

To reconstruct the intrinsic gamma-ray spectrum of Cygnus X-3, the observed flux points reported by the LHAASO Collaboration (open black circles in Fig.~3 of \citealt{cygnus_x3_lhaaso}) were first fitted with a PL spectral model using the \texttt{Gammapy} package. The resulting best-fit parameters were found to be $N_0=(2.7\pm0.8)\times10^{-17}$~TeV$^{-1}$~cm$^{-2}$~s$^{-1}$ and $\Gamma=2.20\pm0.21$, fully consistent with the values reported by the LHAASO Collaboration. The best-fit spectral model was then corrected using the GPSP survival probability profiles derived for the Cygnus X-3 region. The correction was performed by dividing the observed flux by the corresponding survival probability at each energy. Uncertainties associated with the GPSP attenuation profile were propagated into the reconstructed intrinsic spectrum. In addition, the published LHAASO flux points were corrected individually using the same procedure in order to provide a direct comparison between the attenuation corrections obtained with the P17 model and those derived from the GPSP atlas.

The bottom-right panel of Fig.~\ref{cygnus_x3_diag} presents the comparison between the best-fit observed spectrum (gray dashed line) and the attenuation-corrected intrinsic spectra obtained using the GPSP R12 atlas (red dashed line) and the GPSP F98 atlas (blue dashed line), together with their corresponding $1\sigma$ error bands. Consistent with the survival probability profiles, Galactic attenuation begins to influence the spectrum above $\sim$200~TeV and becomes increasingly stronger toward PeV energies. The attenuation correction produces a systematically harder intrinsic spectrum in the sub-PeV to PeV regime and introduces a noticeable spectral-hardening feature. The corrected flux points obtained using the P17 attenuation model (green squares) and the GPSP R12 atlas (red diamonds) are also shown together with the measured LHAASO data (black circles). Although the underlying survival probability profiles show systematic differences, the resulting corrected flux points remain in very good agreement within uncertainties. This indicates that, for the current LHAASO Cygnus X-3 dataset, systematic uncertainties associated with Galactic attenuation modeling remain smaller than the statistical uncertainties of the measurements.

It should also be noted that the uncertainties associated with the corrected spectrum and flux points become slightly larger than those of the original measured data due to the propagation of GPSP-related uncertainties during the correction procedure explained above. The spectral error bands shown in Fig.~\ref{cygnus_x3_diag} include both the statistical measurement uncertainties ($\sigma_{\rm stat}$) and the systematic uncertainties associated with the GPSP atlas ($\sigma_{\rm sys}$), while the latter are mainly driven by source location sampling, MC distance sampling, and interpolation effects. The total uncertainty was calculated by combining the two contributions in quadrature\footnote{A more conservative uncertainty estimate can be obtained using linear error propagation, $\sigma_{\rm total} = \sigma_{\rm stat} + \sigma_{\rm sys}$, resulting in larger uncertainty bands, although the difference remains small in the case of Cygnus X-3 since the statistical uncertainties dominate over the GPSP-related systematics.}, $\sigma_{\rm total}$ = ($\sigma_{\rm stat}^2 + \sigma_{\rm sys}^2)^{1/2}$.

The Cygnus X-3 application presented here demonstrates the capability of the GPSP atlas to perform spatially resolved Galactic attenuation calculations while simultaneously propagating GPSP-related uncertainties into reconstructed intrinsic spectra. More broadly, this example illustrates that line-of-sight-dependent differences between Galactic ISRF models may become increasingly relevant for future UHE observations extending into the multi-PeV regime, particularly once measurement uncertainties become comparable to systematic uncertainties associated with Galactic attenuation modeling.

\subsection{Investigation of Subluminal LIV Scenarios on Cygnus X-3 Spectrum Using the GPSP-LIV Atlas}
\label{cygnus_liv}

\begin{figure*}[ht!]
\centering
\includegraphics[width=18.0cm]{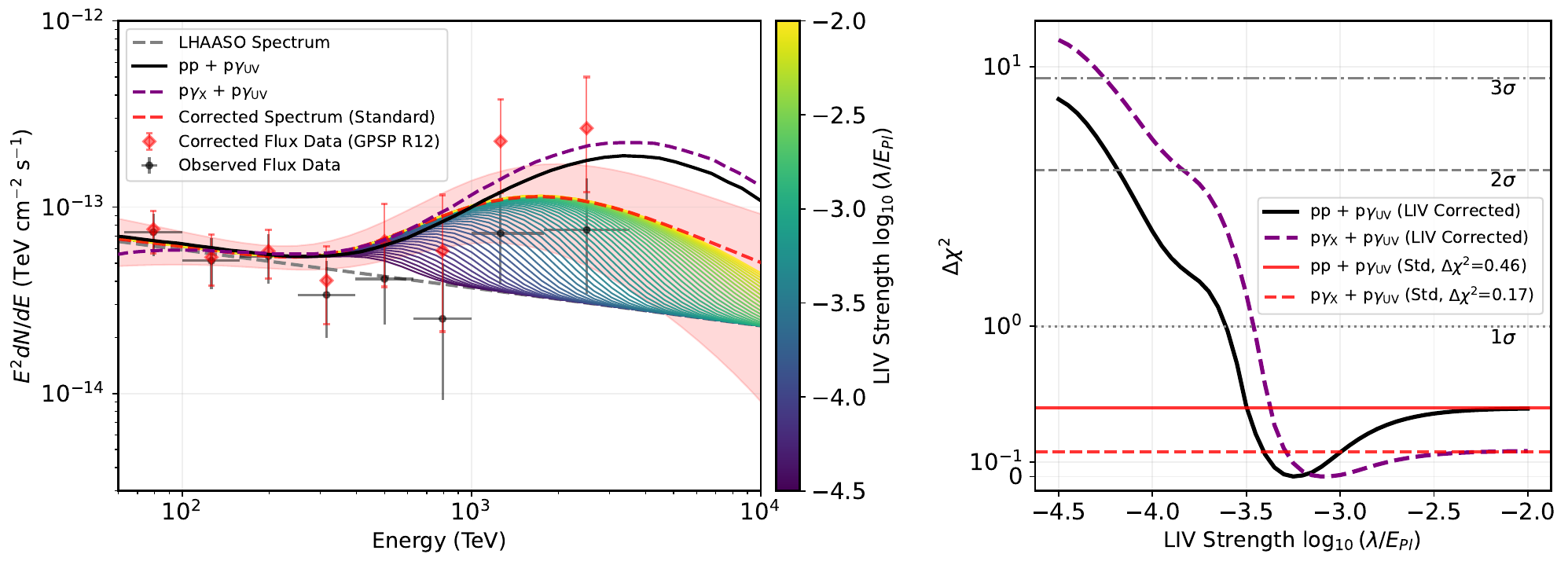}
\caption{Subluminal LIV sensitivity study of Cygnus X-3 using the GPSP-LIV atlas. \textbf{Left panel:} Measured LHAASO flux points (black circles) together with the best-fit observed spectrum (gray dashed line). The standard LI attenuation-corrected spectrum derived from the GPSP R12 atlas is shown by the red dashed line, with corresponding corrected flux points (red diamonds) and the combined statistical and systematic uncertainty band (red shaded region). Colored lines show the reconstructed intrinsic spectra for different values of the subluminal LIV parameter $\log_{10}(\lambda/E_{\rm Pl})$. The intrinsic photohadronic templates from \citet{cygnus_x3_lhaaso} are overlaid for the $pp+p\gamma_{\rm UV}$ (solid black) and $p\gamma_{\rm X}+p\gamma_{\rm UV}$ (dashed purple) scenarios. \textbf{Right panel:} $\Delta\chi^{2}$ profiles derived using the split-normal statistic as a function of $\log_{10}(\lambda/E_{\rm Pl})$ for comparisons with the $pp+p\gamma_{\rm UV}$ (solid black) and $p\gamma_{\rm X}+p\gamma_{\rm UV}$ (dashed purple) templates. Horizontal red lines indicate the corresponding LI reference values, while the gray dotted, dashed, and dotted--dashed lines mark the conventional $1\sigma$, $2\sigma$, and $3\sigma$ reference levels.}
\label{cygnus_x3_liv_test}
\end{figure*}

As discussed in Sect.~\ref{liv_horizons}, Galactic sources emitting gamma rays at $\sim$2--3~PeV energies and particularly located beyond the traditional gamma-ray horizon provide promising targets for testing subluminal LIV scenarios. As evident from the survival probability phase space (bottom-left panel of Fig.~\ref{cygnus_x3_diag}), Cygnus X-3 lies beyond the traditional gamma-ray horizon, even when the lower bound of its distance uncertainty is adopted. The nominal source distance is consistent with the $P_{\rm Surv}=$0.3 contour and approaches $P_{\rm Surv}$$\approx$0.25 when the upper distance estimate is considered. This makes Cygnus X-3 a particularly suitable target for demonstrating the application of the GPSP-LIV atlas framework. More generally, such survival probability phase space maps provide a convenient tool for identifying Galactic sources that may be especially sensitive to LIV-induced transparency effects.

The left panel of Fig.~\ref{cygnus_x3_liv_test} compares the intrinsic spectra reconstructed using the standard LI attenuation absorption from the GPSP R12 atlas (red-dashed line) with those obtained under a range of subluminal LIV strengths of log$_{10}$($\lambda$/$E_{\mathrm{Pl}}$). The measured LHAASO flux points and the corresponding LI-corrected flux points are also shown. For comparison, the intrinsic gamma-ray spectral templates proposed by \citet{cygnus_x3_lhaaso} (see Fig.~4 therein) for the $pp+p\gamma_{\rm UV}$ (black solid line) and $p\gamma_{\rm X}+p\gamma_{\rm UV}$ (dashed purple line) scenarios are overlaid. As the LIV strength increases, pair production attenuation becomes gradually suppressed and the Galaxy becomes increasingly transparent to PeV gamma rays (see survival profiles in Fig.~\ref{liv_phase_space}). Consequently, the reconstructed intrinsic spectra show less pronounced hardening toward the PeV regime than in the standard LI case. Since the strongest attenuation effects occur near $\sim$2--3~PeV, the LIV sensitivity is driven mainly by the highest-energy spectral points of Cygnus X-3. Comparison of the reconstructed LIV-corrected flux points with the published photohadronic model predictions therefore provides a useful framework for assessing the observational compatibility of different LIV scenarios.

In order to demonstrate the application of the GPSP-LIV atlas framework, a simple approach is followed. First, for each tested LIV strength value of $\log_{10}(\lambda/E_{\rm Pl})$, survival probability profiles were extracted from the GPSP-LIV atlas following the same spatial and distance sampling procedure described in Sect.~\ref{cygnus_std_corr}. The resulting survival profiles were then used to correct the measured LHAASO flux points individually. Both the statistical measurement uncertainties and the systematic uncertainties associated with the survival probability profiles were propagated into the reconstructed intrinsic flux points\footnote{It is important to note that the reconstructed intrinsic flux points themselves are LIV-dependent quantities, since they are obtained by correcting the observed spectrum using survival probabilities computed under each assumed LIV scenario.} and combined in quadrature under the assumption that the two contributions are approximately independent. Repeating the correction procedure for all atlas values within the interval $\log_{10}(\lambda/E_{\rm Pl})\in[-4.5,-2.0]$ using a step size of 0.05 therefore produces a family of LIV-corrected intrinsic flux point datasets, which have asymmetric errors.

To evaluate the compatibility of each LIV-corrected dataset with the intrinsic photohadronic spectral templates, a split-normal $\chi^2$ statistic was adopted in order to account for the asymmetric uncertainties of the reconstructed flux points
\begin{equation}
\label{split_chi}
    \chi^2_{\mathrm{split}} = \sum_{i=1}^{N}
    \begin{cases}
        \displaystyle \left( \frac{x_i - \mu_i}{\sigma_{i,+}} \right)^2, & \text{if } \mu_i > x_i
        \\
        \\
        \displaystyle \left( \frac{x_i - \mu_i}{\sigma_{i,-}} \right)^2, & \text{if } \mu_i \le x_i
    \end{cases}
\end{equation}
where $x_i$ is the reconstructed flux value, $\mu_i$ is the photohadronic model prediction, and $\sigma_{i,+}$ and $\sigma_{i,-}$ represent the upper and lower $1\sigma$ flux uncertainties, respectively. This approach approximately accounts for asymmetric confidence intervals by selecting the appropriate uncertainty depending on whether the model prediction lies above or below the measured value. Compared to the common practice of replacing asymmetric uncertainties with their arithmetic average, $\sigma_{\mathrm{err}}=(\sigma_{i,+}+\sigma_{i,-})/2$, the split-normal formulation reduces the statistical bias introduced by enforcing symmetric uncertainties.

For each tested LIV strength, $\chi^2_{\rm split}$ values were computed by comparing the corresponding LIV-corrected flux point datasets with the intrinsic gamma-ray spectral templates associated with the $pp+p\gamma_{\rm UV}$ and $p\gamma_{\rm X}+p\gamma_{\rm UV}$ scenarios using Eq.~\ref{split_chi}. In order to provide a more sensitive comparison metric, the resulting test statistic profiles were expressed as
\begin{equation}
    \Delta\chi^{2} = \chi^2_{\mathrm{split}} - \chi^2_{\mathrm{split,min}}.
\end{equation}
where $\chi^2_{\rm split,min}$ is the minimum value obtained within the explored LIV-parameter range. It should be explicitly noted that this statistical treatment is intended only as a proof-of-principle demonstration rather than a rigorous likelihood analysis. In particular, the reconstructed spectral points are not strictly independent because common survival probability profiles introduce correlated systematic uncertainties that are neglected here. More importantly, the intrinsic photohadronic spectra adopted from \citet{cygnus_x3_lhaaso} are treated as fixed templates, whereas realistic emission models depend on several free parameters, including the proton spectral index, cutoff energy, interaction efficiencies, and target photon-field properties. Consequently, the resulting $\Delta\chi^2$ profiles should not be interpreted as model-independent constraints on LIV, but rather as a demonstration of how the GPSP-LIV atlas can be used to investigate the observational compatibility of different LIV scenarios.

The right panel of Fig.~\ref{cygnus_x3_liv_test} shows the resulting $\Delta\chi^{2}$ profiles obtained for both pp + p$\gamma_{\mathrm{UV}}$ (black solid line) and p$\gamma_{\mathrm{X}}$ + p$\gamma_{\mathrm{UV}}$ (dashed purple line) photohadronic spectral templates, together with the traditional Gaussian-equivalent $1\sigma$--$2\sigma$--$3\sigma$ $\Delta\chi^{2}$ reference levels. As expected, the profiles converge toward the standard LI solutions at weak LIV strengths, providing an internal consistency check for the GPSP-LIV atlas implementation. The figure also indicates that moderate LIV scenarios can lead to slightly improved agreement with the adopted intrinsic photohadronic spectral templates, although the statistical preference remains weak (well below 1$\sigma$) and should not be interpreted as evidence of LIV effects. On the other hand, strong LIV scenarios corresponding approximately to $\log_{10}(\lambda/E_{\mathrm{Pl}})\lesssim-4.2$ produce $\Delta\chi^{2}$ values exceeding the $3\sigma$ reference threshold under the simplified statistical treatment adopted here for the $p\gamma_{\mathrm{X}} + p\gamma_{\mathrm{UV}}$ template.

The primary purpose of this subsection is not to derive robust astrophysical constraints on LIV, but rather to demonstrate the practical application of the GPSP-LIV atlas framework. A rigorous LIV study would require a full likelihood-based analysis in which the underlying photohadronic emission model parameters are allowed to vary and are simultaneously fitted to the LIV-corrected flux point dataset. Such an investigation is beyond the scope of this work. Nevertheless, the analysis presented in this subsection demonstrates how the GPSP-LIV atlas can be systematically incorporated into studies of Galactic PeV gamma-ray sources. In particular, the fine $\log_{10}(\lambda/E_{\rm Pl})$ resolution of the GPSP-LIV atlas allows the construction of smooth statistical test profiles and provides a practical framework for future high-precision LIV investigations.

\section{Summary and Conclusion}
\label{summary}

With the rapidly growing population of Galactic UHE gamma-ray sources revealed by observatories such as LHAASO and HAWC, together with the forthcoming generation of facilities including CTA and SWGO, Galactic $\gamma\gamma$ absorption is becoming an increasingly important component of gamma-ray data analysis, requiring a unified and standardized framework. As observations extend toward the PeV regime, corrections for $\gamma\gamma$ attenuation are expected to play a critical role in reconstructing intrinsic source spectra, identifying Galactic PeVatrons, interpreting diffuse gamma-ray emission, and testing scenarios beyond standard physics.

In this work, a high-resolution GPSP atlas has been developed for modeling the propagation of VHE and UHE gamma rays within the Milky Way. The atlas provides a 4D survival probability framework, $P(E_{\gamma},l,b,d)$, covering energies from 1~TeV to 10~PeV, Galactic latitudes $|b|\leq5^\circ$, and source distances up to 20~kpc. The resulting data product is released in a ready-to-use FITS format compatible with widely used astrophysical analysis frameworks, allowing direct integration into current and future gamma-ray studies.

A key feature of the GPSP framework is the inclusion of two independent GALPROP-based ISRF models, namely the R12 and F98 models. This allows direct assessment of ISRF-related systematic uncertainties and comparisons between different assumptions regarding the Galactic radiation field. The largest model-dependent differences are found in the IR-dominated attenuation regime around $\sim$100--200~TeV, where the absorption is particularly sensitive to the distribution of Galactic dust emission. At PeV energies, the attenuation becomes increasingly dominated by interactions with the isotropic CMB, and the differences between the models gradually diminish.

Several new visualization and analysis techniques have also been introduced. A major development introduced in this work is the "gamma-ray survival phase space" concept, which extends the traditional 1D survival probability representation into a 2D energy--distance framework, providing a direct visualization of Galactic transparency, gamma-ray horizons, and the transition between optically thin and optically thick regimes. This representation is particularly useful for sources with uncertain distances and offers a more complete description of propagation effects than conventional 1D survival profiles.

Using the full framework of the GPSP atlas, top-down Galactic transparency maps and high-resolution Galactic plane survival maps have been constructed. These maps reveal that Galactic gamma-ray attenuation is highly structured and anisotropic, rather than being a simple function of source distance. Strong directional variations are found in the $\sim$100--300~TeV regime, where attenuation is enhanced along Galactic tangent directions and spiral-arm environments due to increased IR photon densities and longer effective propagation paths. At PeV energies, the attenuation progressively transitions into the CMB-dominated regime, where the gamma-ray horizons become increasingly isotropic and reach their minimum distances near $\sim$2--3~PeV.

The GPSP framework has further been extended to subluminal LIV scenarios through the construction of the GPSP-LIV atlas. This represents the first large-scale Galactic photon survival atlas incorporating LIV-modified $\gamma\gamma$ pair production kinematics and cross sections. The results show that subluminal LIV can substantially increase the transparency of the Galaxy to PeV gamma rays by modifying both the pair production threshold and interaction cross section. Consequently, Galactic sources located beyond the conventional gamma-ray horizons, particularly those emitting in the $\sim$2--3~PeV energy range, emerge as promising targets for future LIV studies.

Several practical applications of the GPSP atlas have been presented. The GC diffuse emission region was investigated as a representative example of a heavily absorbed inner-Galaxy environment, illustrating the importance of spatially resolved attenuation calculations for extended sources. The framework was also applied to the recently detected UHE emission from Cygnus X-3, where the observed spectrum was corrected for Galactic $\gamma\gamma$ absorption effects and compared with previous attenuation estimates. Finally, the GPSP-LIV atlas was used to explore potential LIV-induced modifications of the Cygnus X-3 spectrum, demonstrating how future high-statistics PeV observations may be used to investigate subluminal LIV scenarios.

Several aspects of the present GPSP atlas framework can be further improved in future work. In particular, the current atlas calculations employ an isotropic approximation of the ISRF and neglect the development of secondary electromagnetic cascades initiated by pair production interactions. Although these effects are expected to be subdominant for modest optical depths, they may become increasingly relevant for strongly absorbed sources in the inner Galaxy and at energies approaching the PeV regime. Future versions of the GPSP atlas may incorporate fully anisotropic ISRF treatments together with self-consistent electromagnetic cascade calculations in realistic Galactic magnetic fields, allowing a more complete description of gamma-ray propagation and attenuation in the Milky Way.

In summary, the GPSP and GPSP-LIV atlases provide a unified, high-resolution, and publicly available framework for studying Galactic gamma-ray propagation in both standard and LIV scenarios. As gamma-ray astronomy advances into the PeV domain, attenuation corrections are expected to become a standard component of UHE analyses. The resources presented in this work are intended to provide the community with a robust and extensible foundation for future studies of Galactic PeVatrons, diffuse gamma-ray emission, and fundamental-physics effects in the emerging era of Galactic PeV astronomy.

\section{Data Availability}
\label{data_section}
The data products and software developed in this work are publicly released to facilitate reproducibility and integration into existing high-energy astrophysics analysis workflows. The standard GPSP atlas is provided for both the R12 and F98 ISRF models as FITS files, \texttt{GPSP\_Atlas\_R12.fits.gz} ($\sim$11.2~GB) and \texttt{GPSP\_Atlas\_F98.fits.gz} ($\sim$11.2~GB), respectively.~The corresponding GPSP-LIV atlas products are also available for both ISRF models as \texttt{GPSP\_Atlas\_LIV\_R12.fits.gz} ($\sim$5.5~GB) and \texttt{GPSP\_Atlas\_LIV\_F98.fits.gz} ($\sim$5.5~GB). The atlas files, together with the accompanying Python helper modules (\texttt{gpsp\_atlas\_helper.py} and \texttt{gpsp\_liv\_atlas\_helper.py}), are hosted in a permanent Zenodo repository available at \url{https://doi.org/10.5281/zenodo.21403524}. 
\\
In addition to the data products themselves, a dedicated documentation website provides a complete user guide, API reference documentation, worked examples, and ready-to-run Jupyter notebooks demonstrating common analysis workflows. These resources include instructions for loading the atlas FITS files, extracting survival-probability profiles, applying attenuation corrections to gamma-ray spectra, and performing extended-source analyses. The online documentation is available through the project \emph{Read the Docs} website at \url{https://gpsp-atlas.readthedocs.io}.

\begin{acknowledgments}
\modulolinenumbers[10]
\textbf{Acknowledgments}\\ 
\\
E.O.A. gratefully acknowledge financial support from the TÜBİTAK Research Institute for Fundamental Sciences. This research has made use of the CTA instrument response functions provided by the CTA Consortium and Observatory, see \url{https://www.ctao-observatory.org/science/cta-performance/} (version prod5 v0.1; \citep{ctao_sens}) for more details.

\end{acknowledgments}

\bibliographystyle{aasjournal}
\bibliography{gpsp}

@ARTICLE{crab_first_TeV,
       author = {{Weekes}, T.~C. and {Cawley}, M.~F. and {Fegan}, D.~J. and {Gibbs}, K.~G. and {Hillas}, A.~M. and {Kowk}, P.~W. and {Lamb}, R.~C. and {Lewis}, D.~A. and {Macomb}, D. and {Porter}, N.~A. and {Reynolds}, P.~T. and {Vacanti}, G.},
        title = "{Observation of TeV Gamma Rays from the Crab Nebula Using the Atmospheric Cerenkov Imaging Technique}",
      journal = {\apj},
         year = 1989,
        month = jul,
       volume = {342},
        pages = {379},
          doi = {10.1086/167599},
       adsurl = {https://ui.adsabs.harvard.edu/abs/1989ApJ...342..379W}
}

@article{hess_Crab_paper,
	author = {{Aharonian, F.} and {Akhperjanian, A. G.} and {Bazer-Bachi, A. R.} and {Beilicke, M.} and {Benbow, W.} and {Berge, D.} and {Bernl\"ohr, K.} and {Boisson, C.} and {Bolz, O.} and {Borrel, V.} and {Braun, I.} and {Breitling, F.} and {Brown, A. M.} and {B\"uhler, R.} and {B\"usching, I.} and {Carrigan, S.} and {Chadwick, P. M.} and {Chounet, L.-M.} and {Cornils, R.} and {Costamante, L.} and {Degrange, B.} and {Dickinson, H. J.} and {Djannati-Ata\"{\i}, A.} and {Drury, L. O'C.} and {Dubus, G.} and {Egberts, K.} and {Emmanoulopoulos, D.} and {Espigat, P.} and {Feinstein, F.} and {Ferrero, E.} and {Fiasson, A.} and {Fontaine, G.} and {Funk, Seb.} and {Funk, S.} and {Gallant, Y. A.} and {Giebels, B.} and {Glicenstein, J. F.} and {Goret, P.} and {Hadjichristidis, C.} and {Hauser, D.} and {Hauser, M.} and {Heinzelmann, G.} and {Henri, G.} and {Hermann, G.} and {Hinton, J. A.} and {Hofmann, W.} and {Holleran, M.} and {Horns, D.} and {Jacholkowska, A.} and {de Jager, O. C.} and {Kh\'elifi, B.} and {Komin, Nu.} and {Konopelko, A.} and {Kosack, K.} and {Latham, I. J.} and {Le Gallou, R.} and {Lemi\`ere, A.} and {Lemoine-Goumard, M.} and {Lohse, T.} and {Martin, J. M.} and {Martineau-Huynh, O.} and {Marcowith, A.} and {Masterson, C.} and {McComb, T. J. L.} and {de Naurois, M.} and {Nedbal, D.} and {Nolan, S. J.} and {Noutsos, A.} and {Orford, K. J.} and {Osborne, J. L.} and {Ouchrif, M.} and {Panter, M.} and {Pelletier, G.} and {Pita, S.} and {P\"uhlhofer, G.} and {Punch, M.} and {Raubenheimer, B. C.} and {Raue, M.} and {Rayner, S. M.} and {Reimer, A.} and {Reimer, O.} and {Ripken, J.} and {Rob, L.} and {Rolland, L.} and {Rowell, G.} and {Sahakian, V.} and {Saug\'e, L.} and {Schlenker, S.} and {Schlickeiser, R.} and {Schwanke, U.} and {Sol, H.} and {Spangler, D.} and {Spanier, F.} and {Steenkamp, R.} and {Stegmann, C.} and {Superina, G.} and {Tavernet, J.-P.} and {Terrier, R.} and {Th\'eoret, C. G.} and {Tluczykont, M.} and {van Eldik, C.} and {Vasileiadis, G.} and {Venter, C.} and {Vincent, P.} and {V\"olk, H. J.} and {Wagner, S. J.} and {Ward, M.}},
	title = {Observations of the Crab nebula with HESS},
	DOI= "10.1051/0004-6361:20065351",
	url= "https://doi.org/10.1051/0004-6361:20065351",
	journal = {A\&A},
	year = 2006,
	volume = 457,
	number = 3,
	pages = "899-915",
}

@article{hess_gps,
	author = {{H.E.S.S. Collaboration} and {Abdalla, H.} and {Abramowski, A.} and {Aharonian, F.} and {Ait Benkhali, F.} and {Ang\"uner, E. O.} and {Arakawa, M.} and {Arrieta, M.} and {Aubert, P.} and {Backes, M.} and {Balzer, A.} and {Barnard, M.} and {Becherini, Y.} and {Becker Tjus, J.} and {Berge, D.} and {Bernhard, S.} and {Bernl\"ohr, K.} and {Blackwell, R.} and {B\"ottcher, M.} and {Boisson, C.} and {Bolmont, J.} and {Bonnefoy, S.} and {Bordas, P.} and {Bregeon, J.} and {Brun, F.} and {Brun, P.} and {Bryan, M.} and {B\"uchele, M.} and {Bulik, T.} and {Capasso, M.} and {Carrigan, S.} and {Caroff, S.} and {Carosi, A.} and {Casanova, S.} and {Cerruti, M.} and {Chakraborty, N.} and {Chaves, R. C. G.} and {Chen, A.} and {Chevalier, J.} and {Colafrancesco, S.} and {Condon, B.} and {Conrad, J.} and {Davids, I. D.} and {Decock, J.} and {Deil, C.} and {Devin, J.} and {deWilt, P.} and {Dirson, L.} and {Djannati-Ata\"{\i}, A.} and {Domainko, W.} and {Donath, A.} and {Drury, L. O\'{}C.} and {Dutson, K.} and {Dyks, J.} and {Edwards, T.} and {Egberts, K.} and {Eger, P.} and {Emery, G.} and {Ernenwein, J.-P.} and {Eschbach, S.} and {Farnier, C.} and {Fegan, S.} and {Fernandes, M. V.} and {Fiasson, A.} and {Fontaine, G.} and {F\"orster, A.} and {Funk, S.} and {F\"u\ss{}ling, M.} and {Gabici, S.} and {Gallant, Y. A.} and {Garrigoux, T.} and {Gast, H.} and {Gat\'e, F.} and {Giavitto, G.} and {Giebels, B.} and {Glawion, D.} and {Glicenstein, J. F.} and {Gottschall, D.} and {Grondin, M.-H.} and {Hahn, J.} and {Haupt, M.} and {Hawkes, J.} and {Heinzelmann, G.} and {Henri, G.} and {Hermann, G.} and {Hinton, J. A.} and {Hofmann, W.} and {Hoischen, C.} and {Holch, T. L.} and {Holler, M.} and {Horns, D.} and {Ivascenko, A.} and {Iwasaki, H.} and {Jacholkowska, A.} and {Jamrozy, M.} and {Jankowsky, D.} and {Jankowsky, F.} and {Jingo, M.} and {Jouvin, L.} and {Jung-Richardt, I.} and {Kastendieck, M. A.} and {Katarzy\'{}nski, K.} and {Katsuragawa, M.} and {Katz, U.} and {Kerszberg, D.} and {Khangulyan, D.} and {Kh\'elifi, B.} and {King, J.} and {Klepser, S.} and {Klochkov, D.} and {Klu\'{}zniak, W.} and {Komin, Nu.} and {Kosack, K.} and {Krakau, S.} and {Kraus, M.} and {Kr\"uger, P. P.} and {Laffon, H.} and {Lamanna, G.} and {Lau, J.} and {Lees, J.-P.} and {Lefaucheur, J.} and {Lemi\`ere, A.} and {Lemoine-Goumard, M.} and {Lenain, J.-P.} and {Leser, E.} and {Lohse, T.} and {Lorentz, M.} and {Liu, R.} and {L\'opez-Coto, R.} and {Lypova, I.} and {Marandon, V.} and {Malyshev, D.} and {Marcowith, A.} and {Mariaud, C.} and {Marx, R.} and {Maurin, G.} and {Maxted, N.} and {Mayer, M.} and {Meintjes, P.J.} and {Meyer, M.} and {Mitchell, A. M. W.} and {Moderski, R.} and {Mohamed, M.} and {Mohrmann, L.} and {Mor\aa{}, K.} and {Moulin, E.} and {Murach, T.} and {Nakashima, S.} and {de Naurois, M.} and {Ndiyavala, H.} and {Niederwanger, F.} and {Niemiec, J.} and {Oakes, L.} and {O\'{}Brien, P.} and {Odaka, H.} and {Ohm, S.} and {Ostrowski, M.} and {Oya, I.} and {Padovani, M.} and {Panter, M.} and {Parsons, R. D.} and {Paz Arribas, M.} and {Pekeur, N. W.} and {Pelletier, G.} and {Perennes, C.} and {Petrucci, P.-O.} and {Peyaud, B.} and {Piel, Q.} and {Pita, S.} and {Poireau, V.} and {Poon, H.} and {Prokhorov, D.} and {Prokoph, H.} and {P\"uhlhofer, G.} and {Punch, M.} and {Quirrenbach, A.} and {Raab, S.} and {Rauth, R.} and {Reimer, A.} and {Reimer, O.} and {Renaud, M.} and {de los Reyes, R.} and {Rieger, F.} and {Rinchiuso, L.} and {Romoli, C.} and {Rowell, G.} and {Rudak, B.} and {Rulten, C. B.} and {Safi-Harb, S.} and {Sahakian, V.} and {Saito, S.} and {Sanchez, D. A.} and {Santangelo, A.} and {Sasaki, M.} and {Schandri, M.} and {Schlickeiser, R.} and {Sch\"ussler, F.} and {Schulz, A.} and {Schwanke, U.} and {Schwemmer, S.} and {Seglar-Arroyo, M.} and {Settimo, M.} and {Seyffert, A. S.} and {Shafi, N.} and {Shilon, I.} and {Shiningayamwe, K.} and {Simoni, R.} and {Sol, H.} and {Spanier, F.} and {Spir-Jacob, M.} and {Stawarz, L.} and {Steenkamp, R.} and {Stegmann, C.} and {Steppa, C.} and {Sushch, I.} and {Takahashi, T.} and {Tavernet, J.-P.} and {Tavernier, T.} and {Taylor, A. M.} and {Terrier, R.} and {Tibaldo, L.} and {Tiziani, D.} and {Tluczykont, M.} and {Trichard, C.} and {Tsirou, M.} and {Tsuji, N.} and {Tuffs, R.} and {Uchiyama, Y.} and {van der Walt, D. J.} and {van Eldik, C.} and {van Rensburg, C.} and {van Soelen, B.} and {Vasileiadis, G.} and {Veh, J.} and {Venter, C.} and {Viana, A.} and {Vincent, P.} and {Vink, J.} and {Voisin, F.} and {V\"olk, H. J.} and {Vuillaume, T.} and {Wadiasingh, Z.} and {Wagner, S. J.} and {Wagner, P.} and {Wagner, R. M.} and {White, R.} and {Wierzcholska, A.} and {Willmann, P.} and {W\"ornlein, A.} and {Wouters, D.} and {Yang, R.} and {Zaborov, D.} and {Zacharias, M.} and {Zanin, R.} and {Zdziarski, A. A.} and {Zech, A.} and {Zefi, F.} and {Ziegler, A.} and {Zorn, J.} and {Zywucka, N.}},
	title = {The H.E.S.S. Galactic plane survey},
	DOI= "10.1051/0004-6361/201732098",
	url= "https://doi.org/10.1051/0004-6361/201732098",
	journal = {A\&A},
	year = 2018,
	volume = 612,
	pages = "A1",
}

@article{magic_crab,
title = {Performance of the MAGIC stereo system obtained with Crab Nebula data},
journal = {Astroparticle Physics},
volume = {35},
number = {7},
pages = {435-448},
year = {2012},
issn = {0927-6505},
doi = {https://doi.org/10.1016/j.astropartphys.2011.11.007},
url = {https://www.sciencedirect.com/science/article/pii/S0927650511002064},
author = {J. Aleksić and E.A. Alvarez and L.A. Antonelli and P. Antoranz and M. Asensio and M. Backes and J.A. Barrio and D. Bastieri and J. {Becerra González} and W. Bednarek and A. Berdyugin and K. Berger and E. Bernardini and A. Biland and O. Blanch and R.K. Bock and A. Boller and G. Bonnoli and D. {Borla Tridon} and I. Braun and T. Bretz and A. Cañellas and E. Carmona and A. Carosi and P. Colin and E. Colombo and J.L. Contreras and J. Cortina and L. Cossio and S. Covino and F. Dazzi and A. {De Angelis} and G. {De Caneva} and E. {De Cea del Pozo} and B. {De Lotto} and C. {Delgado Mendez} and A. {Diago Ortega} and M. Doert and A. Domı´nguez and D. {Dominis Prester} and D. Dorner and M. Doro and D. Elsaesser and D. Ferenc and M.V. Fonseca and L. Font and C. Fruck and R.J. {Garcı´a López} and M. Garczarczyk and D. Garrido and G. Giavitto and N. Godinović and D. Hadasch and D. Häfner and A. Herrero and D. Hildebrand and D. Höhne-Mönch and J. Hose and D. Hrupec and B. Huber and T. Jogler and H. Kellermann and S. Klepser and T. Krähenbühl and J. Krause and A. {La Barbera} and D. Lelas and E. Leonardo and E. Lindfors and S. Lombardi and M. López and A. López-Oramas and E. Lorenz and M. Makariev and G. Maneva and N. Mankuzhiyil and K. Mannheim and L. Maraschi and M. Mariotti and M. Martı´nez and D. Mazin and M. Meucci and J.M. Miranda and R. Mirzoyan and H. Miyamoto and J. Moldón and A. Moralejo and P. Munar-Adrover and D. Nieto and K. Nilsson and R. Orito and I. Oya and D. Paneque and R. Paoletti and S. Pardo and J.M. Paredes and S. Partini and M. Pasanen and F. Pauss and M.A. Perez-Torres and M. Persic and L. Peruzzo and M. Pilia and J. Pochon and F. Prada and P.G. {Prada Moroni} and E. Prandini and I. Puljak and I. Reichardt and R. Reinthal and W. Rhode and M. Ribó and J. Rico and S. Rügamer and A. Saggion and K. Saito and T.Y. Saito and M. Salvati and K. Satalecka and V. Scalzotto and V. Scapin and C. Schultz and T. Schweizer and M. Shayduk and S.N. Shore and A. Sillanpää and J. Sitarek and I. Snidaric and D. Sobczynska and F. Spanier and S. Spiro and V. Stamatescu and A. Stamerra and B. Steinke and J. Storz and N. Strah and T. Surić and L. Takalo and H. Takami and F. Tavecchio and P. Temnikov and T. Terzić and D. Tescaro and M. Teshima and O. Tibolla and D.F. Torres and A. Treves and M. Uellenbeck and H. Vankov and P. Vogler and R.M. Wagner and Q. Weitzel and V. Zabalza and F. Zandanel and R. Zanin},
}

@article{veritas_main_paper,
title = {The first VERITAS telescope},
journal = {Astroparticle Physics},
volume = {25},
number = {6},
pages = {391-401},
year = {2006},
issn = {0927-6505},
doi = {https://doi.org/10.1016/j.astropartphys.2006.04.002},
url = {https://www.sciencedirect.com/science/article/pii/S092765050600051X},
author = {J. Holder and R.W. Atkins and H.M. Badran and G. Blaylock and S.M. Bradbury and J.H. Buckley and K.L. Byrum and D.A. Carter-Lewis and O. Celik and Y.C.K. Chow and P. Cogan and W. Cui and M.K. Daniel and I. {de la Calle Perez} and C. Dowdall and P. Dowkontt and C. Duke and A.D. Falcone and S.J. Fegan and J.P. Finley and P. Fortin and L.F. Fortson and K. Gibbs and G. Gillanders and O.J. Glidewell and J. Grube and K.J. Gutierrez and G. Gyuk and J. Hall and D. Hanna and E. Hays and D. Horan and S.B. Hughes and T.B. Humensky and A. Imran and I. Jung and P. Kaaret and G.E. Kenny and D. Kieda and J. Kildea and J. Knapp and H. Krawczynski and F. Krennrich and M.J. Lang and S. LeBohec and E. Linton and E.K. Little and G. Maier and H. Manseri and A. Milovanovic and P. Moriarty and R. Mukherjee and P.A. Ogden and R.A. Ong and D. Petry and J.S. Perkins and F. Pizlo and M. Pohl and J. Quinn and K. Ragan and P.T. Reynolds and E.T. Roache and H.J. Rose and M. Schroedter and G.H. Sembroski and G. Sleege and D. Steele and S.P. Swordy and A. Syson and J.A. Toner and L. Valcarcel and V.V. Vassiliev and S.P. Wakely and T.C. Weekes and R.J. White and D.A. Williams and R. Wagner},
}

@article{tibet_as_subPeV,
  title = {First Detection of Photons with Energy beyond 100 TeV from an Astrophysical Source},
  author = {Amenomori, M. and Bao, Y. W. and Bi, X. J. and Chen, D. and Chen, T. L. and Chen, W. Y. and Chen, Xu and Chen, Y. and Cirennima and Cui, S. W. and Danzengluobu and Ding, L. K. and Fang, J. H. and Fang, K. and Feng, C. F. and Feng, Zhaoyang and Feng, Z. Y. and Gao, Qi and Gou, Q. B. and Guo, Y. Q. and He, H. H. and He, Z. T. and Hibino, K. and Hotta, N. and Hu, Haibing and Hu, H. B. and Huang, J. and Jia, H. Y. and Jiang, L. and Jin, H. B. and Kajino, F. and Kasahara, K. and Katayose, Y. and Kato, C. and Kato, S. and Kawata, K. and Kozai, M. and Labaciren and Le, G. M. and Li, A. F. and Li, H. J. and Li, W. J. and Lin, Y. H. and Liu, B. and Liu, C. and Liu, J. S. and Liu, M. Y. and Lou, Y.-Q. and Lu, H. and Meng, X. R. and Mitsui, H. and Munakata, K. and Nakamura, Y. and Nanjo, H. and Nishizawa, M. and Ohnishi, M. and Ohta, I. and Ozawa, S. and Qian, X. L. and Qu, X. B. and Saito, T. and Sakata, M. and Sako, T. K. and Sengoku, Y. and Shao, J. and Shibata, M. and Shiomi, A. and Sugimoto, H. and Takita, M. and Tan, Y. H. and Tateyama, N. and Torii, S. and Tsuchiya, H. and Udo, S. and Wang, H. and Wu, H. R. and Xue, L. and Yagisawa, K. and Yamamoto, Y. and Yang, Z. and Yuan, A. F. and Zhai, L. M. and Zhang, H. M. and Zhang, J. L. and Zhang, X. and Zhang, X. Y. and Zhang, Y. and Zhang, Yi and Zhang, Ying and Zhaxisangzhu and Zhou, X. X.},
  collaboration = {Tibet $\mathrm{AS}\ensuremath{\gamma}$ Collaboration},
  journal = {Phys. Rev. Lett.},
  volume = {123},
  issue = {5},
  pages = {051101},
  numpages = {6},
  year = {2019},
  month = {Jul},
  publisher = {American Physical Society},
  doi = {10.1103/PhysRevLett.123.051101},
  url = {https://link.aps.org/doi/10.1103/PhysRevLett.123.051101}
}

@article{hawc_56_TeV,
  title = {Multiple Galactic Sources with Emission Above 56 TeV Detected by HAWC},
  author = {Abeysekara, A. U. and Albert, A. and Alfaro, R. and Angeles Camacho, J. R. and Arteaga-Vel\'azquez, J. C. and Arunbabu, K. P. and Avila Rojas, D. and Ayala Solares, H. A. and Baghmanyan, V. and Belmont-Moreno, E. and BenZvi, S. Y. and Brisbois, C. and Caballero-Mora, K. S. and Capistr\'an, T. and Carrami\~nana, A. and Casanova, S. and Cotti, U. and Cotzomi, J. and Couti\~no de Le\'on, S. and De la Fuente, E. and de Le\'on, C. and Dichiara, S. and Dingus, B. L. and DuVernois, M. A. and D\'{\i}az-V\'elez, J. C. and Ellsworth, R. W. and Engel, K. and Espinoza, C. and Fleischhack, H. and Fraija, N. and Galv\'an-G\'amez, A. and Garcia, D. and Garc\'{\i}a-Gonz\'alez, J. A. and Garfias, F. and Gonz\'alez, M. M. and Goodman, J. A. and Harding, J. P. and Hernandez, S. and Hinton, J. and Hona, B. and Huang, D. and Hueyotl-Zahuantitla, F. and H\"untemeyer, P. and Iriarte, A. and Jardin-Blicq, A. and Joshi, V. and Kaufmann, S. and Kieda, D. and Lara, A. and Lee, W. H. and Le\'on Vargas, H. and Linnemann, J. T. and Longinotti, A. L. and Luis-Raya, G. and Lundeen, J. and L\'opez-Coto, R. and Malone, K. and Marinelli, S. S. and Martinez, O. and Martinez-Castellanos, I. and Mart\'{\i}nez-Castro, J. and Mart\'{\i}nez-Huerta, H. and Matthews, J. A. and Miranda-Romagnoli, P. and Morales-Soto, J. A. and Moreno, E. and Mostaf\'a, M. and Nayerhoda, A. and Nellen, L. and Newbold, M. and Nisa, M. U. and Noriega-Papaqui, R. and Peisker, A. and P\'erez-P\'erez, E. G. and Pretz, J. and Ren, Z. and Rho, C. D. and Rivi\`ere, C. and Rosa-Gonz\'alez, D. and Rosenberg, M. and Ruiz-Velasco, E. and Salesa Greus, F. and Sandoval, A. and Schneider, M. and Schoorlemmer, H. and Sinnis, G. and Smith, A. J. and Springer, R. W. and Surajbali, P. and Tabachnick, E. and Tanner, M. and Tibolla, O. and Tollefson, K. and Torres, I. and Torres-Escobedo, R. and Villase\~nor, L. and Weisgarber, T. and Wood, J. and Yapici, T. and Zhang, H. and Zhou, H.},
  collaboration = {HAWC Collaboration},
  journal = {Phys. Rev. Lett.},
  volume = {124},
  issue = {2},
  pages = {021102},
  numpages = {7},
  year = {2020},
  month = {Jan},
  publisher = {American Physical Society},
  doi = {10.1103/PhysRevLett.124.021102},
  url = {https://link.aps.org/doi/10.1103/PhysRevLett.124.021102}
}

@article{lhaaso_main_crab,
doi = {10.1088/1674-1137/ad2e82},
url = {https://doi.org/10.1088/1674-1137/ad2e82},
year = {2024},
month = {jun},
publisher = {Chinese Physical Society and the Institute of High Energy Physics of the Chinese Academy of Sciences and the Institute of Modern Physics of the Chinese Academy of Sciences and IOP Publishing Ltd
                        },
volume = {48},
number = {6},
pages = {065001},
author = {Cao, Zhen and Aharonian, F. and An, Q. and Axikegu and Bai, Y.X. and Bao, Y.W. and Bastieri, D. and Bi, X.J. and Bi, Y.J. and Cai, J.T. and Cao, Q. and Cao, W.Y. and Cao, Zhe and Chang, J. and Chang, J.F. and Chen, A.M. and Chen, E.S. and Chen, Liang and Chen, Lin and Chen, Long and Chen, M.J. and Chen, M.L. and Chen, Q.H. and Chen, S.H. and Chen, T.L. and Chen, Y. and Cheng, N. and Cheng, Y.D. and Cui, M.Y. and Cui, S.W. and Cui, X.H. and Cui, Y.D. and Dai, B.Z. and Dai, H.L. and Dai, Z.G. and Danzengluobu and Volpe, D. della and Dong, X.Q. and Duan, K.K. and Fan, J.H. and Fan, Y.Z. and Fang, J. and Fang, K. and Feng, C.F. and Feng, L. and Feng, S.H. and Feng, X.T. and Feng, Y.L. and Gabici, S. and Gao, B. and Gao, C.D. and Gao, L.Q. and Gao, Q. and Gao, W. and Gao, W.K. and Ge, M.M. and Geng, L.S. and Giacinti, G. and Gong, G.H. and Gou, Q.B. and Gu, M.H. and Guo, F.L. and Guo, X.L. and Guo, Y.Q. and Guo, Y.Y. and Han, Y.A. and He, H.H. and He, H.N. and He, J.Y. and He, X.B. and He, Y. and Heller, M. and Hor, Y.K. and Hou, B.W. and Hou, C. and Hou, X. and Hu, H.B. and Hu, Q. and Hu, S.C. and Huang, D.H. and Huang, T.Q. and Huang, W.J. and Huang, X.T. and Huang, X.Y. and Huang, Y. and Huang, Z.C. and Ji, X.L. and Jia, H.Y. and Jia, K. and Jiang, K. and Jiang, X.W. and Jiang, Z.J. and Jin, M. and Kang, M.M. and Ke, T. and Kuleshov, D. and Kurinov, K. and Li, B.B. and Li, Cheng and Li, Cong and Li, D. and Li, F. and Li, H.B. and Li, H.C. and Li, H.Y. and Li, J. and Li, Jian and Li, Jie and Li, K. and Li, W.L. and Li, W.L. and Li, X.R. and Li, Xin and Li, Y.Z. and Li, Zhe and Li, Zhuo and Liang, E.W. and Liang, Y.F. and Lin, S.J. and Liu, B. and Liu, C. and Liu, D. and Liu, H. and Liu, H.D. and Liu, J. and Liu, J.L. and Liu, J.Y. and Liu, M.Y. and Liu, R.Y. and Liu, S.M. and Liu, W. and Liu, Y. and Liu, Y.N. and Lu, R. and Luo, Q. and Lv, H.K. and Ma, B.Q. and Ma, L.L. and Ma, X.H. and Mao, J.R. and Min, Z. and Mitthumsiri, W. and Mu, H.J. and Nan, Y.C. and Neronov, A. and Ou, Z.W. and Pang, B.Y. and Pattarakijwanich, P. and Pei, Z.Y. and Qi, M.Y. and Qi, Y.Q. and Qiao, B.Q. and Qin, J.J. and Ruffolo, D. and Sáiz, A. and Semikoz, D. and Shao, C.Y. and Shao, L. and Shchegolev, O. and Sheng, X.D. and Shu, F.W. and Song, H.C. and Stenkin, Yu.V. and Stepanov, V. and Su, Y. and Sun, Q.N. and Sun, X.N. and Sun, Z.B. and Tam, P.H.T. and Tang, Q.W. and Tang, Z.B. and Tian, W.W. and Wang, C. and Wang, C.B. and Wang, G.W. and Wang, H.G. and Wang, H.H. and Wang, J.C. and Wang, K. and Wang, L.P. and Wang, L.Y. and Wang, P.H. and Wang, R. and Wang, W. and Wang, X.G. and Wang, X.Y. and Wang, Y. and Wang, Y.D. and Wang, Y.J. and Wang, Z.H. and Wang, Z.X. and Wang, Zhen and Wang, Zheng and Wei, D.M. and Wei, J.J. and Wei, Y.J. and Wen, T. and Wu, C.Y. and Wu, H.R. and Wu, S. and Wu, X.F. and Wu, Y.S. and Xi, S.Q. and Xia, J. and Xia, J.J. and Xiang, G.M. and Xiao, D.X. and Xiao, G. and Xin, G.G. and Xin, Y.L. and Xing, Y. and Xiong, Z. and Xu, D.L. and Xu, R.F. and Xu, R.X. and Xu, W.L. and Xue, L. and Yan, D.H. and Yan, J.Z. and Yan, T. and Yang, C.W. and Yang, F. and Yang, F.F. and Yang, H.W. and Yang, J.Y. and Yang, L.L. and Yang, M.J. and Yang, R.Z. and Yang, S.B. and Yao, Y.H. and Yao, Z.G. and Ye, Y.M. and Yin, L.Q. and Yin, N. and You, X.H. and You, Z.Y. and Yu, Y.H. and Yuan, Q. and Yue, H. and Zeng, H.D. and Zeng, T.X. and Zeng, W. and Zha, M. and Zhang, B.B. and Zhang, F. and Zhang, H.M. and Zhang, H.Y. and Zhang, J.L. and Zhang, L.X. and Zhang, Li and Zhang, P.F. and Zhang, P.P. and Zhang, R. and Zhang, S.B. and Zhang, S.R. and Zhang, S.S. and Zhang, X. and Zhang, X.P. and Zhang, Y.F. and Zhang, Yi and Zhang, Yong and Zhao, B. and Zhao, J. and Zhao, L. and Zhao, L.Z. and Zhao, S.P. and Zheng, F. and Zhou, B. and Zhou, H. and Zhou, J.N. and Zhou, M. and Zhou, P. and Zhou, R. and Zhou, X.X. and Zhu, C.G. and Zhu, F.R. and Zhu, H. and Zhu, K.J. and Zuo, X. and (LHAASO Collaboration)},
title = {Optimization of performance of the KM2A full array using the Crab Nebula*},
journal = {Chinese Physics C},
}

@ARTICLE{lhaaso_UHE_paper,
       author = {{Cao}, Zhen and {Aharonian}, F.~A. and {An}, Q. and {Axikegu}, L.~X., Bai and {Bai}, Y.~X. and {Bao}, Y.~W. and {Bastieri}, D. and {Bi}, X.~J. and {Bi}, Y.~J. and {Cai}, H. and {Cai}, J.~T. and {Cao}, Zhe and {Chang}, J. and {Chang}, J.~F. and {Chang}, X.~C. and {Chen}, B.~M. and {Chen}, J. and {Chen}, L. and {Chen}, Liang and {Chen}, Long and {Chen}, M.~J. and {Chen}, M.~L. and {Chen}, Q.~H. and {Chen}, S.~H. and {Chen}, S.~Z. and {Chen}, T.~L. and {Chen}, X.~L. and {Chen}, Y. and {Cheng}, N. and {Cheng}, Y.~D. and {Cui}, S.~W. and {Cui}, X.~H. and {Cui}, Y.~D. and {Dai}, B.~Z. and {Dai}, H.~L. and {Dai}, Z.~G. and {Danzengluobu} and {della Volpe}, D. and {D'Ettorre Piazzoli}, B. and {Dong}, X.~J. and {Fan}, J.~H. and {Fan}, Y.~Z. and {Fan}, Z.~X. and {Fang}, J. and {Fang}, K. and {Feng}, C.~F. and {Feng}, L. and {Feng}, S.~H. and {Feng}, Y.~L. and {Gao}, B. and {Gao}, C.~D. and {Gao}, Q. and {Gao}, W. and {Ge}, M.~M. and {Geng}, L.~S. and {Gong}, G.~H. and {Gou}, Q.~B. and {Gu}, M.~H. and {Guo}, J.~G. and {Guo}, X.~L. and {Guo}, Y.~Q. and {Guo}, Y.~Y. and {Han}, Y.~A. and {He}, H.~H. and {He}, H.~N. and {He}, J.~C. and {He}, S.~L. and {He}, X.~B. and {He}, Y. and {Heller}, M. and {Hor}, Y.~K. and {Hou}, C. and {Hou}, X. and {Hu}, H.~B. and {Hu}, S. and {Hu}, S.~C. and {Hu}, X.~J. and {Huang}, D.~H. and {Huang}, Q.~L. and {Huang}, W.~H. and {Huang}, X.~T. and {Huang}, Z.~C. and {Ji}, F. and {Ji}, X.~L. and {Jia}, H.~Y. and {Jiang}, K. and {Jiang}, Z.~J. and {Jin}, C. and {Kuleshov}, D. and {Levochkin}, K. and {Li}, B.~B. and {Li}, Cong and {Li}, Cheng and {Li}, F. and {Li}, H.~B. and {Li}, H.~C. and {Li}, H.~Y. and {Li}, J. and {Li}, K. and {Li}, W.~L. and {Li}, X. and {Li}, Xin and {Li}, X.~R. and {Li}, Y. and {Li}, Y.~Z. and {Li}, Zhe and {Li}, Zhuo and {Liang}, E.~W. and {Liang}, Y.~F. and {Lin}, S.~J. and {Liu}, B. and {Liu}, C. and {Liu}, D. and {Liu}, H. and {Liu}, H.~D. and {Liu}, J. and {Liu}, J.~L. and {Liu}, J.~S. and {Liu}, J.~Y. and {Liu}, M.~Y. and {Liu}, R.~Y. and {Liu}, S.~M. and {Liu}, W. and {Liu}, Y.~N. and {Liu}, Z.~X. and {Long}, W.~J. and {Lu}, R. and {Lv}, H.~K. and {Ma}, B.~Q. and {Ma}, L.~L. and {Ma}, X.~H. and {Mao}, J.~R. and {Masood}, A. and {Mitthumsiri}, W. and {Montaruli}, T. and {Nan}, Y.~C. and {Pang}, B.~Y. and {Pattarakijwanich}, P. and {Pei}, Z.~Y. and {Qi}, M.~Y. and {Ruffolo}, D. and {Rulev}, V. and {S{\'a}iz}, A. and {Shao}, L. and {Shchegolev}, O. and {Sheng}, X.~D. and {Shi}, J.~R. and {Song}, H.~C. and {Stenkin}, Yu. V. and {Stepanov}, V. and {Sun}, Q.~N. and {Sun}, X.~N. and {Sun}, Z.~B. and {Tam}, P.~H.~T. and {Tang}, Z.~B. and {Tian}, W.~W. and {Wang}, B.~D. and {Wang}, C. and {Wang}, H. and {Wang}, H.~G. and {Wang}, J.~C. and {Wang}, J.~S. and {Wang}, L.~P. and {Wang}, L.~Y. and {Wang}, R.~N. and {Wang}, W. and {Wang}, W. and {Wang}, X.~G. and {Wang}, X.~J. and {Wang}, X.~Y. and {Wang}, Y.~D. and {Wang}, Y.~J. and {Wang}, Y.~P. and {Wang}, Zheng and {Wang}, Zhen and {Wang}, Z.~H. and {Wang}, Z.~X. and {Wei}, D.~M. and {Wei}, J.~J. and {Wei}, Y.~J. and {Wen}, T. and {Wu}, C.~Y. and {Wu}, H.~R. and {Wu}, S. and {Wu}, W.~X. and {Wu}, X.~F. and {Xi}, S.~Q. and {Xia}, J. and {Xia}, J.~J. and {Xiang}, G.~M. and {Xiao}, G. and {Xiao}, H.~B. and {Xin}, G.~G. and {Xin}, Y.~L. and {Xing}, Y. and {Xu}, D.~L. and {Xu}, R.~X. and {Xue}, L. and {Yan}, D.~H. and {Yang}, C.~W.},
        title = "{Ultrahigh-energy photons up to 1.4 petaelectronvolts from 12 {\ensuremath{\gamma}}-ray Galactic sources}",
      journal = {\nat},
         year = 2021,
        month = jun,
       volume = {594},
       number = {7861},
        pages = {33-36},
          doi = {10.1038/s41586-021-03498-z},
       adsurl = {https://ui.adsabs.harvard.edu/abs/2021Natur.594...33C}
}

@article{lhaaso_1st_catalog,
doi = {10.3847/1538-4365/acfd29},
url = {https://doi.org/10.3847/1538-4365/acfd29},
year = {2024},
month = {feb},
publisher = {The American Astronomical Society},
volume = {271},
number = {1},
pages = {25},
author = {Cao, Zhen and Aharonian, F. and An, Q. and Axikegu and Bai, Y. X. and Bao, Y. W. and Bastieri, D. and Bi, X. J. and Bi, Y. J. and Cai, J. T. and Cao, Q. and Cao, W. Y. and Cao, Zhe and Chang, J. and Chang, J. F. and Chen, A. M. and Chen, E. S. and Chen, Liang and Chen, Lin and Chen, Long and Chen, M. J. and Chen, M. L. and Chen, Q. H. and Chen, S. H. and Chen, S. Z. and Chen, T. L. and Chen, Y. and Cheng, N. and Cheng, Y. D. and Cui, M. Y. and Cui, S. W. and Cui, X. H. and Cui, Y. D. and Dai, B. Z. and Dai, H. L. and Dai, Z. G. and Danzengluobu and della Volpe, D. and Dong, X. Q. and Duan, K. K. and Fan, J. H. and Fan, Y. Z. and Fang, J. and Fang, K. and Feng, C. F. and Feng, L. and Feng, S. H. and Feng, X. T. and Feng, Y. L. and Gabici, S. and Gao, B. and Gao, C. D. and Gao, L. Q. and Gao, Q. and Gao, W. and Gao, W. K. and Ge, M. M. and Geng, L. S. and Giacinti, G. and Gong, G. H. and Gou, Q. B. and Gu, M. H. and Guo, F. L. and Guo, X. L. and Guo, Y. Q. and Guo, Y. Y. and Han, Y. A. and He, H. H. and He, H. N. and He, J. Y. and He, X. B. and He, Y. and Heller, M. and Hor, Y. K. and Hou, B. W. and Hou, C. and Hou, X. and Hu, H. B. and Hu, Q. and Hu, S. C. and Huang, D. H. and Huang, T. Q. and Huang, W. J. and Huang, X. T. and Huang, X. Y. and Huang, Y. and Huang, Z. C. and Ji, X. L. and Jia, H. Y. and Jia, K. and Jiang, K. and Jiang, X. W. and Jiang, Z. J. and Jin, M. and Kang, M. M. and Ke, T. and Kuleshov, D. and Kurinov, K. and Li, B. B. and Li, Cheng and Li, Cong and Li, D. and Li, F. and Li, H. B. and Li, H. C. and Li, H. Y. and Li, J. and Li, Jian and Li, Jie and Li, K. and Li, W. L. and Li, W. L. and Li, X. R. and Li, Xin and Li, Y. Z. and Li, Zhe and Li, Zhuo and Liang, E. W. and Liang, Y. F. and Lin, S. J. and Liu, B. and Liu, C. and Liu, D. and Liu, H. and Liu, H. D. and Liu, J. and Liu, J. L. and Liu, J. Y. and Liu, M. Y. and Liu, R. Y. and Liu, S. M. and Liu, W. and Liu, Y. and Liu, Y. N. and Lu, R. and Luo, Q. and Lv, H. K. and Ma, B. Q. and Ma, L. L. and Ma, X. H. and Mao, J. R. and Min, Z. and Mitthumsiri, W. and Mu, H. J. and Nan, Y. C. and Neronov, A. and Ou, Z. W. and Pang, B. Y. and Pattarakijwanich, P. and Pei, Z. Y. and Qi, M. Y. and Qi, Y. Q. and Qiao, B. Q. and Qin, J. J. and Ruffolo, D. and Sáiz, A. and Semikoz, D. and Shao, C. Y. and Shao, L. and Shchegolev, O. and Sheng, X. D. and Shu, F. W. and Song, H. C. and Stenkin, Yu. V. and Stepanov, V. and Su, Y. and Sun, Q. N. and Sun, X. N. and Sun, Z. B. and Tam, P. H. T. and Tang, Q. W. and Tang, Z. B. and Tian, W. W. and Wang, C. and Wang, C. B. and Wang, G. W. and Wang, H. G. and Wang, H. H. and Wang, J. C. and Wang, K. and Wang, L. P. and Wang, L. Y. and Wang, P. H. and Wang, R. and Wang, W. and Wang, X. G. and Wang, X. Y. and Wang, Y. and Wang, Y. D. and Wang, Y. J. and Wang, Z. H. and Wang, Z. X. and Wang, Zhen and Wang, Zheng and Wei, D. M. and Wei, J. J. and Wei, Y. J. and Wen, T. and Wu, C. Y. and Wu, H. R. and Wu, S. and Wu, X. F. and Wu, Y. S. and Xi, S. Q. and Xia, J. and Xia, J. J. and Xiang, G. M. and Xiao, D. X. and Xiao, G. and Xin, G. G. and Xin, Y. L. and Xing, Y. and Xiong, Z. and Xu, D. L. and Xu, R. F. and Xu, R. X. and Xu, W. L. and Xue, L. and Yan, D. H. and Yan, J. Z. and Yan, T. and Yang, C. W. and Yang, F. and Yang, F. F. and Yang, H. W. and Yang, J. Y. and Yang, L. L. and Yang, M. J. and Yang, R. Z. and Yang, S. B. and Yao, Y. H. and Yao, Z. G. and Ye, Y. M. and Yin, L. Q. and Yin, N. and You, X. H. and You, Z. Y. and Yu, Y. H. and Yuan, Q. and Yue, H. and Zeng, H. D. and Zeng, T. X. and Zeng, W. and Zha, M. and Zhang, B. B. and Zhang, F. and Zhang, H. M. and Zhang, H. Y. and Zhang, J. L. and Zhang, L. X. and Zhang, Li and Zhang, P. F. and Zhang, P. P. and Zhang, R. and Zhang, S. B. and Zhang, S. R. and Zhang, S. S. and Zhang, X. and Zhang, X. P. and Zhang, Y. F. and Zhang, Yi and Zhang, Yong and Zhao, B. and Zhao, J. and Zhao, L. and Zhao, L. Z. and Zhao, S. P. and Zheng, F. and Zhou, B. and Zhou, H. and Zhou, J. N. and Zhou, M. and Zhou, P. and Zhou, R. and Zhou, X. X. and Zhu, C. G. and Zhu, F. R. and Zhu, H. and Zhu, K. J. and Zuo, X. and (The LHAASO Collaboration)},
title = {The First LHAASO Catalog of Gamma-Ray Sources},
journal = {The Astrophysical Journal Supplement Series},
}

@article{vernetto,
  title = {Absorption of very high energy gamma rays in the Milky Way},
  author = {Vernetto, Silvia and Lipari, Paolo},
  journal = {Phys. Rev. D},
  volume = {94},
  issue = {6},
  pages = {063009},
  numpages = {16},
  year = {2016},
  month = {Sep},
  publisher = {American Physical Society},
  doi = {10.1103/PhysRevD.94.063009},
  url = {https://link.aps.org/doi/10.1103/PhysRevD.94.063009}
}

@article{Zhang_2026,
doi = {10.3847/1538-4357/ae48f4},
url = {https://doi.org/10.3847/1538-4357/ae48f4},
year = {2026},
month = {mar},
publisher = {The American Astronomical Society},
volume = {1000},
number = {1},
pages = {19},
author = {Zhang, Jianli and Guo, YiQing},
title = {Attenuation of LHAASO PeVatrons by the Interstellar Radiation Field and Cosmic Microwave Background Radiation},
journal = {The Astrophysical Journal},
}

@article{cygnus_x3_lhaaso,
    author = "Cao, Zhen and others",
    collaboration = "LHAASO",
    title = "{Cygnus X-3: A variable petaelectronvolt gamma-ray source}",
    eprint = "2512.16638",
    archivePrefix = "arXiv",
    primaryClass = "astro-ph.HE",
    month = "12",
    year = "2025"
}

@article{cta_pevatrons,
title = {{Sensitivity of the Cherenkov Telescope Array to spectral signatures of hadronic PeVatrons with application to Galactic Supernova Remnants}},
journal = {Astroparticle Physics},
volume = {150},
pages = {102850},
year = {2023},
issn = {0927-6505},
doi = {https://doi.org/10.1016/j.astropartphys.2023.102850},
url = {https://www.sciencedirect.com/science/article/pii/S0927650523000361},
author = {{Acero, F. et al.}},
}

@article{ozi_review,
    author = {Angüner, E. O.},
    title = "{Exploring the high energy frontiers of the Milky Way with ground-based gamma-ray astronomy: PeVatrons and the quest for the origin of Galactic cosmic-rays}",
    journal = {Turkish Journal of Physics},
    year = 2023,
    volume = {47},
    number = {2},
    article = {2},
    doi = {https://doi.org/10.55730/1300-0101.2738},
}

@article{deOnaWilhelmi2024,
    author = "de O\~na Wilhelmi, Emma and L\'opez-Coto, Ruben and Aharonian, Felix and Amato, Elena and Cao, Zhen and Gabici, Stefano and Hinton, Jim",
    title = "{The hunt for PeVatrons as the origin of the most energetic photons observed in the Galaxy}",
    eprint = "2404.16591",
    archivePrefix = "arXiv",
    primaryClass = "astro-ph.HE",
    doi = "10.1038/s41550-024-02224-9",
    journal = "Nature Astron.",
    volume = "8",
    number = "4",
    pages = "425--431",
    year = "2024"
}

@article{pev_annual_rev,
author = "Cao, Zhen and others",
title = "{Ultra-High-Energy Gamma-Ray Astronomy}",
journal = {Annual Review of Nuclear and Particle Science},
volume = {73},
number = {1},
pages = {341-363},
year = {2023},
doi = {10.1146/annurev-nucl-112822-025357},
URL = { https://doi.org/10.1146/annurev-nucl-112822-025357},
eprint = { https://doi.org/10.1146/annurev-nucl-112822-025357},
}

@article{Celli_2020,
doi = {10.3847/1538-4357/abb805},
url = {https://doi.org/10.3847/1538-4357/abb805},
year = {2020},
month = {nov},
publisher = {The American Astronomical Society},
volume = {903},
number = {1},
pages = {61},
author = {Celli, Silvia and Aharonian, Felix and Gabici, Stefano},
title = {Spectral Signatures of PeVatrons},
journal = {The Astrophysical Journal},
}

@article{ozi_GCR_mnras,
    author = {Angüner, E.O. and Spengler, G. and Amato, E. and Casanova, S.},
    title = {Search for the Galactic accelerators of cosmic rays up to the knee with the Pevatron test statistic},
    journal = {Monthly Notices of the Royal Astronomical Society},
    volume = {523},
    number = {3},
    pages = {4097-4112},
    year = {2023},
    month = {06},
    issn = {0035-8711},
    doi = {10.1093/mnras/stad1674},
    url = {https://doi.org/10.1093/mnras/stad1674},
    eprint = {https://academic.oup.com/mnras/article-pdf/523/3/4097/50628891/stad1674.pdf},
}

@ARTICLE{kelner_2006,
       author = {{Kelner}, S.~R. and {Aharonian}, F.~A. and {Bugayov}, V.~V.},
        title = "{Energy spectra of gamma rays, electrons, and neutrinos produced at proton-proton interactions in the very high energy regime}",
      journal = {\prd},
         year = 2006,
        month = aug,
       volume = {74},
       number = {3},
          eid = {034018},
        pages = {034018},
          doi = {10.1103/PhysRevD.74.034018},
archivePrefix = {arXiv},
       eprint = {astro-ph/0606058},
 primaryClass = {astro-ph},
       adsurl = {https://ui.adsabs.harvard.edu/abs/2006PhRvD..74c4018K}
}

@ARTICLE{kafexhiu2014,
       author = {{Kafexhiu}, Ervin and {Aharonian}, Felix and {Taylor}, Andrew M. and {Vila}, Gabriela S.},
        title = "{Parametrization of gamma-ray production cross sections for p-p interactions in a broad proton energy range from the kinematic threshold to PeV energies}",
      journal = {Phys. Rev. D},
         year = 2014,
        month = dec,
       volume = {90},
       number = {12},
          eid = {123014},
        pages = {123014},
          doi = {10.1103/PhysRevD.90.123014},
archivePrefix = {arXiv},
       eprint = {1406.7369},
 primaryClass = {astro-ph.HE},
       adsurl = {https://ui.adsabs.harvard.edu/abs/2014PhRvD..90l3014K}
}

@article{lipari_2018,
  title = {Diffuse Galactic gamma-ray flux at very high energy},
  author = {Lipari, Paolo and Vernetto, Silvia},
  journal = {Phys. Rev. D},
  volume = {98},
  issue = {4},
  pages = {043003},
  numpages = {23},
  year = {2018},
  month = {Aug},
  publisher = {American Physical Society},
  doi = {10.1103/PhysRevD.98.043003},
  url = {https://link.aps.org/doi/10.1103/PhysRevD.98.043003}
}

@article{Vecchiotti_2025,
doi = {10.1088/1475-7516/2025/09/041},
url = {https://doi.org/10.1088/1475-7516/2025/09/041},
year = {2025},
month = {sep},
publisher = {IOP Publishing},
volume = {2025},
number = {09},
pages = {041},
author = {Vecchiotti, V. and Peron, G. and Amato, E. and Menchiari, S. and Morlino, G. and Pagliaroli, G. and Villante, F.L.},
title = {Interpreting the LHAASO Galactic diffuse emission data},
journal = {Journal of Cosmology and Astroparticle Physics},
}

@article{Zhang_2023,
doi = {10.3847/1538-4357/acf842},
url = {https://doi.org/10.3847/1538-4357/acf842},
year = {2023},
month = {oct},
publisher = {The American Astronomical Society},
volume = {957},
number = {1},
pages = {43},
author = {Zhang, Rui and Huang, Xiaoyuan and Xu, Zhi-Hui and Zhao, Shiping and Yuan, Qiang},
title = {Galactic Diffuse γ-Ray Emission from GeV to PeV Energies in Light of Up-to-date Cosmic-Ray Measurements},
journal = {The Astrophysical Journal},
}

@article{He_2024,
doi = {10.3847/1538-4357/ad2a4e},
url = {https://doi.org/10.3847/1538-4357/ad2a4e},
year = {2024},
month = {mar},
publisher = {The American Astronomical Society},
volume = {964},
number = {1},
pages = {28},
author = {He, Xin-Yu and Zhang, Pei-Pei and Yuan, Qiang and Guo, Yi-Qing},
title = {Galactic Diffuse Emission from Radio to Ultra-high-energy γ-Rays in Light of Up-to-date Cosmic-Ray Measurements},
journal = {The Astrophysical Journal},
}

@article{Yan_2023,
    author = "Yan, Kai and Liu, Ruo-Yu and Zhang, Rui and Li, Chao-Ming and Yuan, Qiang and Wang, Xiang-Yu",
    title = "{Insights from LHAASO and IceCube into the origin of the Galactic diffuse teraelectronvolt{\textendash}petaelectronvolt emission}",
    eprint = "2307.12363",
    archivePrefix = "arXiv",
    primaryClass = "astro-ph.HE",
    doi = "10.1038/s41550-024-02221-y",
    journal = "Nature Astron.",
    volume = "8",
    number = "5",
    pages = "628--636",
    year = "2024"
}

@article{lhaaso_GDE,
  title = {Measurement of Ultra-High-Energy Diffuse Gamma-Ray Emission of the Galactic Plane from 10 TeV to 1 PeV with LHAASO-KM2A},
  author = {Cao, Zhen and Aharonian, F. and An, Q. and Axikegu and Bai, Y. X. and Bao, Y. W. and Bastieri, D. and Bi, X. J. and Bi, Y. J. and Cai, J. T. and Cao, Q. and Cao, W. Y. and Cao, Zhe and Chang, J. and Chang, J. F. and Chen, A. M. and Chen, E. S. and Chen, Liang and Chen, Lin and Chen, Long and Chen, M. J. and Chen, M. L. and Chen, Q. H. and Chen, S. H. and Chen, S. Z. and Chen, T. L. and Chen, Y. and Cheng, N. and Cheng, Y. D. and Cui, M. Y. and Cui, S. W. and Cui, X. H. and Cui, Y. D. and Dai, B. Z. and Dai, H. L. and Dai, Z. G. and Danzengluobu and della Volpe, D. and Dong, X. Q. and Duan, K. K. and Fan, J. H. and Fan, Y. Z. and Fang, J. and Fang, K. and Feng, C. F. and Feng, L. and Feng, S. H. and Feng, X. T. and Feng, Y. L. and Gabici, S. and Gao, B. and Gao, C. D. and Gao, L. Q. and Gao, Q. and Gao, W. and Gao, W. K. and Ge, M. M. and Geng, L. S. and Giacinti, G. and Gong, G. H. and Gou, Q. B. and Gu, M. H. and Guo, F. L. and Guo, X. L. and Guo, Y. Q. and Guo, Y. Y. and Han, Y. A. and He, H. H. and He, H. N. and He, J. Y. and He, X. B. and He, Y. and Heller, M. and Hor, Y. K. and Hou, B. W. and Hou, C. and Hou, X. and Hu, H. B. and Hu, Q. and Hu, S. C. and Huang, D. H. and Huang, T. Q. and Huang, W. J. and Huang, X. T. and Huang, X. Y. and Huang, Y. and Huang, Z. C. and Ji, X. L. and Jia, H. Y. and Jia, K. and Jiang, K. and Jiang, X. W. and Jiang, Z. J. and Jin, M. and Kang, M. M. and Ke, T. and Kuleshov, D. and Kurinov, K. and Li, B. B. and Li, Cheng and Li, Cong and Li, D. and Li, F. and Li, H. B. and Li, H. C. and Li, H. Y. and Li, J. and Li, Jian and Li, Jie and Li, K. and Li, W. L. and Li, W. L. and Li, X. R. and Li, Xin and Li, Y. Z. and Li, Zhe and Li, Zhuo and Liang, E. W. and Liang, Y. F. and Lin, S. J. and Liu, B. and Liu, C. and Liu, D. and Liu, H. and Liu, H. D. and Liu, J. and Liu, J. L. and Liu, J. Y. and Liu, M. Y. and Liu, R. Y. and Liu, S. M. and Liu, W. and Liu, Y. and Liu, Y. N. and Lu, R. and Luo, Q. and Lv, H. K. and Ma, B. Q. and Ma, L. L. and Ma, X. H. and Mao, J. R. and Min, Z. and Mitthumsiri, W. and Mu, H. J. and Nan, Y. C. and Neronov, A. and Ou, Z. W. and Pang, B. Y. and Pattarakijwanich, P. and Pei, Z. Y. and Qi, M. Y. and Qi, Y. Q. and Qiao, B. Q. and Qin, J. J. and Ruffolo, D. and S\'aiz, A. and Semikoz, D. and Shao, C. Y. and Shao, L. and Shchegolev, O. and Sheng, X. D. and Shu, F. W. and Song, H. C. and Stenkin, Yu. V. and Stepanov, V. and Su, Y. and Sun, Q. N. and Sun, X. N. and Sun, Z. B. and Tam, P. H. T. and Tang, Q. W. and Tang, Z. B. and Tian, W. W. and Wang, C. and Wang, C. B. and Wang, G. W. and Wang, H. G. and Wang, H. H. and Wang, J. C. and Wang, K. and Wang, L. P. and Wang, L. Y. and Wang, P. H. and Wang, R. and Wang, W. and Wang, X. G. and Wang, X. Y. and Wang, Y. and Wang, Y. D. and Wang, Y. J. and Wang, Z. H. and Wang, Z. X. and Wang, Zhen and Wang, Zheng and Wei, D. M. and Wei, J. J. and Wei, Y. J. and Wen, T. and Wu, C. Y. and Wu, H. R. and Wu, S. and Wu, X. F. and Wu, Y. S. and Xi, S. Q. and Xia, J. and Xia, J. J. and Xiang, G. M. and Xiao, D. X. and Xiao, G. and Xin, G. G. and Xin, Y. L. and Xing, Y. and Xiong, Z. and Xu, D. L. and Xu, R. F. and Xu, R. X. and Xu, W. L. and Xue, L. and Yan, D. H. and Yan, J. Z. and Yan, T. and Yang, C. W. and Yang, F. and Yang, F. F. and Yang, H. W. and Yang, J. Y. and Yang, L. L. and Yang, M. J. and Yang, R. Z. and Yang, S. B. and Yao, Y. H. and Yao, Z. G. and Ye, Y. M. and Yin, L. Q. and Yin, N. and You, X. H. and You, Z. Y. and Yu, Y. H. and Yuan, Q. and Yue, H. and Zeng, H. D. and Zeng, T. X. and Zeng, W. and Zha, M. and Zhang, B. B. and Zhang, F. and Zhang, H. M. and Zhang, H. Y. and Zhang, J. L. and Zhang, L. X. and Zhang, Li and Zhang, P. F. and Zhang, P. P. and Zhang, R. and Zhang, S. B. and Zhang, S. R. and Zhang, S. S. and Zhang, X. and Zhang, X. P. and Zhang, Y. F. and Zhang, Yi and Zhang, Yong and Zhao, B. and Zhao, J. and Zhao, L. and Zhao, L. Z. and Zhao, S. P. and Zheng, F. and Zhou, B. and Zhou, H. and Zhou, J. N. and Zhou, M. and Zhou, P. and Zhou, R. and Zhou, X. X. and Zhu, C. G. and Zhu, F. R. and Zhu, H. and Zhu, K. J. and Zuo, X.},
  collaboration = {LHAASO Collaboration},
  journal = {Phys. Rev. Lett.},
  volume = {131},
  issue = {15},
  pages = {151001},
  numpages = {9},
  year = {2023},
  month = {Oct},
  publisher = {American Physical Society},
  doi = {10.1103/PhysRevLett.131.151001},
  url = {https://link.aps.org/doi/10.1103/PhysRevLett.131.151001}
}

@article{lhaaso_LIV_2022,
  title = {Exploring Lorentz Invariance Violation from Ultrahigh-Energy $\ensuremath{\gamma}$ Rays Observed by LHAASO},
  author = {Cao, Zhen and Aharonian, F. and An, Q. and Axikegu and Bai, L. X. and Bai, Y. X. and Bao, Y. W. and Bastieri, D. and Bi, X. J. and Bi, Y. J. and Cai, H. and Cai, J. T. and Cao, Zhe and Chang, J. and Chang, J. F. and Chen, B. M. and Chen, E. S. and Chen, J. and Chen, Liang and Chen, Liang and Chen, Long and Chen, M. J. and Chen, M. L. and Chen, Q. H. and Chen, S. H. and Chen, S. Z. and Chen, T. L. and Chen, X. L. and Chen, Y. and Cheng, N. and Cheng, Y. D. and Cui, S. W. and Cui, X. H. and Cui, Y. D. and Piazzoli, B. D'Ettorre  and Dai, B. Z. and Dai, H. L. and Dai, Z. G. and Danzengluobu and della Volpe, D. and Dong, X. J. and Duan, K. K. and Fan, J. H. and Fan, Y. Z. and Fan, Z. X. and Fang, J. and Fang, K. and Feng, C. F. and Feng, L. and Feng, S. H. and Feng, Y. L. and Gao, B. and Gao, C. D. and Gao, L. Q. and Gao, Q. and Gao, W. and Ge, M. M. and Geng, L. S. and Gong, G. H. and Gou, Q. B. and Gu, M. H. and Guo, F. L. and Guo, J. G. and Guo, X. L. and Guo, Y. Q. and Guo, Y. Y. and Han, Y. A. and He, H. H. and He, H. N. and He, J. C. and He, S. L. and He, X. B. and He, Y. and Heller, M. and Hor, Y. K. and Hou, C. and Hou, X. and Hu, H. B. and Hu, S. and Hu, S. C. and Hu, X. J. and Huang, D. H. and Huang, Q. L. and Huang, W. H. and Huang, X. T. and Huang, X. Y. and Huang, Z. C. and Ji, F. and Ji, X. L. and Jia, H. Y. and Jiang, K. and Jiang, Z. J. and Jin, C. and Ke, T. and Kuleshov, D. and Levochkin, K. and Li, B. B. and Li, Cheng and Li, Cong and Li, F. and Li, H. B. and Li, H. C. and Li, H. Y. and Li, Jian and Li, Jie and Li, K. and Li, W. L. and Li, X. R. and Li, Xin and Li, Xin and Li, Y. and Li, Y. Z. and Li, Zhe and Li, Zhuo and Liang, E. W. and Liang, Y. F. and Lin, S. J. and Liu, B. and Liu, C. and Liu, D. and Liu, H. and Liu, H. D. and Liu, J. and Liu, J. L. and Liu, J. S. and Liu, J. Y. and Liu, M. Y. and Liu, R. Y. and Liu, S. M. and Liu, W. and Liu, Y. and Liu, Y. N. and Liu, Z. X. and Long, W. J. and Lu, R. and Lv, H. K. and Ma, B. Q. and Ma, L. L. and Ma, X. H. and Mao, J. R. and Masood, A. and Min, Z. and Mitthumsiri, W. and Montaruli, T. and Nan, Y. C. and Pang, B. Y. and Pattarakijwanich, P. and Pei, Z. Y. and Qi, M. Y. and Qi, Y. Q. and Qiao, B. Q. and Qin, J. J. and Ruffolo, D. and Rulev, V. and S\'aiz, A. and Shao, L. and Shchegolev, O. and Sheng, X. D. and Shi, J. R. and Song, H. C. and Stenkin, Yu. V. and Stepanov, V. and Su, Y. and Sun, Q. N. and Sun, X. N. and Sun, Z. B. and Tam, P. H. T. and Tang, Z. B. and Tian, W. W. and Wang, B. D. and Wang, C. and Wang, H. and Wang, H. G. and Wang, J. C. and Wang, J. S. and Wang, L. P. and Wang, L. Y. and Wang, R. N. and Wang, W. and Wang, W. and Wang, X. G. and Wang, X. J. and Wang, X. Y. and Wang, Y. and Wang, Y. D. and Wang, Y. J. and Wang, Y. P. and Wang, Z. H. and Wang, Z. X. and Wang, Zhen and Wang, Zheng and Wei, D. M. and Wei, J. J. and Wei, Y. J. and Wen, T. and Wu, C. Y. and Wu, H. R. and Wu, S. and Wu, W. X. and Wu, X. F. and Xi, S. Q. and Xia, J. and Xia, J. J. and Xiang, G. M. and Xiao, D. X. and Xiao, G. and Xiao, H. B. and Xin, G. G. and Xin, Y. L. and Xing, Y. and Xu, D. L. and Xu, R. X. and Xue, L. and Yan, D. H. and Yan, J. Z. and Yang, C. W. and Yang, F. F. and Yang, J. Y. and Yang, L. L. and Yang, M. J. and Yang, R. Z. and Yang, S. B. and Yao, Y. H. and Yao, Z. G. and Ye, Y. M. and Yin, L. Q. and Yin, N. and You, X. H. and You, Z. Y. and Yu, Y. H. and Yuan, Q. and Zeng, H. D. and Zeng, T. X. and Zeng, W. and Zeng, Z. K. and Zha, M. and Zhai, X. X. and Zhang, B. B. and Zhang, H. M. and Zhang, H. Y. and Zhang, J. L. and Zhang, J. W. and Zhang, L. X. and Zhang, Li and Zhang, Lu and Zhang, P. F. and Zhang, P. P. and Zhang, R. and Zhang, S. R. and Zhang, S. S. and Zhang, X. and Zhang, X. P. and Zhang, Y. F. and Zhang, Y. L. and Zhang, Yi and Zhang, Yong and Zhao, B. and Zhao, J. and Zhao, L. and Zhao, L. Z. and Zhao, S. P. and Zheng, F. and Zheng, Y. and Zhou, B. and Zhou, H. and Zhou, J. N. and Zhou, P. and Zhou, R. and Zhou, X. X. and Zhu, C. G. and Zhu, F. R. and Zhu, H. and Zhu, K. J. and Zuo, X.},
  collaboration = {LHAASO Collaboration},
  journal = {Phys. Rev. Lett.},
  volume = {128},
  issue = {5},
  pages = {051102},
  numpages = {7},
  year = {2022},
  month = {Feb},
  publisher = {American Physical Society},
  doi = {10.1103/PhysRevLett.128.051102},
  url = {https://link.aps.org/doi/10.1103/PhysRevLett.128.051102}
}

@article{lhaaso_LIV_2024,
  title = {Stringent Tests of Lorentz Invariance Violation from LHAASO Observations of GRB 221009A},
  author = {Cao, Zhen and Aharonian, F. and Axikegu and Bai, Y. X. and Bao, Y. W. and Bastieri, D. and Bi, X. J. and Bi, Y. J. and Bian, W. and Bukevich, A. V. and Cao, Q. and Cao, W. Y. and Cao, Zhe and Chang, J. and Chang, J. F. and Chen, A. M. and Chen, E. S. and Chen, H. X. and Chen, Liang and Chen, Lin and Chen, Long and Chen, M. J. and Chen, M. L. and Chen, Q. H. and Chen, S. and Chen, S. H. and Chen, S. Z. and Chen, T. L. and Chen, Y. and Cheng, N. and Cheng, Y. D. and Cui, M. Y. and Cui, S. W. and Cui, X. H. and Cui, Y. D. and Dai, B. Z. and Dai, H. L. and Dai, Z. G. and Danzengluobu and Dong, X. Q. and Duan, K. K. and Fan, J. H. and Fan, Y. Z. and Fang, J. and Fang, J. H. and Fang, K. and Feng, C. F. and Feng, H. and Feng, L. and Feng, S. H. and Feng, X. T. and Feng, Y. and Feng, Y. L. and Gabici, S. and Gao, B. and Gao, C. D. and Gao, Q. and Gao, W. and Gao, W. K. and Ge, M. M. and Geng, L. S. and Giacinti, G. and Gong, G. H. and Gou, Q. B. and Gu, M. H. and Guo, F. L. and Guo, X. L. and Guo, Y. Q. and Guo, Y. Y. and Han, Y. A. and Hasan, M. and He, H. H. and He, H. N. and He, J. Y. and He, Y. and Hor, Y. K. and Hou, B. W. and Hou, C. and Hou, X. and Hu, H. B. and Hu, Q. and Hu, S. C. and Huang, D. H. and Huang, T. Q. and Huang, W. J. and Huang, X. T. and Huang, X. Y. and Huang, Y. and Ji, X. L. and Jia, H. Y. and Jia, K. and Jiang, K. and Jiang, X. W. and Jiang, Z. J. and Jin, M. and Kang, M. M. and Karpikov, I. and Kuleshov, D. and Kurinov, K. and Li, B. B. and Li, C. M. and Li, Cheng and Li, Cong and Li, D. and Li, F. and Li, H. B. and Li, H. C. and Li, Jian and Li, Jie and Li, K. and Li, S. D. and Li, W. L. and Li, W. L. and Li, X. R. and Li, Xin and Li, Y. Z. and Li, Zhe and Li, Zhuo and Liang, E. W. and Liang, Y. F. and Lin, S. J. and Liu, B. and Liu, C. and Liu, D. and Liu, D. B. and Liu, H. and Liu, H. D. and Liu, J. and Liu, J. L. and Liu, M. Y. and Liu, R. Y. and Liu, S. M. and Liu, W. and Liu, Y. and Liu, Y. N. and Luo, Q. and Luo, Y. and Lv, H. K. and Ma, B. Q. and Ma, L. L. and Ma, X. H. and Mao, J. R. and Min, Z. and Mitthumsiri, W. and Mu, H. J. and Nan, Y. C. and Neronov, A. and Ou, L. J. and Pattarakijwanich, P. and Pei, Z. Y. and Qi, J. C. and Qi, M. Y. and Qiao, B. Q. and Qin, J. J. and Raza, A. and Ruffolo, D. and S\'aiz, A. and Saeed, M. and Semikoz, D. and Shao, L. and Shchegolev, O. and Sheng, X. D. and Shu, F. W. and Song, H. C. and Stenkin, Yu. V. and Stepanov, V. and Su, Y. and Sun, D. X. and Sun, Q. N. and Sun, X. N. and Sun, Z. B. and Takata, J. and Tam, P. H. T. and Tang, Q. W. and Tang, R. and Tang, Z. B. and Tian, W. W. and Wang, C. and Wang, C. B. and Wang, G. W. and Wang, H. G. and Wang, H. H. and Wang, J. C. and Wang, Kai and Wang, Kai and Wang, L. P. and Wang, L. Y. and Wang, P. H. and Wang, R. and Wang, W. and Wang, X. G. and Wang, X. Y. and Wang, Y. and Wang, Y. D. and Wang, Y. J. and Wang, Z. H. and Wang, Z. X. and Wang, Zhen and Wang, Zheng and Wei, D. M. and Wei, J. J. and Wei, Y. J. and Wen, T. and Wu, C. Y. and Wu, H. R. and Wu, Q. W. and Wu, S. and Wu, X. F. and Wu, Y. S. and Xi, S. Q. and Xia, J. and Xiang, G. M. and Xiao, D. X. and Xiao, G. and Xin, Y. L. and Xing, Y. and Xiong, D. R. and Xiong, Z. and Xu, D. L. and Xu, R. F. and Xu, R. X. and Xu, W. L. and Xue, L. and Yan, D. H. and Yan, J. Z. and Yan, T. and Yang, C. W. and Yang, C. Y. and Yang, F. and Yang, F. F. and Yang, L. L. and Yang, M. J. and Yang, R. Z. and Yang, W. X. and Yao, Y. H. and Yao, Z. G. and Yin, L. Q. and Yin, N. and You, X. H. and You, Z. Y. and Yu, Y. H. and Yuan, Q. and Yue, H. and Zeng, H. D. and Zeng, T. X. and Zeng, W. and Zha, M. and Zhang, B. B. and Zhang, F. and Zhang, H. and Zhang, H. M. and Zhang, H. Y. and Zhang, J. L. and Zhang, Li and Zhang, P. F. and Zhang, P. P. and Zhang, R. and Zhang, S. B. and Zhang, S. R. and Zhang, S. S. and Zhang, X. and Zhang, X. P. and Zhang, Y. F. and Zhang, Yi and Zhang, Yong and Zhao, B. and Zhao, J. and Zhao, L. and Zhao, L. Z. and Zhao, S. P. and Zhao, X. H. and Zheng, F. and Zhong, W. J. and Zhou, B. and Zhou, H. and Zhou, J. N. and Zhou, M. and Zhou, P. and Zhou, R. and Zhou, X. X. and Zhou, X. X. and Zhu, B. Y. and Zhu, C. G. and Zhu, F. R. and Zhu, H. and Zhu, K. J. and Zou, Y. C. and Zuo, X.},
  collaboration = {The LHAASO Collaboration},
  journal = {Phys. Rev. Lett.},
  volume = {133},
  issue = {7},
  pages = {071501},
  numpages = {7},
  year = {2024},
  month = {Aug},
  publisher = {American Physical Society},
  doi = {10.1103/PhysRevLett.133.071501},
  url = {https://link.aps.org/doi/10.1103/PhysRevLett.133.071501}
}

@article{LIV_1989,
  title = {Spontaneous breaking of Lorentz symmetry in string theory},
  author = {Kosteleck\'y, V. Alan and Samuel, Stuart},
  journal = {Phys. Rev. D},
  volume = {39},
  issue = {2},
  pages = {683--685},
  numpages = {0},
  year = {1989},
  month = {Jan},
  publisher = {American Physical Society},
  doi = {10.1103/PhysRevD.39.683},
  url = {https://link.aps.org/doi/10.1103/PhysRevD.39.683}
}

@article{LIV_2001,
  title = {Noncommutative Field Theory and Lorentz Violation},
  author = {Carroll, Sean M. and Harvey, Jeffrey A. and Kosteleck\'y, V. Alan and Lane, Charles D. and Okamoto, Takemi},
  journal = {Phys. Rev. Lett.},
  volume = {87},
  issue = {14},
  pages = {141601},
  numpages = {4},
  year = {2001},
  month = {Sep},
  publisher = {American Physical Society},
  doi = {10.1103/PhysRevLett.87.141601},
  url = {https://link.aps.org/doi/10.1103/PhysRevLett.87.141601}
}

@ARTICLE{test_of_QG_1998,
       author = {{Amelino-Camelia}, G. and {Ellis}, John and {Mavromatos}, N.~E. and {Nanopoulos}, D.~V. and {Sarkar}, Subir},
        title = "{Tests of quantum gravity from observations of {\ensuremath{\gamma}}-ray bursts}",
      journal = {\nat},
         year = 1998,
        month = jun,
       volume = {393},
       number = {6687},
        pages = {763-765},
          doi = {10.1038/31647},
archivePrefix = {arXiv},
       eprint = {astro-ph/9712103},
 primaryClass = {astro-ph},
       adsurl = {https://ui.adsabs.harvard.edu/abs/1998Natur.393..763A}
}

@ARTICLE{qg_mms_review,
       author = {{Addazi}, A. and {Alvarez-Muniz}, J. and {Alves Batista}, R. and {Amelino-Camelia}, G. and {Antonelli}, V. and {Arzano}, M. and {Asorey}, M. and {Atteia}, J.-L. and {Bahamonde}, S. and {Bajardi}, F. and {Ballesteros}, A. and {Baret}, B. and {Barreiros}, D.~M. and {Basilakos}, S. and {Benisty}, D. and {Birnholtz}, O. and {Blanco-Pillado}, J.~J. and {Blas}, D. and {Bolmont}, J. and {Boncioli}, D. and {Bosso}, P. and {Calcagni}, G. and {Capozziello}, S. and {Carmona}, J.~M. and {Cerci}, S. and {Chernyakova}, M. and {Clesse}, S. and {Coelho}, J.~A.~B. and {Colak}, S.~M. and {Cortes}, J.~L. and {Das}, S. and {D'Esposito}, V. and {Demirci}, M. and {Di Luca}, M.~G. and {di Matteo}, A. and {Dimitrijevic}, D. and {Djordjevic}, G. and {Prester}, D. Dominis and {Eichhorn}, A. and {Ellis}, J. and {Escamilla-Rivera}, C. and {Fabiano}, G. and {Franchino-Vi{\~n}as}, S.~A. and {Frassino}, A.~M. and {Frattulillo}, D. and {Funk}, S. and {Fuster}, A. and {Gamboa}, J. and {Gent}, A. and {Gergely}, L. {\'A}. and {Giammarchi}, M. and {Giesel}, K. and {Glicenstein}, J.-F. and {Gracia-Bond{\'\i}a}, J. and {Gracia-Ruiz}, R. and {Gubitosi}, G. and {Guendelman}, E.~I. and {Gutierrez-Sagredo}, I. and {Haegel}, L. and {Heefer}, S. and {Held}, A. and {Herranz}, F.~J. and {Hinderer}, T. and {Illana}, J.~I. and {Ioannisian}, A. and {Jetzer}, P. and {Joaquim}, F.~R. and {Kampert}, K.-H. and {Uysal}, A. Karasu and {Katori}, T. and {Kazarian}, N. and {Kerszberg}, D. and {Kowalski-Glikman}, J. and {Kuroyanagi}, S. and {L{\"a}mmerzahl}, C. and {Said}, J. Levi and {Liberati}, S. and {Lim}, E. and {Lobo}, I.~P. and {L{\'o}pez-Moya}, M. and {Luciano}, G.~G. and {Manganaro}, M. and {Marcian{\`o}}, A. and {Mart{\'\i}n-Moruno}, P. and {Martinez}, Manel and {Martinez}, Mario and {Mart{\'\i}nez-Huerta}, H. and {Mart{\'\i}nez-Mirav{\'e}}, P. and {Masip}, M. and {Mattingly}, D. and {Mavromatos}, N. and {Mazumdar}, A. and {M{\'e}ndez}, F. and {Mercati}, F. and {Micanovic}, S. and {Mielczarek}, J. and {Miller}, A.~L. and {Milosevic}, M. and {Minic}, D. and {Miramonti}, L. and {Mitsou}, V.~A. and {Moniz}, P. and {Mukherjee}, S. and {Nardini}, G. and {Navas}, S. and {Niechciol}, M. and {Nielsen}, A.~B. and {Obers}, N.~A. and {Oikonomou}, F. and {Oriti}, D. and {Paganini}, C.~F. and {Palomares-Ruiz}, S. and {Pasechnik}, R. and {Pasic}, V. and {P{\'e}rez de los Heros}, C. and {Pfeifer}, C. and {Pieroni}, M. and {Piran}, T. and {Platania}, A. and {Rastgoo}, S. and {Relancio}, J.~J. and {Reyes}, M.~A. and {Ricciardone}, A. and {Risse}, M. and {Frias}, M.~D. Rodriguez and {Rosati}, G. and {Rubiera-Garcia}, D. and {Sahlmann}, H. and {Sakellariadou}, M. and {Salamida}, F. and {Saridakis}, E.~N. and {Satunin}, P. and {Schiffer}, M. and {Sch{\"u}ssler}, F. and {Sigl}, G. and {Sitarek}, J. and {Peracaula}, J. Sol{\`a} and {Sopuerta}, C.~F. and {Sotiriou}, T.~P. and {Spurio}, M. and {Staicova}, D. and {Stergioulas}, N. and {Stoica}, S. and {Stri{\v{s}}kovi{\'c}}, J. and {Stuttard}, T. and {Cerci}, D. Sunar and {Tavakoli}, Y. and {Ternes}, C.~A. and {Terzi{\'c}}, T. and {Thiemann}, T. and {Tinyakov}, P. and {Torri}, M.~D.~C. and {T{\'o}rtola}, M. and {Trimarelli}, C. and {Trze{\'s}niewski}, T. and {Tureanu}, A. and {Urban}, F.~R. and {Vagenas}, E.~C. and {Vernieri}, D. and {Vitagliano}, V. and {Wallet}, J.-C. and {Zornoza}, J.~D.},
        title = "{Quantum gravity phenomenology at the dawn of the multi-messenger era-A review}",
      journal = {Progress in Particle and Nuclear Physics},
         year = 2022,
        month = jul,
       volume = {125},
          eid = {103948},
        pages = {103948},
          doi = {10.1016/j.ppnp.2022.103948},
archivePrefix = {arXiv},
       eprint = {2111.05659},
 primaryClass = {hep-ph},
       adsurl = {https://ui.adsabs.harvard.edu/abs/2022PrPNP.12503948A}
}

@article{superluminal_LIV_2017,
  title = {Restrictions from Lorentz invariance violation on cosmic ray propagation},
  author = {Mart\'{\i}nez-Huerta, H. and P\'erez-Lorenzana, A.},
  journal = {Phys. Rev. D},
  volume = {95},
  issue = {6},
  pages = {063001},
  numpages = {14},
  year = {2017},
  month = {Mar},
  publisher = {American Physical Society},
  doi = {10.1103/PhysRevD.95.063001},
  url = {https://link.aps.org/doi/10.1103/PhysRevD.95.063001}
}

@article{superluminal_LIV_2016,
doi = {10.1088/1742-6596/761/1/012035},
url = {https://doi.org/10.1088/1742-6596/761/1/012035},
year = {2016},
month = {oct},
publisher = {IOP Publishing},
volume = {761},
number = {1},
pages = {012035},
author = {Martínez-Huerta, H. and Pérez-Lorenzana, A.},
title = {Vacuum Cherenkov radiation and photon decay rates from generic Lorentz Invariance Violation},
journal = {Journal of Physics: Conference Series},
}

@article{carmona_2024,
  title = {Approaches to photon absorption in a Lorentz invariance violation scenario},
  author = {Carmona, J. M. and Cort\'es, J. L. and Rescic, F. and Reyes, M. A. and Terzi\ifmmode \acute{c}\else \'{c}\fi{}, T. and Vrban, F. I.},
  journal = {Phys. Rev. D},
  volume = {110},
  issue = {6},
  pages = {063035},
  numpages = {13},
  year = {2024},
  month = {Sep},
  publisher = {American Physical Society},
  doi = {10.1103/PhysRevD.110.063035},
  url = {https://link.aps.org/doi/10.1103/PhysRevD.110.063035}
}

@article{Abdalla_2019,
doi = {10.3847/1538-4357/aaf1c4},
url = {https://doi.org/10.3847/1538-4357/aaf1c4},
year = {2019},
month = {jan},
publisher = {The American Astronomical Society},
volume = {870},
number = {2},
pages = {93},
author = {Abdalla, H. and Aharonian, F. and Benkhali, F. Ait and Angüner, E. O. and Arakawa, M. and Arcaro, C. and Armand, C. and Arrieta, M. and Backes, M. and Barnard, M. and Becherini, Y. and Tjus, J. Becker and Berge, D. and Bernhard, S. and Bernlöhr, K. and Blackwell, R. and Böttcher, M. and Boisson, C. and Bolmont, J. and Bonnefoy, S. and Bordas, P. and Bregeon, J. and Brun, F. and Brun, P. and Bryan, M. and Büchele, M. and Bulik, T. and Bylund, T. and Capasso, M. and Caroff, S. and Carosi, A. and Cerruti, M. and Chakraborty, N. and Chandra, S. and Chaves, R. C. G. and Chen, A. and Colafrancesco, S. and Condon, B. and Davids, I. D. and Deil, C. and Devin, J. and deWilt, P. and Dirson, L. and Djannati-Ataï, A. and Dmytriiev, A. and Donath, A. and Doroshenko, V. and Drury, L. O’C. and Dyks, J. and Egberts, K. and Emery, G. and Ernenwein, J.-P. and Eschbach, S. and Fegan, S. and Fiasson, A. and Fontaine, G. and Funk, S. and Füßling, M. and Gabici, S. and Gallant, Y. A. and Gaté, F. and Giavitto, G. and Glawion, D. and Glicenstein, J. F. and Gottschall, D. and Grondin, M.-H. and Hahn, J. and Haupt, M. and Heinzelmann, G. and Henri, G. and Hermann, G. and Hinton, J. A. and Hofmann, W. and Hoischen, C. and Holch, T. L. and Holler, M. and Horns, D. and Huber, D. and Iwasaki, H. and Jacholkowska, A. and Jamrozy, M. and Jankowsky, D. and Jankowsky, F. and Jouvin, L. and Jung-Richardt, I. and Kastendieck, M. A. and Katarzyński, K. and Katsuragawa, M. and Katz, U. and Kerszberg, D. and Khangulyan, D. and Khélifi, B. and King, J. and Klepser, S. and Kluźniak, W. and Komin, Nu. and Kosack, K. and Krakau, S. and Kraus, M. and Krüger, P. P. and Lamanna, G. and Lau, J. and Lefaucheur, J. and Lemière, A. and Lemoine-Goumard, M. and Lenain, J.-P. and Leser, E. and Lohse, T. and Lorentz, M. and López-Coto, R. and Lypova, I. and Malyshev, D. and Marandon, V. and Marcowith, A. and Mariaud, C. and Martí-Devesa, G. and Marx, R. and Maurin, G. and Meintjes, P. J. and Mitchell, A. M. W. and Moderski, R. and Mohamed, M. and Mohrmann, L. and Moulin, E. and Murach, T. and Nakashima, S. and Naurois, M. de and Ndiyavala, H. and Niederwanger, F. and Niemiec, J. and Oakes, L. and O’Brien, P. and Odaka, H. and Ohm, S. and Ostrowski, M. and Oya, I. and Padovani, M. and Panter, M. and Parsons, R. D. and Perennes, C. and Petrucci, P.-O. and Peyaud, B. and Piel, Q. and Pita, S. and Poireau, V. and Priyana Noel, A. and Prokhorov, D. and Prokoph, H. and Pühlhofer, G. and Punch, M. and Quirrenbach, A. and Raab, S. and Rauth, R. and Reimer, A. and Reimer, O. and Renaud, M. and Rieger, F. and Rinchiuso, L. and Romoli, C. and Rowell, G. and Rudak, B. and Ruiz-Velasco, E. and Sahakian, V. and Saito, S. and Sanchez, D. A. and Santangelo, A. and Sasaki, M. and Schlickeiser, R. and Schüssler, F. and Schulz, A. and Schwanke, U. and Schwemmer, S. and Seglar-Arroyo, M. and Senniappan, M. and Seyffert, A. S. and Shafi, N. and Shilon, I. and Shiningayamwe, K. and Simoni, R. and Sinha, A. and Sol, H. and Spanier, F. and Specovius, A. and Spir-Jacob, M. and Stawarz, Ł. and Steenkamp, R. and Stegmann, C. and Steppa, C. and Takahashi, T. and Tavernet, J.-P. and Tavernier, T. and Taylor, A. M. and Terrier, R. and Tibaldo, L. and Tiziani, D. and Tluczykont, M. and Trichard, C. and Tsirou, M. and Tsuji, N. and Tuffs, R. and Uchiyama, Y. and van der Walt, D. J. and van Eldik, C. and van Rensburg, C. and van Soelen, B. and Vasileiadis, G. and Veh, J. and Venter, C. and Vincent, P. and Vink, J. and Voisin, F. and Völk, H. J. and Vuillaume, T. and Wadiasingh, Z. and Wagner, S. J. and Wagner, R. M. and White, R. and Wierzcholska, A. and Yang, R. and Zaborov, D. and Zacharias, M. and Zanin, R. and Zdziarski, A. A. and Zech, A. and Zefi, F. and Ziegler, A. and Zorn, J. and Żywucka, N. and (H.E.S.S. Collaboration)},
title = {The 2014 TeV γ-Ray Flare of Mrk 501 Seen with H.E.S.S.: Temporal and Spectral Constraints on Lorentz Invariance Violation},
journal = {The Astrophysical Journal},
}

@article{Li_2023,
doi = {10.1088/1475-7516/2023/10/061},
url = {https://doi.org/10.1088/1475-7516/2023/10/061},
year = {2023},
month = {oct},
publisher = {IOP Publishing},
volume = {2023},
number = {10},
pages = {061},
author = {Li, Hao and Ma, Bo-Qiang},
title = {Revisiting Lorentz invariance violation from GRB 221009A},
journal = {Journal of Cosmology and Astroparticle Physics},
}

@article{swgo_science,
      title={Science Prospects for the Southern Wide-field Gamma-ray Observatory: SWGO}, 
      author={{SWGO Collaboration}},
      year={2025},
      eprint={2506.01786},
      archivePrefix={arXiv},
      primaryClass={astro-ph.HE},
      url={https://arxiv.org/abs/2506.01786}, 
}

@article{swgo_pevatrons,
title = {The expected potential of hadronic PeVatron searches with spectral γ-ray data from the Southern Wide-field Gamma-ray Observatory},
journal = {Astroparticle Physics},
volume = {158},
pages = {102936},
year = {2024},
issn = {0927-6505},
doi = {https://doi.org/10.1016/j.astropartphys.2024.102936},
url = {https://www.sciencedirect.com/science/article/pii/S0927650524000136},
author = {Ekrem Oğuzhan Angüner and Tülün Ergin},
}

@article{cta_gps,
doi = {10.1088/1475-7516/2024/10/081},
url = {https://doi.org/10.1088/1475-7516/2024/10/081},
year = {2024},
month = {oct},
publisher = {IOP Publishing},
volume = {2024},
number = {10},
pages = {081},
author = {Abe, S. and Abhir, J. and Abhishek, A. and Acero, F. and Acharyya, A. and Adam, R. and Aguasca-Cabot, A. and Agudo, I. and Aguirre-Santaella, A. and Alfaro, J. and Alvarez-Crespo, N. and Alves Batista, R. and Amans, J.-P. and Amato, E. and Ambrosi, G. and Ambrosino, F. and Angüner, E.O. and Aramo, C. and Arcaro, C. and Arrabito, L. and Asano, K. and Ascasíbar, Y. and Aschersleben, J. and Augusto Stuani, L. and Backes, M. and Balazs, C. and Balbo, M. and Ballet, J. and Baquero Larriva, A. and Barbosa Martins, V. and Barres de Almeida, U. and Barrio, J.A. and Batković, I. and Batzofin, R. and Baxter, J. and Becerra González, J. and Beck, G. and Beiske, L. and Belmont, R. and Benbow, W. and Bernardini, E. and Bernete, J. and Bernlöhr, K. and Berti, A. and Bertucci, B. and Beshley, V. and Bhattacharjee, P. and Bhattacharyya, S. and Bi, B. and Biederbeck, N. and Biland, A. and Bissaldi, E. and Biteau, J. and Blanch, O. and Blazek, J. and Bocchino, F. and Boisson, C. and Bolmont, J. and Bonneau Arbeletche, L. and Bonnoli, G. and Bonollo, A. and Bordas, P. and Bosnjak, Z. and Bottacini, E. and Braiding, C. and Bronzini, E. and Brose, R. and Brown, A.M. and Brun, F. and Brunelli, G. and Bucciantini, N. and Bulgarelli, A. and Burelli, I. and Burmistrov, L. and Burton, M. and Burtovoi, A. and Bylund, T. and Calisse, P.G. and Campoy-Ordaz, A. and Cantlay, B.K. and Caproni, A. and Capuzzo-Dolcetta, R. and Caraveo, P. and Caroff, S. and Carosi, A. and Carosi, R. and Carquin, E. and Carrasco, M.-S. and Cascone, E. and Cassol, F. and Castrejon, N. and Castro-Tirado, A.J. and Cerasole, D. and Cerruti, M. and Chadwick, P.M. and Chambery, P. and Chaty, S. and Chen, A.W. and Chernyakova, M. and Chiavassa, A. and Chytka, L. and Cifuentes, A. and Coimbra Araujo, C.H. and Conforti, V. and Conte, F. and Contreras, J.L. and Cortina, J. and Costa, A. and Costantini, H. and Cotter, G. and Crestan, S. and Cristofari, P. and Cuevas, O. and Curtis-Ginsberg, Z. and D'Aì, A. and D'Amico, G. and D'Ammando, F. and Dadina, M. and Dalchenko, M. and David, L. and Dazzi, F. and de Bony de Lavergne, M. and De Caprio, V. and De Frondat Laadim, F. and de Gouveia Dal Pino, E.M. and De Lotto, B. and De Lucia, M. and de Martino, D. and de Menezes, R. and de Naurois, M. and de Ona Wilhelmi, E. and de Souza, V. and del Peral, L. and Delgado Giler, A.G. and Delgado, C. and Dell'aiera, M. and Della Valle, M. and della Volpe, D. and Depaoli, D. and Di Girolamo, T. and Di Piano, A. and Di Pierro, F. and Di Tria, R. and Di Venere, L. and Díaz, C. and Diebold, S. and Dinesh, A. and Djannati-Ataï, A. and Djuvsland, J. and Domínguez, A. and Dominik, R.M. and Donini, A. and Dörner, J. and Doro, M. and dos Anjos, R.D.C. and Dournaux, J.-L. and Duangchan, C. and Dubos, C. and Dubus, G. and Duffy, S. and Dumora, D. and Dwarkadas, V.V. and Ebr, J. and Eckner, C. and Egberts, K. and Einecke, S. and Elsässer, D. and Emery, G. and Errando, M. and Escanuela, C. and Escarate, P. and Escobar Godoy, M. and Escudero, J. and Esposito, P. and Evoli, C. and Falceta-Goncalves, D. and Fattorini, A. and Fegan, S. and Feijen, K. and Feng, Q. and Ferrand, G. and Ferrarotto, F. and Fiandrini, E. and Fiasson, A. and Filipovic, M. and Fioretti, V. and Fiori, M. and Flores, H. and Foffano, L. and Font Guiteras, L. and Fontaine, G. and Fröse, S. and Fukazawa, Y. and Fukui, Y. and Funk, S. and Furniss, A. and Gaggero, D. and Galanti, G. and Galaz, G. and Gallant, Y.A. and Gallozzi, S. and Gammaldi, V. and Garczarczyk, M. and Gasbarra, C. and Gasparrini, D. and Gaug, M. and Ghalumyan, A. and Giarrusso, M. and Giesbrecht, J. and Giglietto, N. and Giordano, F. and Giuffrida, R. and Giuliani, A. and Glicenstein, J.-F. and Glombitza, J. and Godinovic, N. and Goldoni, P. and González, J.M. and Goulart Coelho, J. and Granot, J. and Grasso, D. and Grau, R. and Gréaux, L. and Green, D. and Green, J.G. and Greenshaw, T. and Grenier, I. and Grolleron, G. and Grondin, M.-H. and Gueta, O. and Gunji, S. and Hackfeld, J. and Hadasch, D. and Hanlon, W. and Hara, S. and Harvey, V.M. and Hassan, T. and Hayashi, K. and Heckmann, L. and Heller, M. and Hermann, G. and Hernández Cadena, S. and Hervet, O. and Hinton, J. and Hiroshima, N. and Hnatyk, B. and Hnatyk, R. and Hofmann, W. and Holder, J. and Holler, M. and Horan, D. and Horvath, P. and Hovatta, T. and Hrabovsky, M. and Iarlori, M. and Inada, T. and Incardona, F. and Inoue, S. and Iocco, F. and Iori, M. and Jamrozy, M. and Janecek, P. and Jankowsky, F. and Jarnot, C. and Jean, P. and Jiménez Martínez, I. and Jin, W. and Juramy-Gilles, C. and Jurysek, J. and Kagaya, M. and Kalekin, O. and Kantzas, D. and Karas, V. and Katagiri, H. and Kataoka, J. and Kaufmann, S. and Kazanas, D. and Kerszberg, D. and Khélifi, B. and Kieda, D.B. and Kissmann, R. and Kleiner, T. and Kluge, G. and Kluźniak, W. and Knödlseder, J. and Kobayashi, Y. and Kohri, K. and Komin, N. and Kornecki, P. and Kosack, K. and Kostunin, D. and Kowal, G. and Kubo, H. and Kushida, J. and La Barbera, A. and La Palombara, N. and Láinez, M. and Lamastra, A. and Lapington, J. and Laporte, P. and Lazarević, S. and Lazendic-Galloway, J. and Lemoine-Goumard, M. and Lenain, J.-P. and Leone, F. and Leto, G. and Leuschner, F. and Lindfors, E. and Linhoff, M. and Liodakis, I. and Lombardi, S. and Longo, F. and López-Coto, R. and López-Moya, M. and López-Oramas, A. and Loporchio, S. and Lozano Bahilo, J. and Lucarelli, F. and Luque-Escamilla, P.L. and Lyard, E. and Macias, O. and Mackey, J. and Maier, G. and Malyshev, D. and Mandat, D. and Manicò, G. and Marcowith, A. and Marinos, P. and Mariotti, M. and Markoff, S. and Marquez, P. and Marsella, G. and Martí, J. and Martin, P. and Martínez, G.A. and Martínez, M. and Martinez, O. and Marty, C. and Mas-Aguilar, A. and Mastropietro, M. and Maurin, G. and Mazin, D. and McKeague, S. and Mello, A.J.T.S. and Menchiari, S. and Mereghetti, S. and Mestre, E. and Meunier, J.-L. and Meyer, D.M.-A. and Miceli, D. and Miceli, M. and Michailidis, M. and Michałowski, J. and Miener, T. and Miranda, J.M. and Mitchell, A. and Mizuno, T. and Moderski, R. and Mohrmann, L. and Molero, M. and Molfese, C. and Molina, E. and Montaruli, T. and Moralejo, A. and Morcuende, D. and Morik, K. and Morlino, G. and Morselli, A. and Moulin, E. and Moya Zamanillo, V. and Mukherjee, R. and Munari, K. and Murach, T. and Muraczewski, A. and Muraishi, H. and Nagataki, S. and Nakamori, T. and Nemmen, R. and Nickel, L. and Niemiec, J. and Nieto, D. and Nievas Rosillo, M. and Nikołajuk, M. and Nikolić, L. and Noda, K. and Nosek, D. and Novosyadlyj, B. and Novotny, V. and Nozaki, S. and Ohishi, M. and Ohtani, Y. and Okumura, A. and Olive, J.-F. and Olmi, B. and Ong, R.A. and Orienti, M. and Orito, R. and Orlandini, M. and Orlando, E. and Orlando, S. and Ostrowski, M. and Oya, I. and Pagano, I. and Pagliaro, A. and Palatiello, M. and Panebianco, G. and Paneque, D. and Pantaleo, F.R. and Paoletti, R. and Paredes, J.M. and Parmiggiani, N. and Patel, S.R. and Patricelli, B. and Pavlović, D. and Pech, M. and Pecimotika, M. and Peresano, M. and Pérez-Romero, J. and Pérez-Torres, M.A. and Peron, G. and Persic, M. and Petrucci, P.-O. and Petruk, O. and Piano, G. and Pierre, E. and Pietropaolo, E. and Pihet, M. and Pintore, F. and Pittori, C. and Plard, C. and Podobnik, F. and Pohl, M. and Pons, E. and Ponti, G. and Prandini, E. and Principe, G. and Priyadarshi, C. and Produit, N. and Prokhorov, D. and Pueschel, E. and Pühlhofer, G. and Pumo, M.L. and Punch, M. and Queiroz, F. and Quirrenbach, A. and Rando, R. and Ravel, T. and Razzaque, S. and Regeard, M. and Reichherzer, P. and Reimer, A. and Reimer, O. and Remy, Q. and Renaud, M. and Reposeur, T. and Rhode, W. and Ribeiro, D. and Ribó, M. and Richtler, T. and Rico, J. and Rieger, F. and Rigoselli, M. and Rizi, V. and Roache, E. and Rodriguez Fernandez, G. and Rodríguez-Vázquez, J.J. and Romano, P. and Romeo, G. and Rosado, J. and Rosales de Leon, A. and Rowell, G. and Rudak, B. and Ruiter, A.J. and Rulten, C.B. and Russo, F. and Sadeh, I. and Saha, L. and Saito, T. and Salzmann, H. and Sánchez-Conde, M. and Sangiorgi, P. and Sano, H. and Santander, M. and Santangelo, A. and Santos-Lima, R. and Sapienza, V. and Šarić, T. and Sarkar, S. and Saturni, F.G. and Scherer, A. and Schiavone, F. and Schipani, P. and Schleicher, B. and Schovanek, P. and Schubert, J.L. and Schussler, F. and Schwanke, U. and Schwefer, G. and Seglar Arroyo, M. and Seitenzahl, I. and Sergijenko, O. and Servillat, M. and Sguera, V. and Sharma, P. and Siejkowski, H. and Siqueira, C. and Sizun, P. and Sliusar, V. and Slowikowska, A. and Sol, H. and Spencer, S.T. and Spiga, D. and Stamerra, A. and Stanič, S. and Starling, R. and Stawarz, Ł. and Steinmassl, S. and Steppa, C. and Stolarczyk, T. and Suda, Y. and Suomijärvi, T. and Tajima, H. and Takeishi, R. and Tanaka, S.J. and Tavecchio, F. and Tavernier, T. and Terada, Y. and Terrier, R. and Teshima, M. and Tian, W.W. and Tibaldo, L. and Tibolla, O. and Torradeflot, F. and Torres, D.F. and Tothill, N. and Toussenel, F. and Touzard, V. and Travnicek, P. and Tripodo, G. and Trois, A. and Tsiahina, A. and Tutone, A. and Umana, G. and Vaclavek, L. and Vacula, M. and Vallania, P. and van Eldik, C. and Vassiliev, V. and Vazquez Acosta, M.L. and Vecchi, M. and Ventura, S. and Vercellone, S. and Verna, G. and Viana, A. and Viaux, N. and Vigliano, A. and Vignatti, J. and Vigorito, C.F. and Villanueva, J. and Vink, J. and Vitale, V. and Vodeb, V. and Voisin, V. and Vorobiov, S. and Voutsinas, G. and Vovk, I. and Vuillaume, T. and Waegebaert, V. and Wagner, S.J. and Walter, R. and Wechakama, M. and White, R. and Wierzcholska, A. and Williams, D.A. and Wohlleben, F. and Yamazaki, R. and Yang, L. and Yoshida, T. and Yoshikoshi, T. and Zacharias, M. and Zaharijas, G. and Zampieri, L. and Zanin, R. and Zavrtanik, D. and Zavrtanik, M. and Zdziarski, A.A. and Zech, A. and Zhdanov, V.I. and Ziętara, K. and Živec, M. and Zuriaga-Puig, J. and De la Torre Luque, P. and Guillemot, L. and Smith, D.A. and The CTA Consortium},
title = {Prospects for a survey of the galactic plane with the Cherenkov Telescope Array},
journal = {Journal of Cosmology and Astroparticle Physics},
}

@article{science_with_CTA,
author = {{CTA~Collaboration}},
title = {Science with the Cherenkov Telescope Array},
publisher = {WORLD SCIENTIFIC},
year = {2019},
doi = {10.1142/10986},
address = {},
edition   = {},
URL = {https://www.worldscientific.com/doi/abs/10.1142/10986},
eprint = {https://www.worldscientific.com/doi/pdf/10.1142/10986}
}

@article{Porter_2022,
doi = {10.3847/1538-4365/ac80f6},
url = {https://doi.org/10.3847/1538-4365/ac80f6},
year = {2022},
month = {sep},
publisher = {The American Astronomical Society},
volume = {262},
number = {1},
pages = {30},
author = {Porter, T. A. and Jóhannesson, G. and Moskalenko, I. V.},
title = {The GALPROP Cosmic-ray Propagation and Nonthermal Emissions Framework: Release v57},
journal = {The Astrophysical Journal Supplement Series},
}

@article{Robitaille_2012,
	author = {{Robitaille, T.P.} and {Churchwell, E.} and {Benjamin, R.A.} and {Whitney, B.A.} and {Wood, K.} and {Babler, B.L.} and {Meade, M.R.}},
	title = {A self-consistent model of Galactic stellar and dust infrared
          emission and the abundance of polycyclic aromatic hydrocarbons},
	DOI= "10.1051/0004-6361/201219073",
	url= "https://doi.org/10.1051/0004-6361/201219073",
	journal = {A\&A},
	year = 2012,
	volume = 545,
	pages = "A39",
	month = "",
}

@article{Moskalenko_2006,
doi = {10.1086/503524},
url = {https://doi.org/10.1086/503524},
year = {2006},
month = {mar},
publisher = {},
volume = {640},
number = {2},
pages = {L155},
author = {Moskalenko, Igor V. and Porter, Troy A. and Strong, Andrew W.},
title = {Attenuation of Very High Energy Gamma Rays by the Milky Way Interstellar Radiation Field},
journal = {The Astrophysical Journal},
}

@article{Freudenreich_1998,
doi = {10.1086/305065},
url = {https://doi.org/10.1086/305065},
year = {1998},
month = {jan},
publisher = {},
volume = {492},
number = {2},
pages = {495},
author = {Freudenreich, H. T.},
title = {A COBE Model of the Galactic Bar and Disk},
journal = {The Astrophysical Journal},
}

@article{breit_wheeler_CS,
  title = {Collision of Two Light Quanta},
  author = {Breit, G. and Wheeler, John A.},
  journal = {Phys. Rev.},
  volume = {46},
  issue = {12},
  pages = {1087--1091},
  numpages = {0},
  year = {1934},
  month = {Dec},
  publisher = {American Physical Society},
  doi = {10.1103/PhysRev.46.1087},
  url = {https://link.aps.org/doi/10.1103/PhysRev.46.1087}
}

@article{Franceschini2008,
	author = {{Franceschini, A.} and {Rodighiero, G.} and {Vaccari, M.}},
	title = {Extragalactic optical-infrared background radiation, 
 its time evolution and the cosmic photon-photon opacity
},
	DOI= "10.1051/0004-6361:200809691",
	url= "https://doi.org/10.1051/0004-6361:200809691",
	journal = {A\&A},
	year = 2008,
	volume = 487,
	number = 3,
	pages = "837-852",
}

@article{Dominguez2011,
    author = {Domínguez, A. and Primack, J. R. and Rosario, D. J. and Prada, F. and Gilmore, R. C. and Faber, S. M. and Koo, D. C. and Somerville, R. S. and Pérez-Torres, M. A. and Pérez-González, P. and Huang, J.-S. and Davis, M. and Guhathakurta, P. and Barmby, P. and Conselice, C. J. and Lozano, M. and Newman, J. A. and Cooper, M. C.},
    title = {Extragalactic background light inferred from AEGIS galaxy-SED-type fractions},
    journal = {Monthly Notices of the Royal Astronomical Society},
    volume = {410},
    number = {4},
    pages = {2556-2578},
    year = {2011},
    month = {01},
    issn = {0035-8711},
    doi = {10.1111/j.1365-2966.2010.17631.x},
    url = {https://doi.org/10.1111/j.1365-2966.2010.17631.x},
    eprint = {https://academic.oup.com/mnras/article-pdf/410/4/2556/6295256/mnras0410-2556.pdf},
}

@article{Fixsen2009,
doi = {10.1088/0004-637X/707/2/916},
url = {https://doi.org/10.1088/0004-637X/707/2/916},
year = {2009},
month = {nov},
publisher = {The American Astronomical Society},
volume = {707},
number = {2},
pages = {916},
author = {Fixsen, D. J.},
title = {THE TEMPERATURE OF THE COSMIC MICROWAVE BACKGROUND},
journal = {The Astrophysical Journal},
}

@article{Gould1967,
  title = {Opacity of the Universe to High-Energy Photons},
  author = {Gould, Robert J. and Schr\'eder, G\'erard P.},
  journal = {Phys. Rev.},
  volume = {155},
  issue = {5},
  pages = {1408--1411},
  numpages = {0},
  year = {1967},
  month = {Mar},
  publisher = {American Physical Society},
  doi = {10.1103/PhysRev.155.1408},
  url = {https://link.aps.org/doi/10.1103/PhysRev.155.1408}
}

@article{Draine2003,
   author = "Draine, B.T.",
   title = "Interstellar Dust Grains", 
   journal= "Annual Review of Astronomy and Astrophysics",
   year = "2003",
   volume = "41",
   number = "Volume 41, 2003",
   pages = "241-289",
   doi = "https://doi.org/10.1146/annurev.astro.41.011802.094840",
   url = "https://www.annualreviews.org/content/journals/10.1146/annurev.astro.41.011802.094840",
   publisher = "Annual Reviews",
   issn = "1545-4282",
   type = "Journal Article",
  }

@article{Porter_2017,
doi = {10.3847/1538-4357/aa844d},
url = {https://doi.org/10.3847/1538-4357/aa844d},
year = {2017},
month = {aug},
publisher = {The American Astronomical Society},
volume = {846},
number = {1},
pages = {67},
author = {Porter, T. A. and Jóhannesson, G. and Moskalenko, I. V.},
title = {High-energy Gamma Rays from the Milky Way: Three-dimensional Spatial Models for the Cosmic-Ray and Radiation Field Densities in the Interstellar Medium},
journal = {The Astrophysical Journal},
}

@article{Popescu_2017,
    author = {Popescu, C. C. and Yang, R. and Tuffs, R. J. and Natale, G. and Rushton, M. and Aharonian, F.},
    title = {A radiation transfer model for the Milky Way: I. Radiation fields and application to high-energy astrophysics★},
    journal = {Monthly Notices of the Royal Astronomical Society},
    volume = {470},
    number = {3},
    pages = {2539-2558},
    year = {2017},
    month = {05},
    issn = {0035-8711},
    doi = {10.1093/mnras/stx1282},
    url = {https://doi.org/10.1093/mnras/stx1282},
    eprint = {https://academic.oup.com/mnras/article-pdf/470/3/2539/18245516/stx1282.pdf},
}

@article{gammapy,
 author = {{Donath}, Axel and {Terrier}, R\'egis and {Remy}, Quentin and {Sinha}, Atreyee and {Nigro}, Cosimo and {Pintore}, Fabio and {Kh\'elifi}, Bruno and {Olivera-Nieto}, Laura and {Ruiz}, Jose Enrique and
 {Br\"ugge}, Kai and {Linhoff}, Maximilian and {Contreras}, Jose Luis and {Acero}, Fabio and
 {Aguasca-Cabot}, Arnau and {Berge}, David and {Bhattacharjee}, Pooja and {Buchner}, Johannes and
 {Boisson}, Catherine and {Carreto Fidalgo}, David and {Chen}, Andrew and {de Bony de Lavergne}, Mathieu and
 {de Miranda Cardoso}, Jos\'e Vinicius and {Deil}, Christoph and {F\"u\ss{}ling}, Matthias and
 {Funk}, Stefan and {Giunti}, Luca and {Hinton}, Jim and {Jouvin}, L\'ea and {King}, Johannes and
 {Lefaucheur}, Julien and {Lemoine-Goumard}, Marianne and {Lenain}, Jean-Philippe and {L\'opez-Coto}, Rub\'en
 and {Mohrmann}, Lars and {Morcuende}, Daniel and {Panny}, Sebastian and {Regeard}, Maxime and {Saha}, Lab
 and {Siejkowski}, Hubert and {Siemiginowska}, Aneta and {Sip"ocz}, Brigitta M. and {Unbehaun}, Tim
 and {van Eldik}, Christopher and {Vuillaume}, Thomas and {Zanin}, Roberta},
 title = {Gammapy: A Python package for gamma-ray astronomy},
 DOI= "10.1051/0004-6361/202346488",
 url= "https://doi.org/10.1051/0004-6361/202346488",
 journal = {A\&A},
 year = 2023,
 volume = 678,
 pages = "A157",
 }

@article{Dominguez_2013,
doi = {10.1088/0004-637X/770/1/77},
url = {https://doi.org/10.1088/0004-637X/770/1/77},
year = {2013},
month = {may},
publisher = {The American Astronomical Society},
volume = {770},
number = {1},
pages = {77},
author = {Domínguez, A. and Finke, J. D. and Prada, F. and Primack, J. R. and Kitaura, F. S. and Siana, B. and Paneque, D.},
title = {DETECTION OF THE COSMIC γ-RAY HORIZON FROM MULTIWAVELENGTH OBSERVATIONS OF BLAZARS},
journal = {The Astrophysical Journal},
}

@article{Arsioli_2025,
    author = {Arsioli, Bruno and Chang, Yu-Ling and Ighina, Luca},
    title = {Mapping the cosmic gamma-ray horizon: the 1CGH catalogue of Fermi-LAT detections above 10 GeV},
    journal = {Monthly Notices of the Royal Astronomical Society},
    volume = {539},
    number = {2},
    pages = {1458-1470},
    year = {2025},
    month = {05},
    issn = {0035-8711},
    doi = {10.1093/mnras/staf329},
    url = {https://doi.org/10.1093/mnras/staf329},
    eprint = {https://academic.oup.com/mnras/article-pdf/539/2/1458/63010864/staf329.pdf},
}

@article{Vallee_2014,
doi = {10.1088/0067-0049/215/1/1},
url = {https://doi.org/10.1088/0067-0049/215/1/1},
year = {2014},
month = {oct},
publisher = {The American Astronomical Society},
volume = {215},
number = {1},
pages = {1},
author = {Vallée, Jacques P.},
title = {CATALOG OF OBSERVED TANGENTS TO THE SPIRAL ARMS IN THE MILKY WAY GALAXY},
journal = {The Astrophysical Journal Supplement Series},
}

@article{Hou_2014,
	author = {{Hou, L. G.} and {Han, J. L.}},
	title = {The observed spiral structure of the Milky Way},
	DOI= "10.1051/0004-6361/201424039",
	url= "https://doi.org/10.1051/0004-6361/201424039",
	journal = {A\&A},
	year = 2014,
	volume = 569,
	pages = "A125",
	month = "",
}

@article{Protheroe_1996,
title = {A new estimate of the extragalactic radio background and implications for ultra-high-energy γ-ray propagation},
journal = {Astroparticle Physics},
volume = {6},
number = {1},
pages = {45-54},
year = {1996},
issn = {0927-6505},
doi = {https://doi.org/10.1016/S0927-6505(96)00041-2},
url = {https://www.sciencedirect.com/science/article/pii/S0927650596000412},
author = {R.J. Protheroe and P.L. Biermann},
}

@article{hess_gc_pevatron,
    author = "Abramowski, A. and others",
    collaboration = "H.E.S.S.",
    title = "{Acceleration of petaelectronvolt protons in the Galactic Centre}",
    eprint = "1603.07730",
    archivePrefix = "arXiv",
    primaryClass = "astro-ph.HE",
    doi = "10.1038/nature17147",
    journal = "Nature",
    volume = "531",
    pages = "476",
    year = "2016"
}

@article{Kachelriess2025,
    author = "Kachelriess, M. and Lammert, E.",
    title = "{Cygnus X-3 as a PeVatron and the LHAASO 2025 data}",
    eprint = "2512.18786",
    archivePrefix = "arXiv",
    primaryClass = "astro-ph.HE",
    month = "12",
    year = "2025"
}

@article{Zdziarski2026,
    author = "Zdziarski, Andrzej A. and Dmytriiev, Anton and Koljonen, Karri I. I.",
    title = "{The counterjet dominates the production of PeV photons from Cyg X-3}",
    eprint = "2603.27805",
    archivePrefix = "arXiv",
    primaryClass = "astro-ph.HE",
    month = "3",
    year = "2026"
}

@ARTICLE{Giacconi_1967,
       author = {{Giacconi}, R. and {Gorenstein}, P. and {Gursky}, H. and {Waters}, J.~R.},
        title = "{An X-Ray Survey of the Cygnus Region}",
      journal = {\apjl},
         year = 1967,
        month = jun,
       volume = {148},
        pages = {L119},
          doi = {10.1086/180028},
       adsurl = {https://ui.adsabs.harvard.edu/abs/1967ApJ...148L.119G} 
}

@ARTICLE{Becklin_1973,
       author = {{Becklin}, E.~E. and {Neugebauer}, G. and {Hawkins}, F.~J. and {Mason}, K.~O. and {Sanford}, P.~W. and {Matthews}, K. and {Wynn-Williams}, C.~G.},
        title = "{Infrared and X-ray Variability of Cyg X-3}",
      journal = {\nat},
         year = 1973,
        month = oct,
       volume = {245},
       number = {5424},
        pages = {302-304},
          doi = {10.1038/245302a0},
       adsurl = {https://ui.adsabs.harvard.edu/abs/1973Natur.245..302B}
}

@ARTICLE{Molnar_1984,
       author = {{Molnar}, L.~A. and {Reid}, M.~J. and {Grindlay}, J.~E.},
        title = "{Low-level radio flares from Cygnus X-3}",
      journal = {\nat},
         year = 1984,
        month = aug,
       volume = {310},
       number = {5979},
        pages = {662-665},
          doi = {10.1038/310662a0},
       adsurl = {https://ui.adsabs.harvard.edu/abs/1984Natur.310..662M},
}

@ARTICLE{Kerkwijk_1992,
       author = {{van Kerkwijk}, M.~H. and {Charles}, P.~A. and {Geballe}, T.~R. and {King}, D.~L. and {Miley}, G.~K. and {Molnar}, L.~A. and {van den Heuvel}, E.~P.~J. and {van der Klis}, M. and {van Paradijs}, J.},
        title = "{Infrared helium emission lines from Cygnus X-3 suggesting a Wolf-Rayet star companion}",
      journal = {\nat},
         year = 1992,
        month = feb,
       volume = {355},
       number = {6362},
        pages = {703-705},
          doi = {10.1038/355703a0},
       adsurl = {https://ui.adsabs.harvard.edu/abs/1992Natur.355..703V},
}

@ARTICLE{Tavani_2009,
       author = {{Tavani}, M. and {Bulgarelli}, A. and {Piano}, G. and {Sabatini}, S. and {Striani}, E. and {Evangelista}, Y. and {Trois}, A. and {Pooley}, G. and {Trushkin}, S. and {Nizhelskij}, N.~A. and {McCollough}, M. and {Koljonen}, K.~I.~I. and {Pucella}, G. and {Giuliani}, A. and {Chen}, A.~W. and {Costa}, E. and {Vittorini}, V. and {Trifoglio}, M. and {Gianotti}, F. and {Argan}, A. and {Barbiellini}, G. and {Caraveo}, P. and {Cattaneo}, P.~W. and {Cocco}, V. and {Contessi}, T. and {D'Ammando}, F. and {Del Monte}, E. and {de Paris}, G. and {Di Cocco}, G. and {di Persio}, G. and {Donnarumma}, I. and {Feroci}, M. and {Ferrari}, A. and {Fuschino}, F. and {Galli}, M. and {Labanti}, C. and {Lapshov}, I. and {Lazzarotto}, F. and {Lipari}, P. and {Longo}, F. and {Mattaini}, E. and {Marisaldi}, M. and {Mastropietro}, M. and {Mauri}, A. and {Mereghetti}, S. and {Morelli}, E. and {Morselli}, A. and {Pacciani}, L. and {Pellizzoni}, A. and {Perotti}, F. and {Picozza}, P. and {Pilia}, M. and {Prest}, M. and {Rapisarda}, M. and {Rappoldi}, A. and {Rossi}, E. and {Rubini}, A. and {Scalise}, E. and {Soffitta}, P. and {Vallazza}, E. and {Vercellone}, S. and {Zambra}, A. and {Zanello}, D. and {Pittori}, C. and {Verrecchia}, F. and {Giommi}, P. and {Colafrancesco}, S. and {Santolamazza}, P. and {Antonelli}, A. and {Salotti}, L.},
        title = "{Extreme particle acceleration in the microquasar CygnusX-3}",
      journal = {\nat},
         year = 2009,
        month = dec,
       volume = {462},
       number = {7273},
        pages = {620-623},
          doi = {10.1038/nature08578},
archivePrefix = {arXiv},
       eprint = {0910.5344},
 primaryClass = {astro-ph.HE},
       adsurl = {https://ui.adsabs.harvard.edu/abs/2009Natur.462..620T},
}

@ARTICLE{Mioduszewski_2001,
       author = {{Mioduszewski}, Amy J. and {Rupen}, Michael P. and {Hjellming}, Robert M. and {Pooley}, Guy G. and {Waltman}, Elizabeth B.},
        title = "{A One-sided Highly Relativistic Jet from Cygnus X-3}",
      journal = {\apj},
         year = 2001,
        month = jun,
       volume = {553},
       number = {2},
        pages = {766-775},
          doi = {10.1086/320965},
archivePrefix = {arXiv},
       eprint = {astro-ph/0102018},
 primaryClass = {astro-ph},
       adsurl = {https://ui.adsabs.harvard.edu/abs/2001ApJ...553..766M},
}

@ARTICLE{Marti_2001,
       author = {{Mart{\'\i}}, J. and {Paredes}, J.~M. and {Peracaula}, M.},
        title = "{Development of a two-sided relativistic jet in Cygnus X-3}",
      journal = {\aap},
         year = 2001,
        month = aug,
       volume = {375},
        pages = {476-484},
          doi = {10.1051/0004-6361:20010907},
       adsurl = {https://ui.adsabs.harvard.edu/abs/2001A&A...375..476M}
}

@ARTICLE{MJ_2004,
       author = {{Miller-Jones}, James C.~A. and {Blundell}, Katherine M. and {Rupen}, Michael P. and {Mioduszewski}, Amy J. and {Duffy}, Peter and {Beasley}, Anthony J.},
        title = "{Time-sequenced Multi-Radio Frequency Observations of Cygnus X-3 in Flare}",
      journal = {\apj},
         year = 2004,
        month = jan,
       volume = {600},
       number = {1},
        pages = {368-389},
          doi = {10.1086/379706},
archivePrefix = {arXiv},
       eprint = {astro-ph/0311277},
 primaryClass = {astro-ph},
       adsurl = {https://ui.adsabs.harvard.edu/abs/2004ApJ...600..368M}
}

@dataset{ctao_sens,
  author       = {Cherenkov Telescope Array Observatory and
                  Cherenkov Telescope Array Consortium},
  title        = {CTAO Instrument Response Functions - prod5 version
                   v0.1
                  },
  month        = sep,
  year         = 2021,
  publisher    = {Zenodo},
  version      = {v0.1},
  doi          = {10.5281/zenodo.5499840},
  url          = {https://doi.org/10.5281/zenodo.5499840},
}

@article{porter_2018_pev,
  title = {Galactic PeVatrons and helping to find them: Effects of galactic absorption on the observed spectra of very high energy $\ensuremath{\gamma}$-ray sources},
  author = {Porter, T. A. and Rowell, G. P. and J\'ohannesson, G. and Moskalenko, I. V.},
  journal = {Phys. Rev. D},
  volume = {98},
  issue = {4},
  pages = {041302(R)},
  numpages = {6},
  year = {2018},
  month = {Aug},
  publisher = {American Physical Society},
  doi = {10.1103/PhysRevD.98.041302},
  url = {https://link.aps.org/doi/10.1103/PhysRevD.98.041302}
}

@ARTICLE{Aharonian_1994,
       author = {{Aharonian}, F.~A. and {Coppi}, P.~S. and {Voelk}, H.~J.},
        title = "{Very High Energy Gamma Rays from Active Galactic Nuclei: Cascading on the Cosmic Background Radiation Fields and the Formation of Pair Halos}",
      journal = {\apjl},
         year = 1994,
        month = mar,
       volume = {423},
        pages = {L5},
          doi = {10.1086/187222},
archivePrefix = {arXiv},
       eprint = {astro-ph/9312045},
 primaryClass = {astro-ph},
       adsurl = {https://ui.adsabs.harvard.edu/abs/1994ApJ...423L...5A}
}

@article{DiMarco2025,
  title = {Revisiting the propagation of highly-energetic gamma rays in the Galaxy},
  author = {Di Marco, Gaetano and Alves Batista, Rafael and S\'anchez-Conde, Miguel A.},
  journal = {Phys. Rev. D},
  volume = {111},
  issue = {8},
  pages = {083004},
  numpages = {16},
  year = {2025},
  month = {Apr},
  publisher = {American Physical Society},
  doi = {10.1103/PhysRevD.111.083004},
  url = {https://link.aps.org/doi/10.1103/PhysRevD.111.083004}
}

@article{Elyiv2009,
  title = {Gamma-ray induced cascades and magnetic fields in the intergalactic medium},
  author = {Elyiv, A. and Neronov, A. and Semikoz, D. V.},
  journal = {Phys. Rev. D},
  volume = {80},
  issue = {2},
  pages = {023010},
  numpages = {11},
  year = {2009},
  month = {Jul},
  publisher = {American Physical Society},
  doi = {10.1103/PhysRevD.80.023010},
  url = {https://link.aps.org/doi/10.1103/PhysRevD.80.023010}
}

@article{Zhang2006,
 author = {{Zhang, J.-L.} and {Bi, X.-J.} and {Hu, H.-B.}},
 title = {Very high energy gamma-ray absorption by the galactic interstellar radiation field},
 DOI= "10.1051/0004-6361:20054422",
 url= "https://doi.org/10.1051/0004-6361:20054422",
 journal = {A\&A},
 year = 2006,
 volume = 449,
 number = 2,
 pages = "641-643",
}

%% This command is needed to show the entire author+affiliation list when
%% the collaboration and author truncation commands are used.  It has to
%% go at the end of the manuscript.\textbf{}
%\allauthors

%% Include this line if you are using the \added, \replaced, \deleted
%% commands to see a summary list of all changes at the end of the article.
%\listofchanges

\end{document}